\documentclass[fleqn,usenatbib]{mnras}

\usepackage[T1]{fontenc}

\DeclareRobustCommand{\VAN}[3]{#2}
\let\VANthebibliography\thebibliography
\def\thebibliography{\DeclareRobustCommand{\VAN}[3]{##3}\VANthebibliography}

\usepackage{graphicx}	% Including figure files
\usepackage{amsmath}	% Advanced maths commands
\usepackage{amssymb}	% Extra maths symbols
\usepackage[dvipsnames]{xcolor} 

\newcommand{\ud}{\mathrm{d}}

\title[Magnetorotational Explosions]{Three dimensional magnetorotational core-collapse supernova explosions of a 39 solar mass progenitor star}

\author[Powell et al.]{
Jade Powell,$^{1}$\thanks{E-mail: dr.jade.powell@gmail.com}
Bernhard M\"uller,$^{2}$
David R. Aguilera-Dena,$^{3}$
Norbert Langer$^{4}$$^{5}$
\\
$^{1}$ Centre for Astrophysics and Supercomputing, Swinburne University of Technology, Hawthorn, VIC 3122, Australia.\\
$^{2}$ School of Physics and Astronomy, Monash University, VIC 3800, Australia. \\
$^{3}$ Department of Physics, Institute of Astrophysics, FORTH, University of Crete, Voutes, University Campus, GR-71003 Heraklion, Greece.\\
$^{4}$ Argelander-Institut für Astronomie, Universität Bonn, Auf dem H\"ugel 71, 53121, Bonn, Germany. \\
$^{5}$ Max-Planck-Institut für Radioastronomie, Auf dem H\"ugel 69, 53121, Bonn, 
Germany.
}

\pubyear{2022}

\begin{document}
\label{firstpage}
\pagerange{\pageref{firstpage}--\pageref{lastpage}}
\maketitle

\begin{abstract}
We perform three-dimensional simulations of magnetorotational supernovae  using a $39\,M_{\odot}$ progenitor star with two different initial magnetic field strengths of $10^{10}$\,G and $10^{12}$\,G in the core. Both models rapidly undergo shock revival and their  explosion energies asymptote within a few hundred milliseconds to values of $\gtrsim 2\times10^{51}$\,erg after conservatively correcting for
the binding energy of the envelope.
Magnetically collimated, non-relativistic jets form in both models, though the jets are subject to non-axisymmetric instabilities. The jets do not appear crucial for driving the explosion, as they only emerge once the shock has already expanded considerably. Our simulations predict moderate neutron star kicks of about $150\, \mathrm{km}\,\mathrm{s}^{-1}$, no spin-kick alignment, and rapid early spin-down that would result in birth periods of about $20\, \mathrm{ms}$, too slow to power an energetic gamma-ray burst jet. More than $0.2\,M_\odot$ of iron-group material are ejected, but we estimate that the mass of ejected $^{56}\mathrm{Ni}$ will be considerably smaller as the bulk of this material is neutron-rich. Explosive burning does not contribute appreciable amounts of $^{56}\mathrm{Ni}$ because the burned material originates from the slightly neutron-rich silicon shell. The iron-group ejecta also show no pronounced bipolar geometry by the end of the simulations. The models thus do not immediately fit the characteristics of observed hypernovae, but may be representative of other transients with moderately high explosion energies. The gravitational-wave emission reaches high frequencies of up to 2000\,Hz and amplitudes of over 100\,cm. The gravitational-wave emission is detectable out to distances of $\sim4$\,Mpc in the planned Cosmic Explorer detector. 
\end{abstract}


%%%%%%%%%%%%%%%%%%%%%%%%%%%%%%%%%%%%%%%%%%%%%%%%%%
%%%%%%%%%%%%%%%%% BODY OF PAPER %%%%%%%%%%%%%%%%%%

\section{Introduction}

Core-collapse supernovae (CCSNe), the explosions of stars massive enough to develop an iron core, are some of the most violent events in the Universe. They result in a brilliant optical outburst that can reach the brightness of an entire galaxy for weeks, and they emit an enormous amount of energy, several $10^{53}$\,erg, in the form of neutrinos as well as $\sim10^{46}$\,erg in the form of gravitational waves (GWs), making them a perfect target for multi-messenger astronomy.

In regular CCSNe, shock revival is likely achieved by neutrino heating.
In recent years, significant progress has been made in 3D modelling of non-rotating neutrino-driven CCSNe, which has considerably enhanced our understanding of CCSN explosion properties, multi-messenger observables, and neutron star and black hole birth masses, spins and kicks (see recent reviews \citealt{kalogera_19, 2020arXiv201004356A,mueller_20,burrows_21}). However, even though the neutrino-driven mechanism can explain explosions with ejecta kinetic energies up to $\mathord{\sim}10^{51}$\,erg, there is a class of considerably more energetic ``hypernovae'' with energies of about $10^{52}$\,erg; observationally these are classified as broad-lined Ic supernovae (Type Ic-BL; \citealp{woosley_06,gal-yam_17}). These explosions, which sometimes are accompanied by long gamma-ray bursts (GRB), are likely explained by some form of magnetohydrodynamic (MHD) mechanism that taps the energy of a rapidly spinning millisecond magnetar
\citep{bisnovatyi_76,duncan_92,usov_92,akiyama_03}
or a black hole accretion disk in the collapsar scenario
\citep{macfadyen_99}.

Hypernovae are of interest beyond supernova theory. Since the hypernova rate is significantly higher at low metallacity, the explosions may have played a key role for the enrichment of the Milky Way with heavy elements in its early history as major production sites for elements like zinc \citep{nomoto_06,kobayashi_06} and possible rapid-neutron capture process elements \citep{2012ApJ...750L..22W,moesta_18,grimmett_21,reichert_22}. 

MHD explosions with detailed neutrino transport have been investigated in a large number of 2D simulations \citep{2007ApJ...664..416B, obergaulinger_17, 2021MNRAS.tmp...72R, 2021MNRAS.501.2764G, Jardine_2021}. However, the assumption of 2D axisymmetry is a significant concern for accurate explosion dynamics and nucleosynthesis predictions, as it it prohibits the development of any kink instability of MHD jets \citep{moesta_14b} and also inhibits  Kelvin-Helmholtz instability between outflows and downflows at high Mach number.

In recent years, 3D MHD simulations of  magnetorotational explosions have become available \citep{2012ApJ...750L..22W, moesta_14b, moesta_18, 2020ApJ...896..102K, 2020MNRAS.492.4613O, 2020ApJ...896..102K,bugli_21}. However, few of the existing 
3D MHD models (some with relatively coarse spatial resolution) include detailed multi-group neutrino transport as yet \citep{2020ApJ...896..102K,2020MNRAS.492.4613O,bugli_21}, which is critical for correctly predicting the composition of the magnetically-driven outflows. Although a first nucleosynthesis study based on 3D hypernova models with detailed neutrino transport is now available
\citep{reichert_22}, further simulations are desirable. 
A better understanding of seed fields in the progenitor is also needed to fully understand magnetorotational explosions. Currently, 3D progenitor models with magnetic fields are only available for non-rotating progenitors \citep{2021arXiv210100213V}. A few studies of the impact of different magnetic field configurations have also been carried out by \citet{2020MNRAS.492...58B, 10.1093/mnras/staa3273}.   

From the perspective of gravitational-wave (GW) astronomy,
it is noteworthy that the majority of the recent long-duration 2D and 3D CCSN MHD simulations have not discussed predictions of the GW emission during the post-bounce phase with the exception of the 2D study of \citet{Jardine_2021}. 
In recent years, the Advanced LIGO \citep{2015CQGra..32g4001L} and Virgo \citep{2015CQGra..32b4001A} GW detectors have detected GWs from compact binary sources in increasing numbers \citep{2019PhRvX...9c1040A, 2020arXiv201014527A, 2021arXiv211103606T}. As the detectors sensitivity increases over the coming years, the chance of detecting GWs from another source, like a CCSN, significantly increases.  

Different from ordinary CCSNe, the GW signals from MHD explosions may be loud enough to be detected outside the Milky Way by current generation detectors \citep{  2016PhRvD..93d2002G, 2020PhRvD.101h4002A, Szczepanczyk_2021}. However, those detection studies only include the core-bounce part of MHD signals, as longer duration GW signals from MHD simulations were not available at the time, resulting in their detection distances being only a lower limit. The GW core-bounce signal has been extensively studied and may inform us of the progenitor rotation rate and nuclear equation of state properties  \citep{obergaulinger_06, 2008PhRvD..78f4056D, Scheidegger2008, takiwaki_11, 2014PhRvD..90d4001A, 2015MNRAS.450..414F, 2017PhRvD..95f3019R, 2021PhRvD.103b4025E}. However, the full GW signal in the MHD case must be predicted in order for us to perform more accurate detection studies and understand the relationship between the GW  signal morphology and the astrophysics of the source. 2D simulations \citep{Jardine_2021} have provided a first glimpse at the GW signal from the post-bounce phase over several hundreds of milliseconds, but suffer from the inherent limitations as discused before.
On even longer time scales, anelastic 3D simulations have addressed GW emission due to PNS convection and revealed interesting signal features \citep{2021arXiv210312445R}. 
The GW emission from the very dynamical early explosion phase of magnetorotational CCSNe also needs to be studied based on 3D MHD simulations of the first few hundred milliseconds after shock revival.

There are already cogent reasons to anticipate that the GW emission from the post-bounce phase of magnetorotational explosions is quantitatively important and may differ in important respects from the post-bounce GW emission in neutrino-driven CCSNe.
In the GW signatures of 3D neutrino-driven explosion models, common features have been identified and linked to the explosion and PNS properties through quantitative relations for their characteristics frequencies. The GW f/g-mode frequency \citep{mueller_13,2018ApJ...861...10M,torres_forne_18,sotani_16}, which produces a conspicuous emission band at a few $100\, \mathrm{Hz}$ to about $1000\, \mathrm{Hz}$ , can be described by universal relations that depend on the mass and frequency of the PNS \citep{torres-forne_universal_2019}. Similarly, low-frequency emission in models with strong activity of the standing accretion shock instability \citep[SASI:][]{0004-637X-584-2-971, 2006ApJ...642..401B, 2007ApJ...654.1006F} will reflect the SASI frequency, which is determined by the shock radius and the radius of the PNS \citep{2014ApJ...788...82M}.
3D simulations have already demonstrated
that rotation alone already quantitatively affects
the post-bounce GW emission \citep{andresen_19, pajkos_19, 2014PhRvD..89d4011K, shibagaki_20,2020MNRAS.494.4665P} and
can break the familiar relation between
neutron star parameters and the f/g-mode frequency
\citep{2020MNRAS.494.4665P}. For rapid
rotation, qualitatively new emission features
from non-axisymmetric instabilities can arise
\citep{2014PhRvD..89d4011K, shibagaki_20}. 
When both rotation and magnetic fields are taken
into account, \citet{Jardine_2021} found a significant impact on the time-frequency structure of the GW signal both for models with rapid rotation or strong initial field strengths in a small 2D parameter study. 
As the next step, 3D simulations are needed
to thoroughly understand the effect of rotation and magnetic fields on the post-bounce GW emission and to potentially identify new signal features.

In this paper, we perform two 3D simulations using the  MHD version of the \textsc{CoCoNuT-FMT} code with a pseudo-relativistic potential to mimic the effects of relativistic gravity. We simulate magnetorotational explosions of a rapidly rotating $39\,M_{\odot}$ progenitor star  \citep{2018ApJ...858..115A}
with two different magnetic field strengths ($10^{12}$\,G and $10^{10}$\,G at the centre of the star). We compare the results of our models to the previous 3D hydrodynamic simulation of the $39\,M_{\odot}$ progenitor star  without magnetic fields in \citet{2020MNRAS.494.4665P}, and the 2D simulations of the same $39\,M_{\odot}$ progenitor model in \citet{Jardine_2021}. 

In Section~\ref{sec:sim}, we give a brief description of the progenitor model and the MHD code used in our simulations. In Section~\ref{sec:dynamics}, we describe the impact of the rotation and magnetic fields on the explosion dynamics and jets. 
The remnant properties are discussed in Section~\ref{sec:pns} with a particular view to the millisecond-magnetar scenario for long gamma-ray bursts. We study the ejecta composition and geometry of our models in Section~\ref{sec:ejecta}. 
In Section~\ref{sec:grav_waves}, we discuss the GW emission and the prospects for detection. A summary and conclusions are given in Section~\ref{sec:conclusion}. 

%%%%%%%%%%%%%%%%%%%%%%%%%%%%%%%%%%%%%%%%%%%%%%%%%%%
%%%%%%%%%%%%%%%%%%%%%%%%%%%%%%%%%%%%%%%%%%%%%%%%%%%
\section{Progenitor model and simulation methodology}
\label{sec:sim} 

We perform two 3D simulations of the B-series model with an initial mass of $39\,M_{\odot}$ from  \citet{2018ApJ...858..115A} evolved using the stellar evolution code \textsc{MESA} \citep{2011ApJS..192....3P,paxton_13,paxton_15,paxton_18}. 
Similar models are also presented in \citep{aguilera_20}.
Due to rapid rotation, this model evolved quasi-chemically homogeneously
and developed a large CO core of $\mathord{\sim}25\, M_\odot$.
Mass loss reduces the pre-collapse mass of the progenitor star to $22.05\,M_{\odot}$.
It is rapidly rotating with an initial surface rotational velocity of $600\,\mathrm{km\,s}^{-1}$, 
a pre-collapse core rotation rate (angular velocity) of
$0.54\, \mathrm{rad}\, \mathrm{s}^{-1}$, and a low metallicity of $1/50\,Z_{\odot}$. Despite its large CO core mass, the progenitor only has a 
moderately high
pre-collapse compactness parameter \citep{oconnor_11} of
$\xi_{2.5}=0.36$.

The $39\,M_{\odot}$ progenitor model was previously simulated in 3D by \citet{2020MNRAS.494.4665P} with the same rotation profile, but neglecting magnetic fields. Their simulation was performed with the general relativistic version of the neutrino hydrodynamics code CoCoNuT-FMT \citep{M_ller_2010,2015MNRAS.448.2141M}. The model developed a strong neutrino-driven explosion without the aid of magnetic fields. Where relevant, we include some of the results from this non-magnetic model in our paper for a  comparison to our new models. We label this reference model m39\_B0. 

In our two new 3D simulations, we add initial magnetic fields to the model. We impose a dipolar field with two different field strengths, peaking at $10^{12}$\,G and $10^{10}$\,G for both the poloidal and toroidal component at the centre of the star. Following  \citet{Suwa2007,Varma2021}, the field is derived from the vector potential
$\mathbf{A}$ 
\begin{align}
\label{eq:magini}
    \left(A^{r}, A^{\theta}, A^{\phi}\right)=
\frac{1}{2\left(r^{3}+r_{0}^{3}\right)}
    \left(B_0 r_{0}^{3} r \cos \theta, 0, B_0 r_{0}^{3} r \sin \theta\right),
\end{align}
where $B_0$ is the maximum poloidal and toroidal field strength, and the radius parameter is set to $r_0=10^8\,\mathrm{cm}$. We refer to the models with $B_0=10^{12}\,\mathrm{G}$ and $B_0=10^{10}\,\mathrm{G}$ as m39\_B12 and m39\_B10, respectively. 
The dynamo model used for the progenitor evolution predicts toroidal field
strengths of up to $10^{11}\, \mathrm{G}$
and poloidal field strengths of
up $10^{10}\, \mathrm{G}$ at mass
coordinates of $1.8\texttt{-}2\,M_\odot$. Throughout most of the core, however,
the field strengths are significantly lower
with typical values of 
$10^{9}\texttt{-}10^{10}\, \mathrm{G}$ for the toroidal field and as low as
$10^6 \, \mathrm{G}$ for the poloidal field.
The choice of relatively large fields with a higher poloidal-to-toroidal field ratio in the simulations is motivated by several considerations. The high numerical
resolution required to self-consistently follow
the emergence of the requisite strong poloidal field
from realistic, toroidally-dominated progenitor fields 
for jet-driven explosions by post-collapse amplification
processes \citep{moesta_15} 
is not currently affordable
in global long-time simulations with neutrino transport.
It is therefore common practice to mimic the assumed strong-field conditions resulting from dynamo processes by
imposing strong seed fields as initial conditions. Second,
the simple dynamo models in spherically symmetric stellar evolution codes only provide effective mean-field values for the poloidal and toroidal field, but not the detailed field geometry needed for 3D simulations
(although more sophisticated models of magnetorotational stellar evolution may provide more information, see
\citealp{takahashi_21}). Finally, 3D simulations of magnetoconvection in CCSN progenitors suggest
that current stellar evolution models may underestimate
magnetic field strengths \citep{2021arXiv210100213V}.
Given the uncertainties about the magnetic fields in the progenitor stars and post-collapse field amplification processes, the current models remain scenarios for the case 
of rapid explosions due to fast field amplification or strong progenitor fields, rather than genuine predictions based on first principles.

The m39\_B12 model was simulated up to 0.68\,s after core bounce, and the m39\_B10 model was simulated up to 0.33\,s after core bounce due to less available computater time. The collapse was followed in axisymmetry, and the models were mapped to 3D shortly after bounce.

We perform our simulations using the MHD version of \textsc{CoCoNuT-FMT} as described in \citet{2020MNRAS.498L.109M}. The code solves the Newtonian MHD equations using the HLLC solver \citep{Gurski_2004, 2005JCoPh.208..315M} and hyperbolic divergence cleaning \citep{2002JCoPh.175..645D}. 
The  MHD equations 
for the density $\rho$, magnetic field $\mathbf{B}$, total energy density $e$, velocity $\mathbf{v}$ and Lagrangian multiplier $\psi$ are expressed in Equations~\eqref{eq:cont}--\eqref{eq:divclean}
in Gaussian units including divergence
cleaning terms as
\begin{align}
\partial_{t} \rho+\nabla \cdot(\rho \mathbf{v}) &=0,   \label{eq:cont}\\
\label{eq:mhd2}
\partial_{t}(\rho \mathbf{v})+\nabla \cdot\left[\rho \mathbf{v} \mathbf{v}^{\mathrm{T}}+\left(P+\frac{\mathbf{B}^{2}}{8\pi}\right) \mathcal{I}-\frac{\mathbf{B} \mathbf{B}^{\mathrm{T}}}{4\pi}\right] &=\rho \mathbf{g}+\mathbf{Q}_\mathrm{m}
-\frac{(\nabla \cdot \mathbf{B}) \mathbf{B}}{4\pi}, \\
\partial_{t} \mathbf{B}+\nabla \cdot\left(\mathbf{v} \mathbf{B}^{\mathrm{T}}-\mathbf{B} \mathbf{v}^{\mathrm{T}}+\psi \mathcal{I}\right) &=0,  \\
\label{eq:mhd4}
\partial_{t} e+\nabla \cdot\left[\left(e+P+\frac{\mathbf{B}^{2}}{8\pi} \right) \mathbf{v}-\mathbf{B}(\mathbf{v} \cdot \mathbf{B})\right] &=
\rho \mathbf{v}\cdot \mathbf{g}+Q_\mathrm{e}\\
&+\mathbf{Q}_\mathrm{m}\cdot\mathbf{v}
-\frac{\mathbf{B} \cdot \nabla (c_\mathrm{h} \psi)}{4\pi},  
\nonumber
\\
\partial_{t} \psi+c_\mathrm{h} \nabla \cdot \mathbf{B} &=-\frac{\psi}{\tau} . 
\label{eq:divclean}
\end{align}
Here $c_\mathrm{h}$ denotes the hyperbolic cleaning speed, $P$ the gas pressure, $\tau$ the damping time for
the Lagrangian multiplier, and $Q_\mathrm{e}$ and $\mathbf{Q}_\mathrm{m}$ are the neutrino energy
and momentum source terms.
The cleaning speed $c_\mathrm{h}$ is identified with the fast magnetosonic velocity, and the damping
time is set to eight times the magnetosonic crossing time of a cell.
The effective potential of \citet{mueller_08} is used to approximate
the effects of relativistic gravity.
In comparing to the reference model m39\_B0, the use of a pseudo-Newtonian treatment as opposed
to the general relativistic reference
simulation of model m39\_B0 can lead to
small systematic differences, e.g.,
in proto-neutron star structure and hence in global proto-neutron star parameters like its
total mass.

 Neutrinos are treated using the \textsc{FMT} (Fast Multi-group Transport) method of \citet{2015MNRAS.448.2141M}. 
 The effects of relativistic gravity are approximately taken into account by means of a modified gravitational potential \citep{mueller_08}, and are also included as gravitational redshift in the neutrino transport.
 The GW emission is extracted by the time-integrated quadrupole formula \citep{finn_89, finn_90, blanchet_90}. 

The models have a grid resolution of $550\times 128 \times 256$  zones in radius, latitude and longitude. The grid reaches out to $10^5$\,km, and 21 energy zones are used in the transport solver. To save computer time, we follow the collapse phase in axisymmetry (2D) and map to 3D shortly after bounce, imposing small random seed perturbation to trigger the growth of non-axisymmetric modes.

We use the \citet{1991NuPhA.535..331L} equation of state, with a bulk incompressibility parameter of K=220\,MeV. At low densities, we use an equation of state accounting for photons, electrons, positrons and an ideal gas of nuclei together with a flashing treatment for nuclear reactions \citep{2002A&A...396..361R}.
The flashing scheme burns $^{12}\mathrm{C}$ to $^{24}\mathrm{Mg}$ above temperatures of $2.5\, \mathrm{GK}$; $^{16}\mathrm{Mg}$, $^{20}\mathrm{Ne}$ and $^{24}\mathrm{Mg}$
to $^{28}\mathrm{Si}$ above $2.5\, \mathrm{GK}$ and nuclei below the
iron group to $^{56}\mathrm{Ni}$ above $4.5\, \mathrm{GK}$. Above $5.8\, \mathrm{GK}$
(i.e., $0.5\,\mathrm{MeV}$), the code uses a nuclear statistical equilibrium (NSE) table comprising neutron, protons,
$^{4}\mathrm{He}$,
$^{12}\mathrm{C}$,
$^{16}\mathrm{O}$,
$^{20}\mathrm{Ne}$,
$^{24}\mathrm{Mg}$,
$^{28}\mathrm{Si}$,
$^{32}\mathrm{S}$,
$^{36}\mathrm{Ar}$,
$^{40}\mathrm{Ca}$,
$^{44}\mathrm{Ti}$,
$^{48}\mathrm{Cr}$,
$^{54}\mathrm{Mn}$,
$^{56}\mathrm{Fe}$,
$^{60}\mathrm{Fe}$,
$^{56}\mathrm{Ni}$,
and (as representative ``dummy'' nuclei for
very neutron-rich conditions) 
$^{70}\mathrm{70}$ and
$^{200}\mathrm{200}$ \citep{Buras2006b}. The inclusion of several iron-group nuclei with a different neutron-to-proton-ration
allows us to roughly estimate whether ejeected iron-group elements are likely in the form of
$^{56}\mathrm{Ni}$ (for electron fraction 
$Y_\mathrm{e}\gtrsim 0.49$, \citealp{hartmann_85,wanajo_18}) or in the form
of more neutron-rich nuclei.

%%%%%%%%%%%%%%%%%%%%%%%%%%%%%%%%%%%%%%%%%%%%%%%%%%%
%%%%%%%%%%%%%%%%%%%%%%%%%%%%%%%%%%%%%%%%%%%%%%%%%%%
\section{Explosion dynamics and morphology}
\label{sec:dynamics} 

\begin{figure*}
\includegraphics[width=\columnwidth]{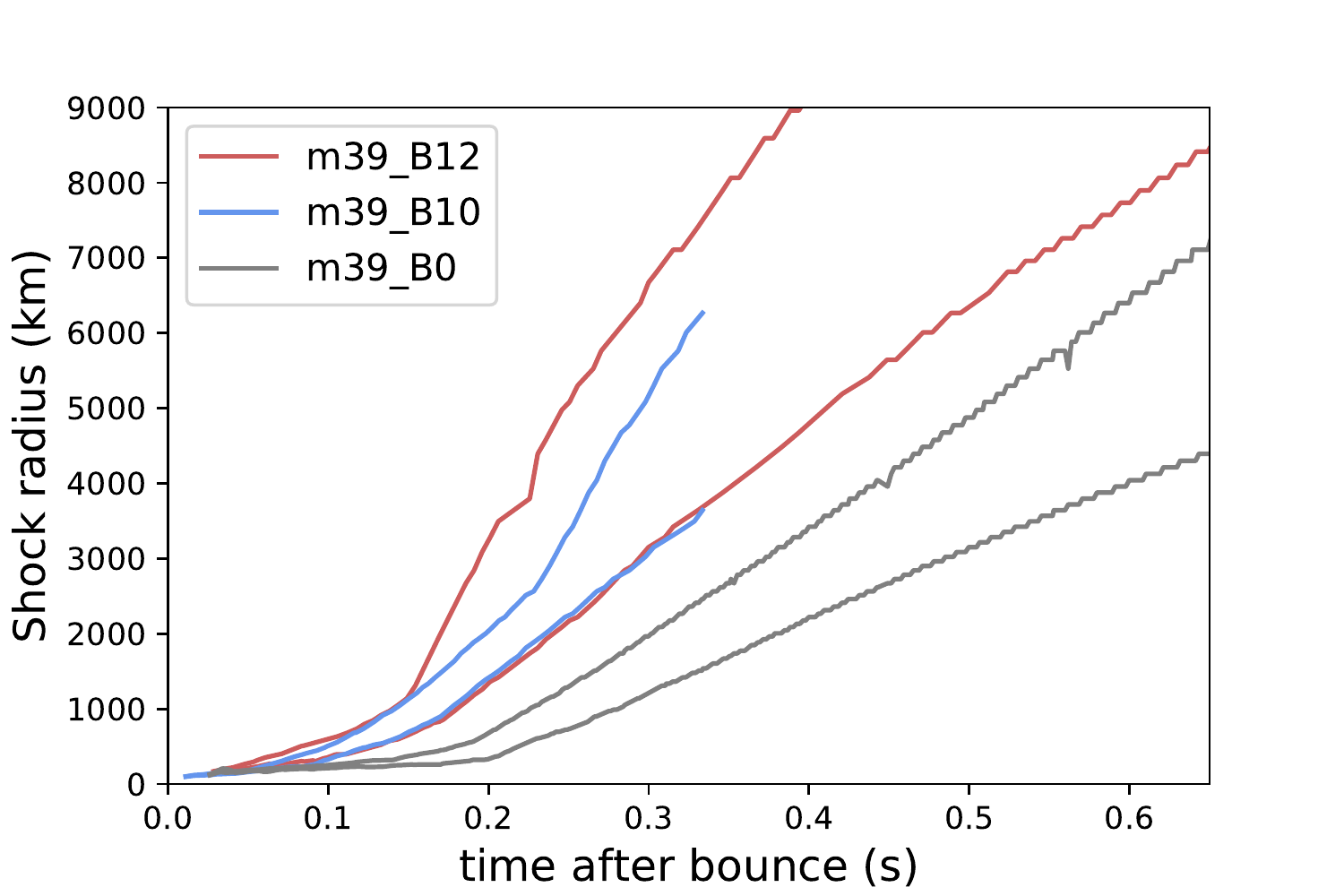}
\includegraphics[width=\columnwidth]{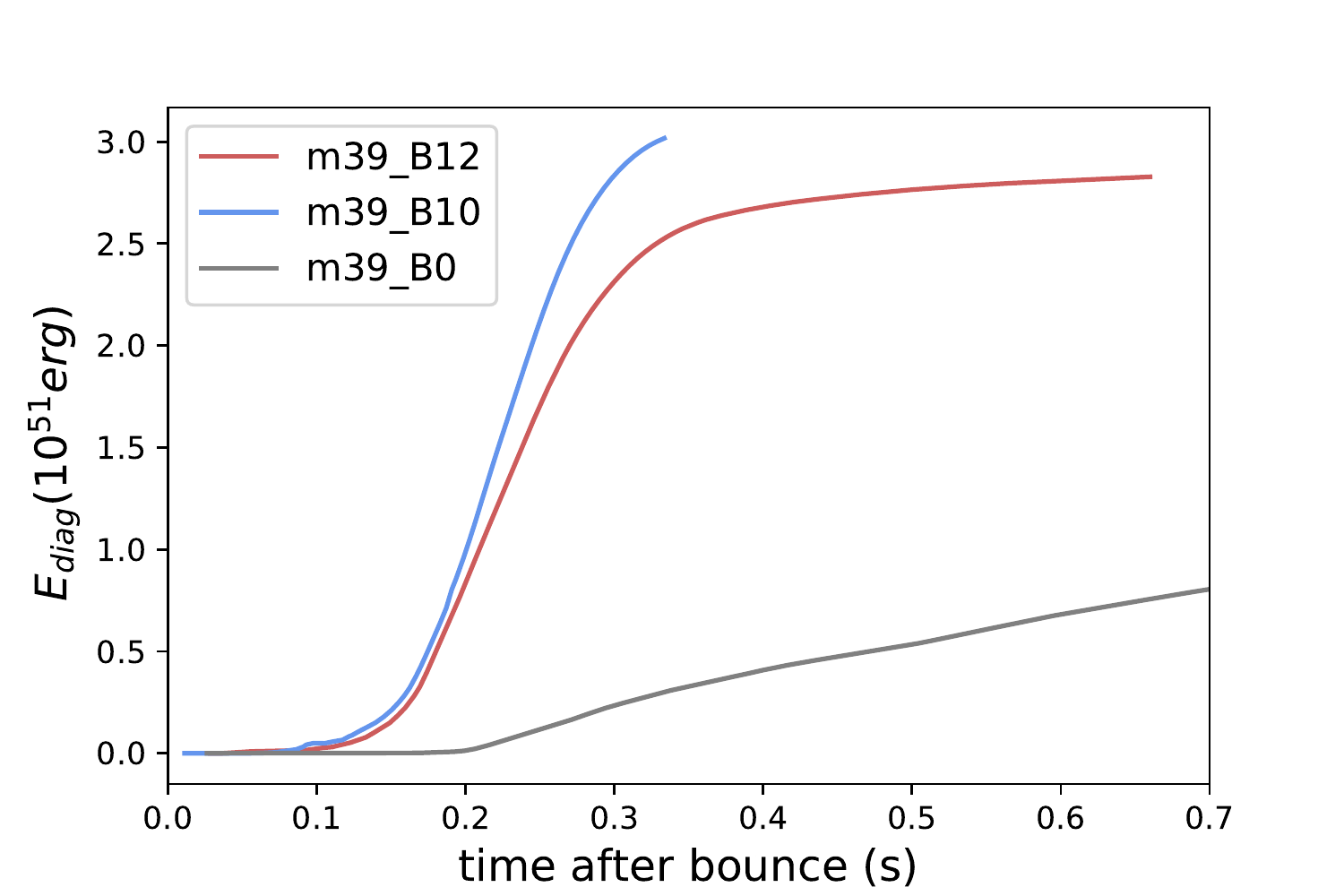}
\caption{Left: Maximum and minimum shock radius of models
m39\_B12, m39\_B10, and m39\_B0. Right: Diagnostic explosion energy of all three models. Both magnetic models achieve shock revival within  $\mathord{\sim} 100\,\mathrm{ms}$ after bounce, with shock expansion accelerating considerably
around $150\texttt{-}200\,\mathrm{ms}$ due to the emergence of bipolar jets, whereas shock revival is delayed to $\mathord{\sim} 200\,\mathrm{ms}$ in the non-magnetised model. Both magnetic models have almost reach their final explosion energies by $\mathord{\sim}350$\,ms after bounce. Model m39\_B10 produces a slightly more energetic explosion 
than m39\_B12.}
\label{fig:shock_energy}
\end{figure*}

The minimum and maximum shock radius of the two magnetised models
and their non-magnetised counterpart are shown in Figure~\ref{fig:shock_energy}. The shock is revived at a similar early time for both magnetised models, and about 100\,ms earlier than the m39\_B0 model from \citet{2020MNRAS.494.4665P}. Earlier explosion times for models with stronger magnetic fields are consistent with the findings of previous work \citep{obergaulinger_14,2022MNRAS.512.2489O}. 
Early shock revival in model m39\_B10 is noteworthy, because the behaviour for moderately high initial fields is markedly
different from 2D models of the same progenitor \citep{Jardine_2021}.
In the 2D models of \citet{Jardine_2021}, the m39\_B12 model undergoes shock revival at a similarly early time, but their 2D version of m39\_B10 undergoes shock revival much later at $\sim 400$\,ms after bounce, i.e., later than in 3D and later than the corresponding non-magnetised model. 

We compute the diagnostic explosion energy
$E_\mathrm{diag}$ following \citet{Buras2006,varma_22}
by integrating the total thermal, kinetic, magnetic, and potential energy $e_\mathrm{tot}$ over the region that is nominally 
unbound,
\begin{equation}
E_\mathrm{diag} = \int\limits_{e_\mathrm{tot}>0} \rho e_\mathrm{tot}
\,\ud V.
\end{equation}
The explosion energies are shown in Figure~\ref{fig:shock_energy}. The magnetic models
m39\_B12 and m39\_B10 reach a significantly higher explosion energy than the neutrino-driven model m39\_B0. 
The m39\_B12 model reaches energies of $2.8 \times 10^{51}$\,erg by the end of the simulation, and model m39\_B10 reaches $3.0\times 10^{51}$\,erg. 
Moreover, whereas $E_\mathrm{diag}$ gradually increases
in model m39\_B0 over the duration of the entire simulation
 and clearly has not saturated yet, 
 the explosion energy increases  rapidly in models m39\_B10 and m39\_B12 around $\sim 200$\,ms after bounce and then starts to plateau at $\sim 300$\,ms (although  m39\_B12 has just about reached this point).
Our results put models m39\_B12 and m39\_B10 in the realm of hypernovae, albeit still quite far from the most energetic ones with  energies of $\mathord{\sim}10^{52}$\,erg. However, the final explosion energy is still uncertain because the material ahead of the shock still has a binding energy (``overburden'') of $\mathord{\sim}10^{51}\, \mathrm{erg}$. There is no straightforward means of estimating the final explosion energy, although subtracting the overburden from the diagnostic explosion energy
has worked well for long-time models of fallback explosions
\citep{chan_20}.

\begin{figure*}
\includegraphics[width=\columnwidth]{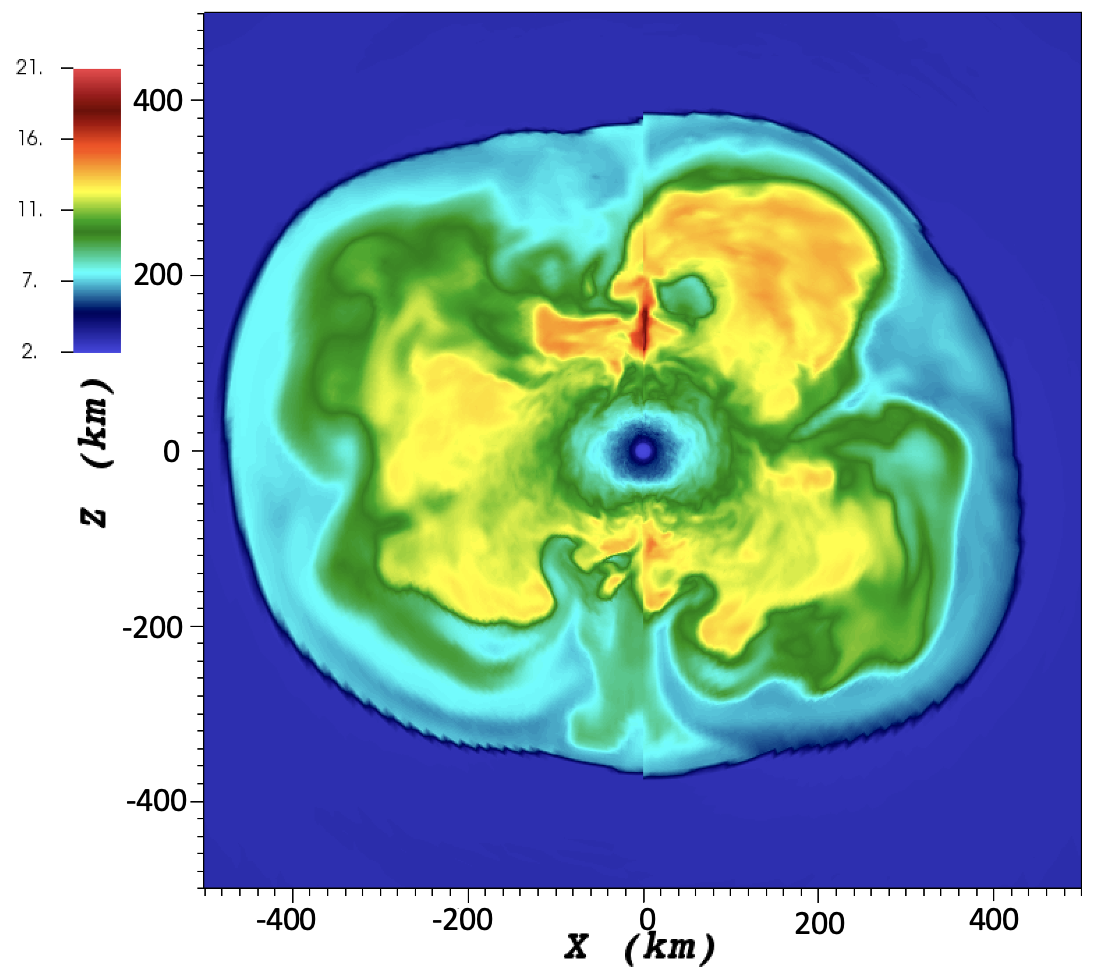}
\includegraphics[width=\columnwidth]{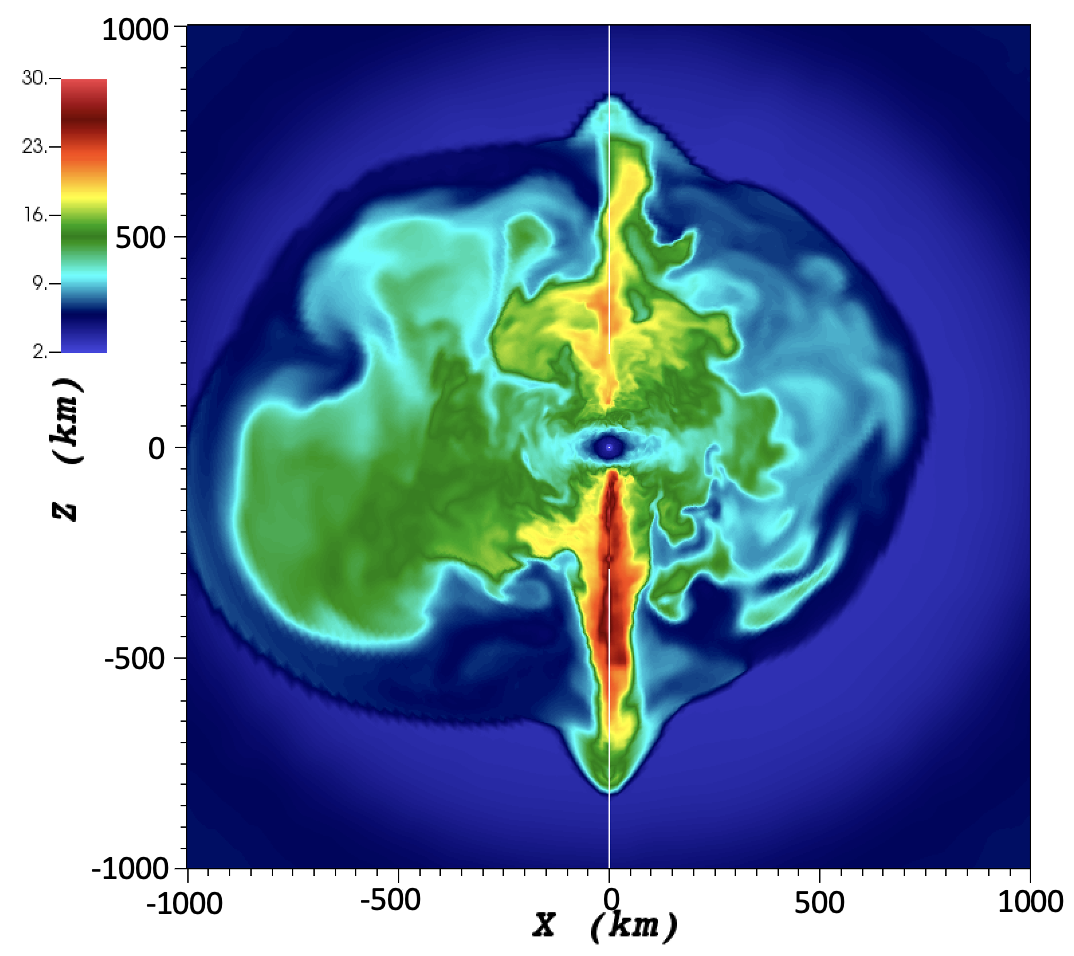}
\includegraphics[width=\columnwidth]{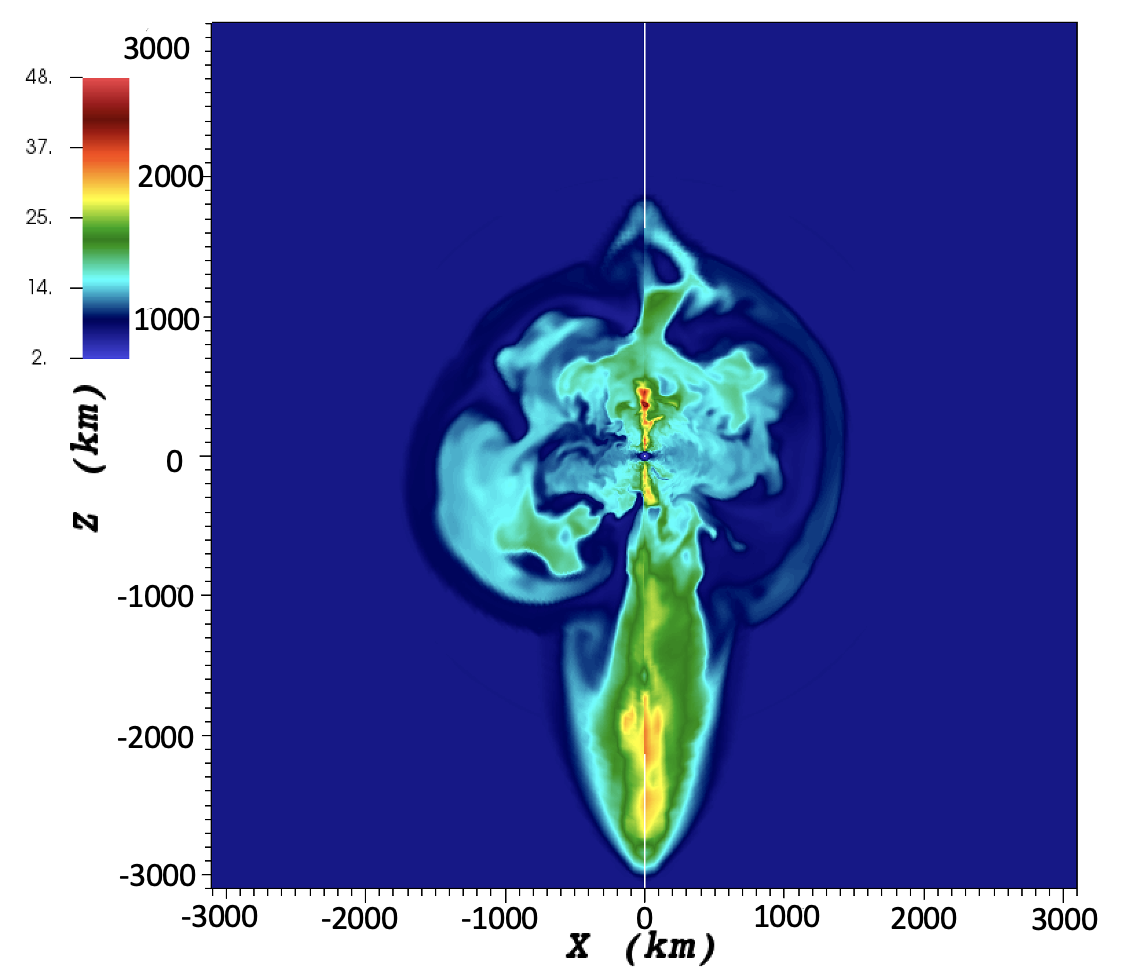}
\includegraphics[width=\columnwidth]{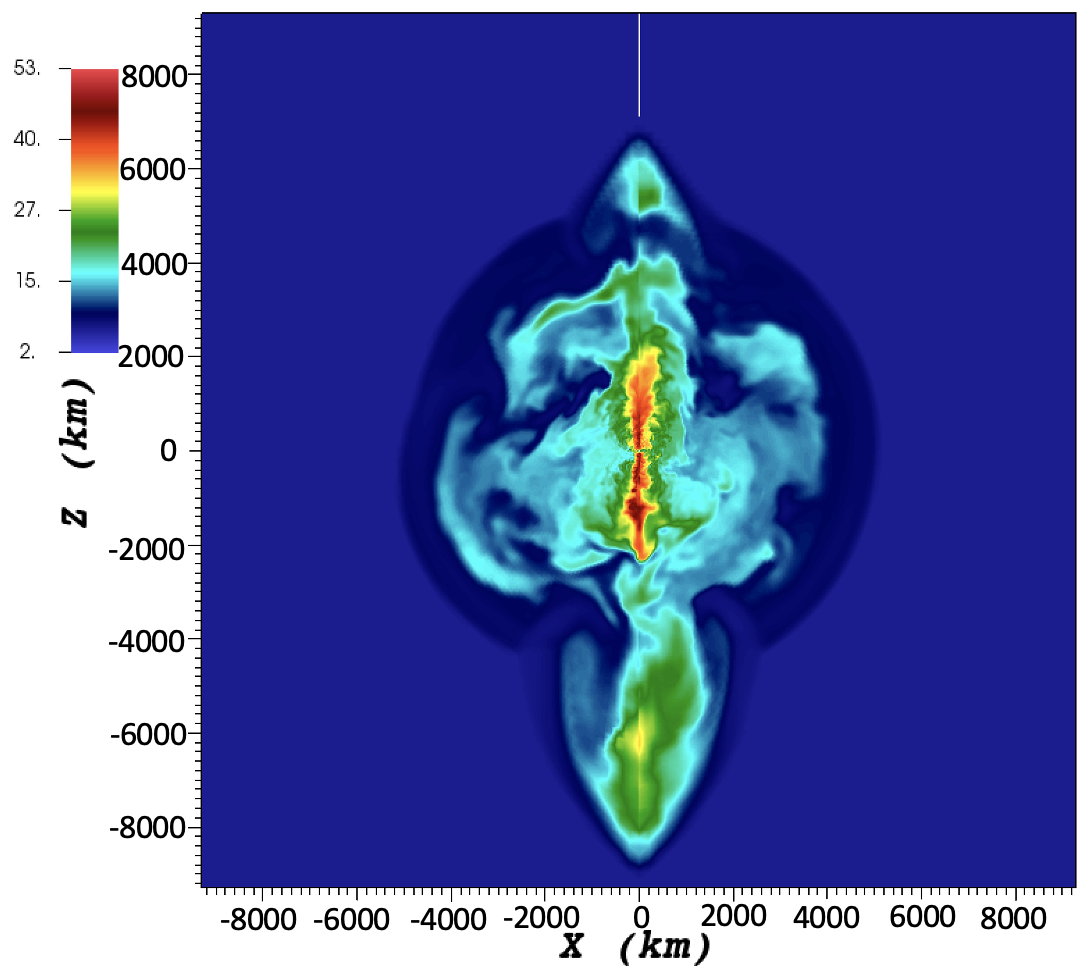}
\caption{Meridional slices showing the specific entropy (in units of $k_\mathrm{B}/\mathrm{nucleon}$) in model m39\_B12 at 100\,ms (top left), 150\,ms (top right), 200\,ms (bottom left) and 400\,ms (bottom right) after core bounce.  }
\label{fig:entropy_b12}
\end{figure*}

\begin{figure*}
\includegraphics[width=\columnwidth]{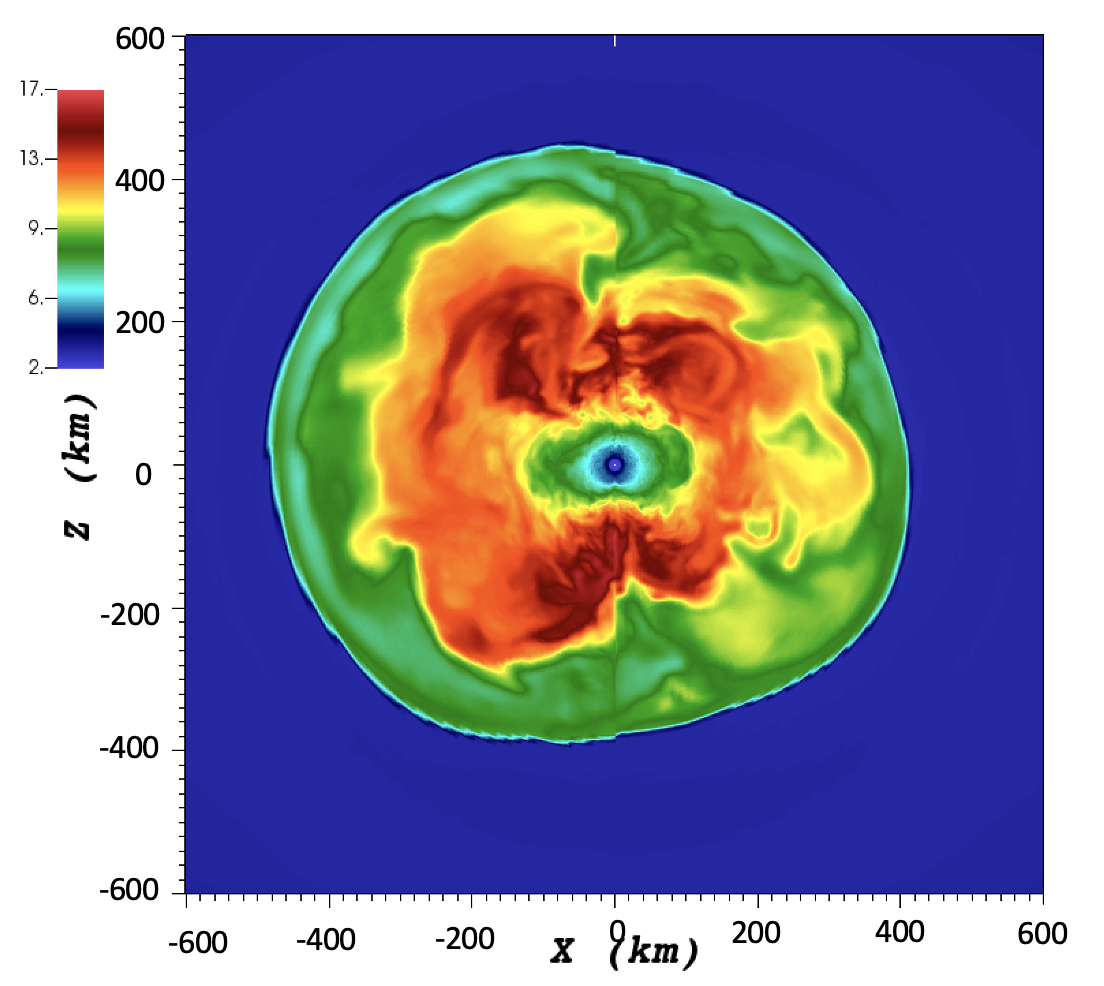}
\includegraphics[width=\columnwidth]{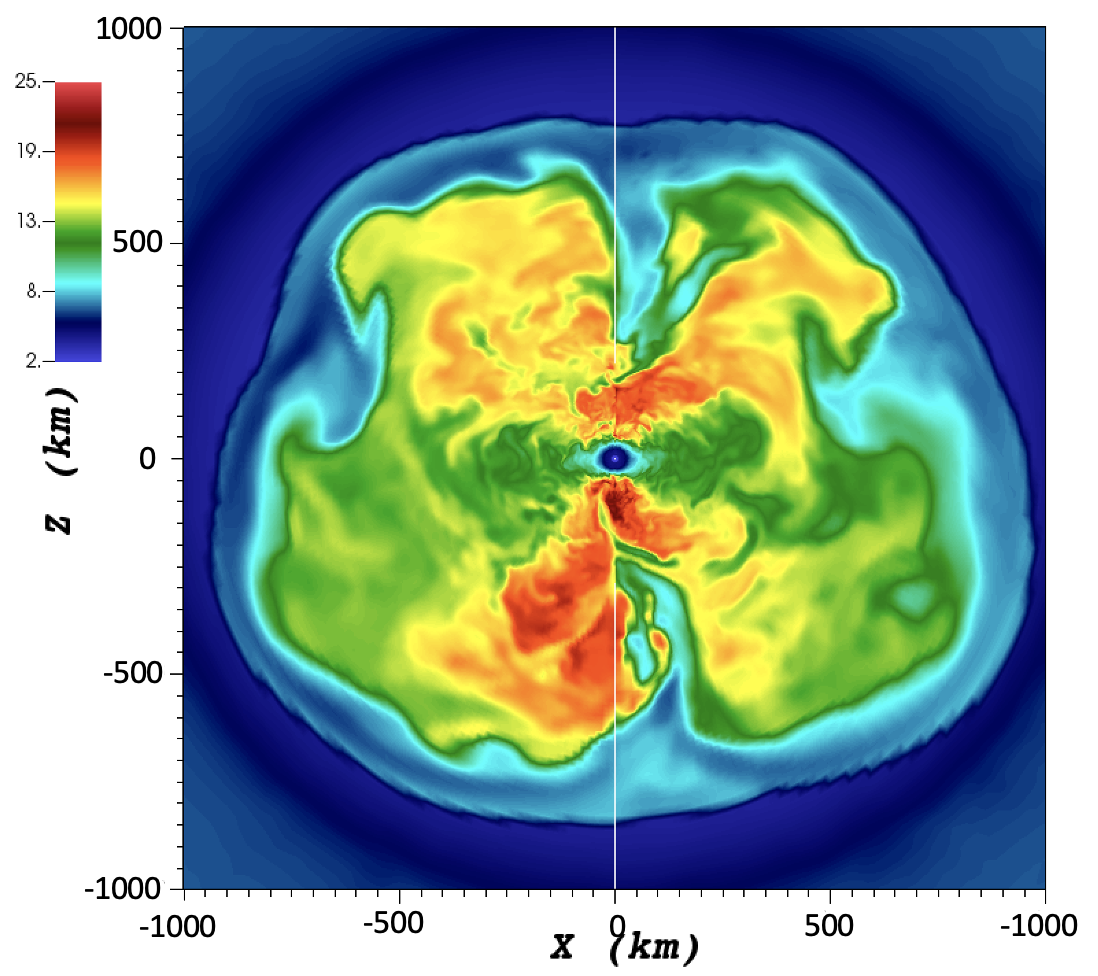}
\includegraphics[width=\columnwidth]{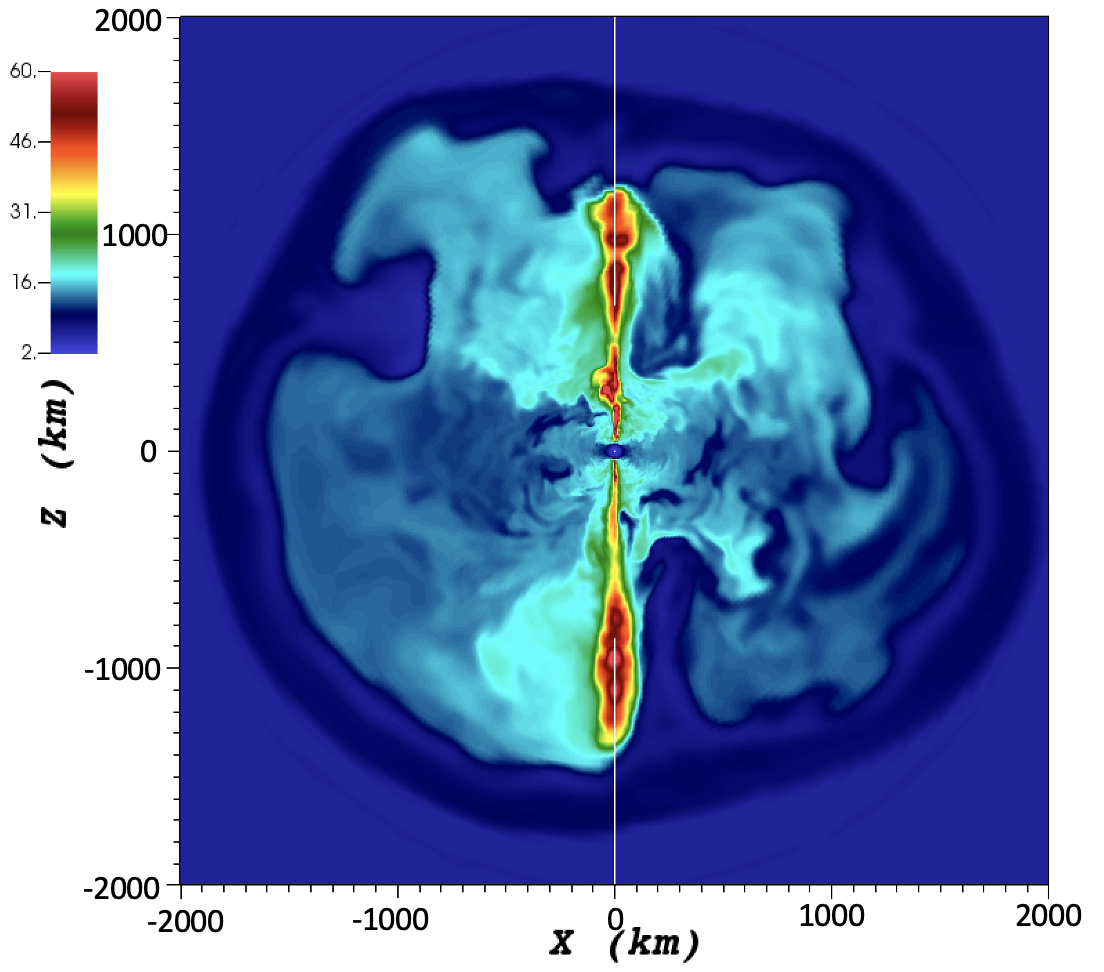}
\includegraphics[width=\columnwidth]{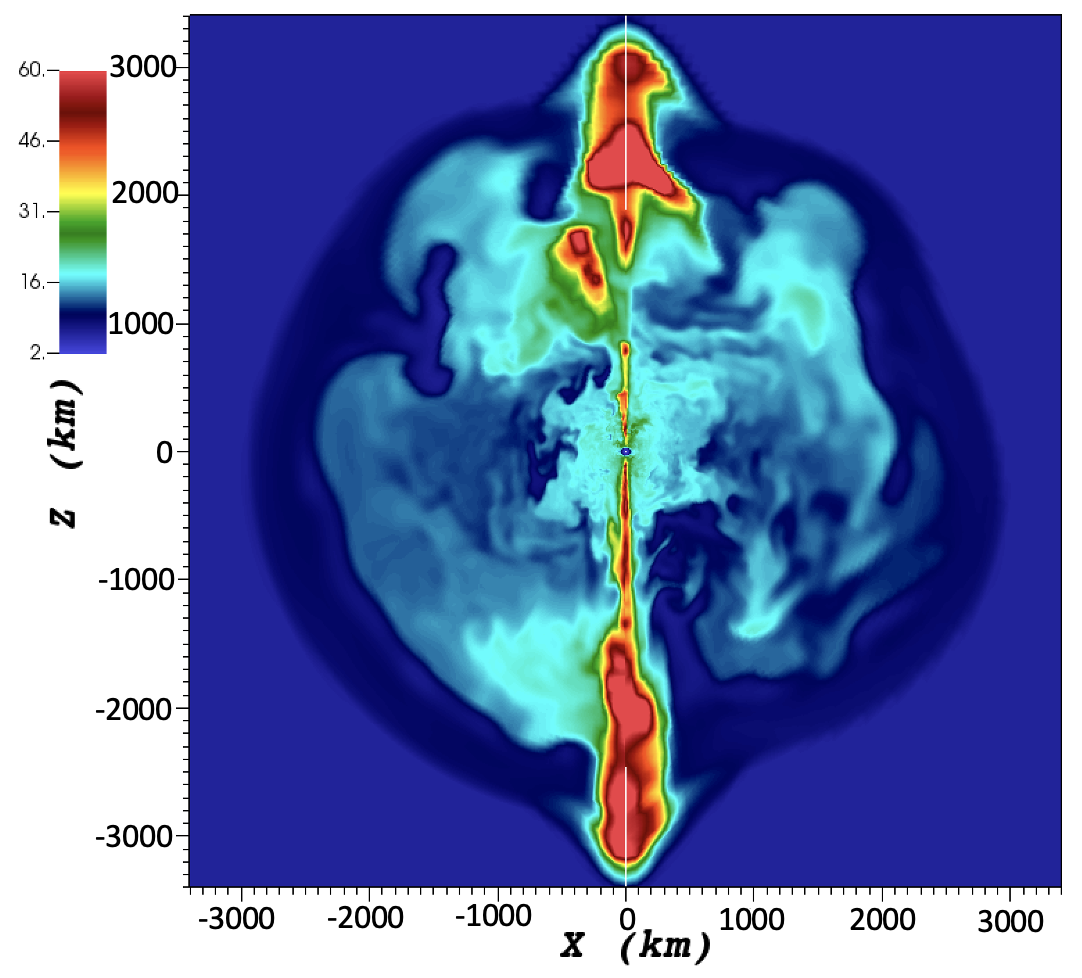}
\caption{Meridional slices showing the specific entropy (in units of $k_\mathrm{B}/\mathrm{nucleon}$) in  model m39\_1e10 at 100\,ms, 150\,ms, 200\,ms and 250\,ms after core bounce. Note that this model only develops a strong jet when the shock has already expanded to a radius of about $1000\, \mathrm{km}$.}
\label{fig:entropy_b10}
\end{figure*}

Although the inclusion of magnetic fields clearly has a significant impact
on shock revival, models m39\_B12 and m39\_B10 do not conform to the classical picture of magnetorotational explosions driven by bipolar jets initially. Instead, the shock already expands significantly in both models before jets truly start to emerge, at which point shock expansion accelerates. The development of the explosion morphology is illustrated by Figures~\ref{fig:entropy_b12} and \ref{fig:entropy_b10}, which show 2D slices of the specific entropy for model m39\_B12 and model m39\_B10, respectively.

In model m39\_B12, the shock expands to about $400\, \mathrm{km}$ before
jets develop along both directions of the rotation axis  around $150\, \mathrm{ms}$, i.e., at the time corresponding to the visible break in the shock trajectory. The jets remain rather unstable and show intermittent activity. By 200\,ms after bounce, the jet in the North direction disappears and a strong jet is only visible in the South direction. The jet direction vacillates visibly, indicating kink instability \citep{1993ApJ...419..111E, 1998ApJ...493..291B}
as in previous 3D simulations of magnetorotational CCSNe
\citep{moesta_14b,2020ApJ...896..102K, 2020MNRAS.492.4613O, 2020arXiv200807205O,bugli_21,bugli_22}, which will be analysed further below. The intermittency of the jet
and the instability of the jet direction may be unrelated phenomena, however (see below).

In model m39\_B10, jet formation is delayed considerably longer; no jet is visible yet at 150\,ms. At 200\,ms after core bounce, the model develops a strong jet in both directions. Although the jets are very narrow at their base, and somewhat close to being constricted by surrounding material at radii of several hundred km, they remain more stable than in m39\_B12 and grow continuously until the end of the simulation. The variations in jet strength are most reminiscent of sausage instability than of kink instability in this model, but the extreme narrowness of the jets (with a width of only a few grid cells) calls for caution in interpreting the jet morphology.

Following previous studies \citep{moesta_14b,2020ApJ...896..102K, 2020MNRAS.492.4613O, 2020arXiv200807205O,bugli_21},  we further quantify the development of jet instabilities by determining the displacement of the barycentre of the  magnetic energy from the z-axis,
\begin{equation}
    \langle \mathbf{x} \rangle =
    \frac{\int \mathbf{x}\, B^2 \,\ud V}{{\int  B^2 \,\ud V}},
\end{equation}
where the integration is restricted to a region around the axis
and outside the neutron star (such as to cover only one lobe of the jet).
Following closely the method of \citet{2020ApJ...896..102K}, we
integrate over a region
within 50\,km of the axis and $r=50\,\mathrm{km}$, i.e., slices at at constant radial coordinate through the jet rather than an extended volume containing the entire jet.
To interpret the nature of the instability, we also examine the  toroidal-to-poloidal field ratio
$\langle B_\mathrm{T}/B_\mathrm{P}\rangle$, which determines the 
growth rate and critical wavelength $\lambda$ of the kink mode under for idealised setups \citep{kruskal_58,1998ApJ...493..291B,moesta_14b}, 
\begin{equation}
    \lambda
    \gtrsim \frac{2\pi a}{
    \langle B_\mathrm{T}/B_\mathrm{P}\rangle},
\end{equation}
where $a$ is the width of the jet (pinch). We quantify this by
integrating $B_\mathrm{T}^2=B_\varphi^2$ and $B_\mathrm{P}^2=B_r^2+B_\theta^2$ over
the same area,
\begin{equation}
\langle B_\mathrm{T}/B_\mathrm{P}\rangle  =  \frac{\int  B_\varphi^2 \,\ud V}{{\int  (B_r^2+B_\theta^2) \,\ud V}}.
\end{equation}

\begin{figure*}
\includegraphics[width=\columnwidth]{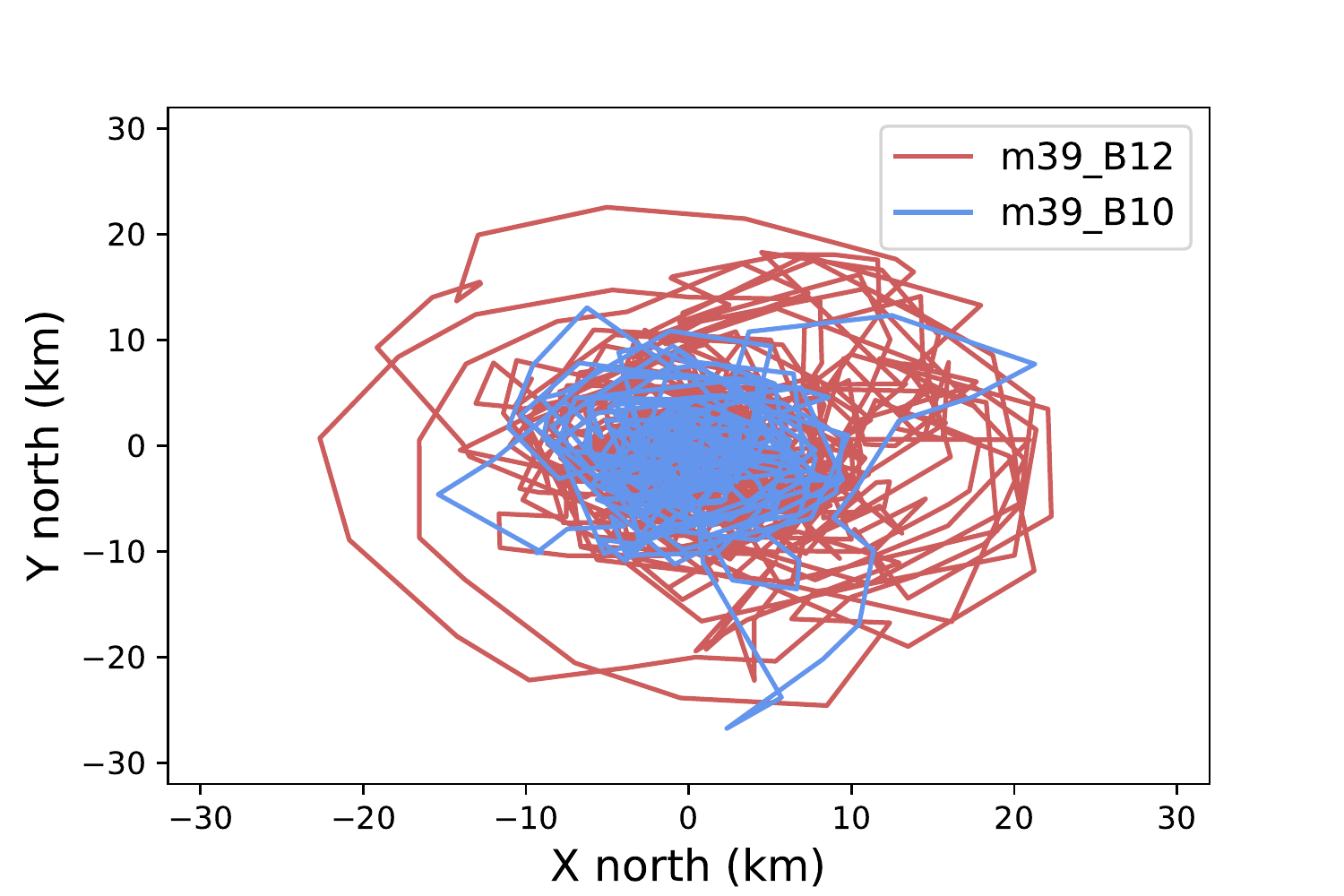}
\includegraphics[width=\columnwidth]{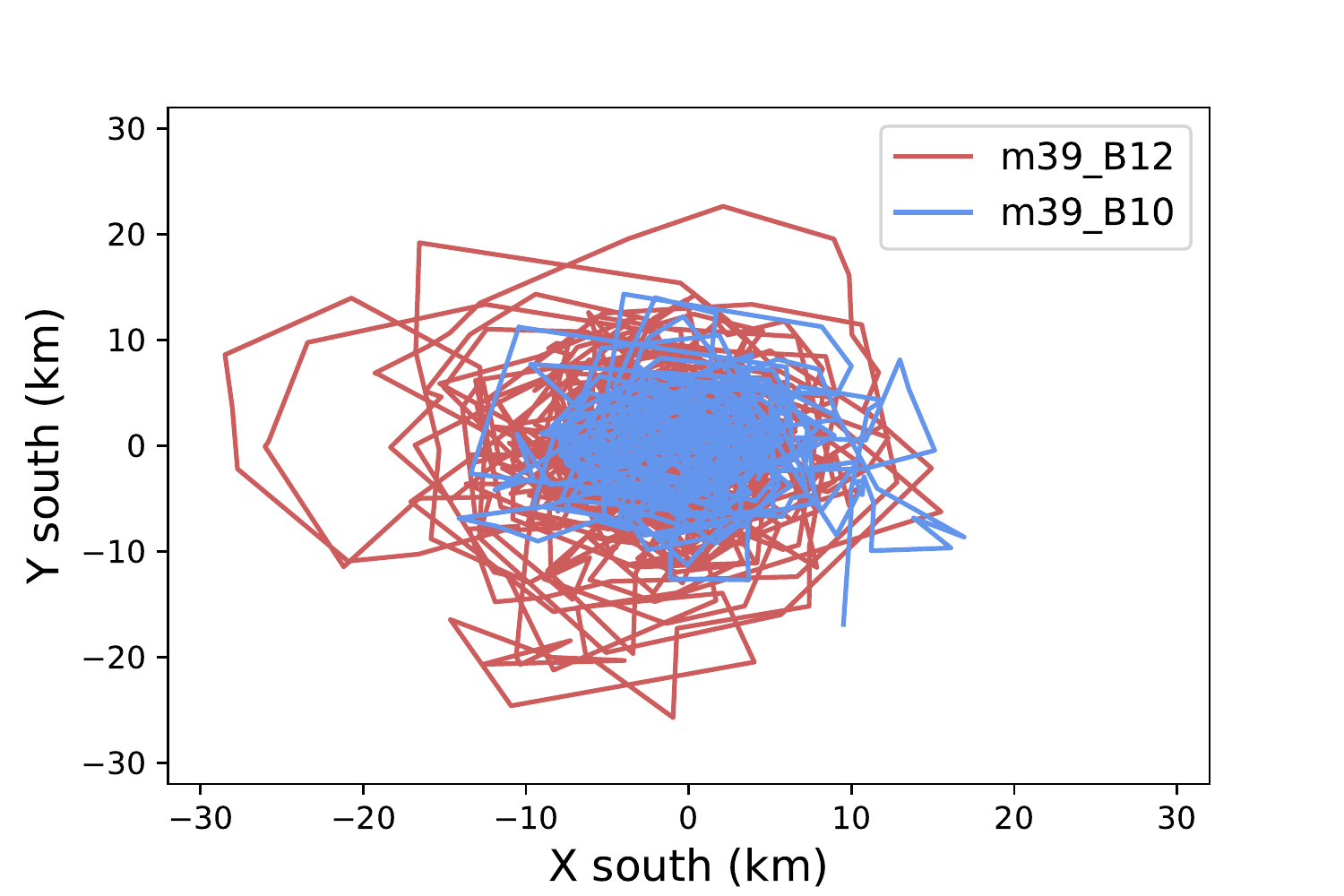}
\includegraphics[width=\columnwidth]{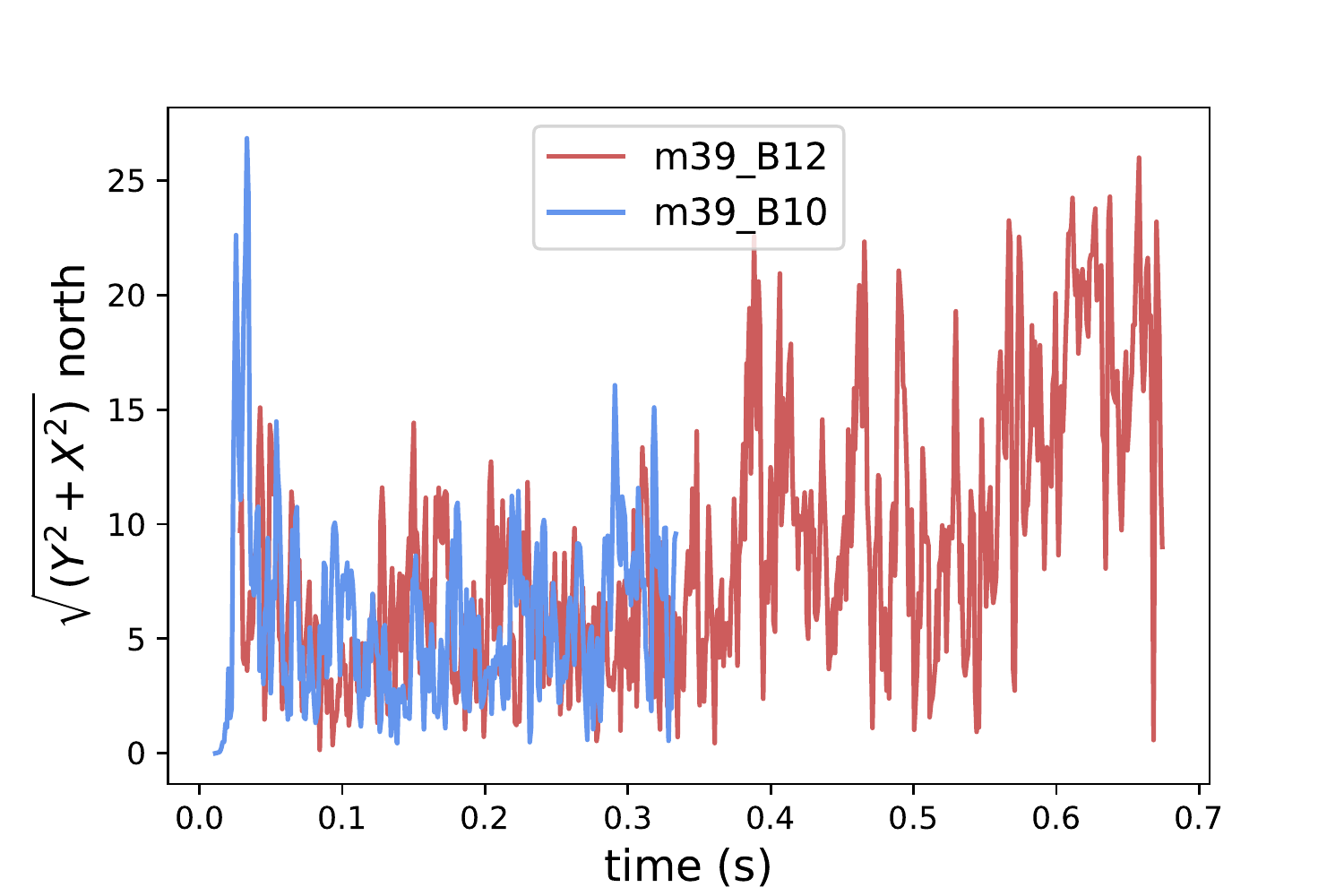}
\includegraphics[width=\columnwidth]{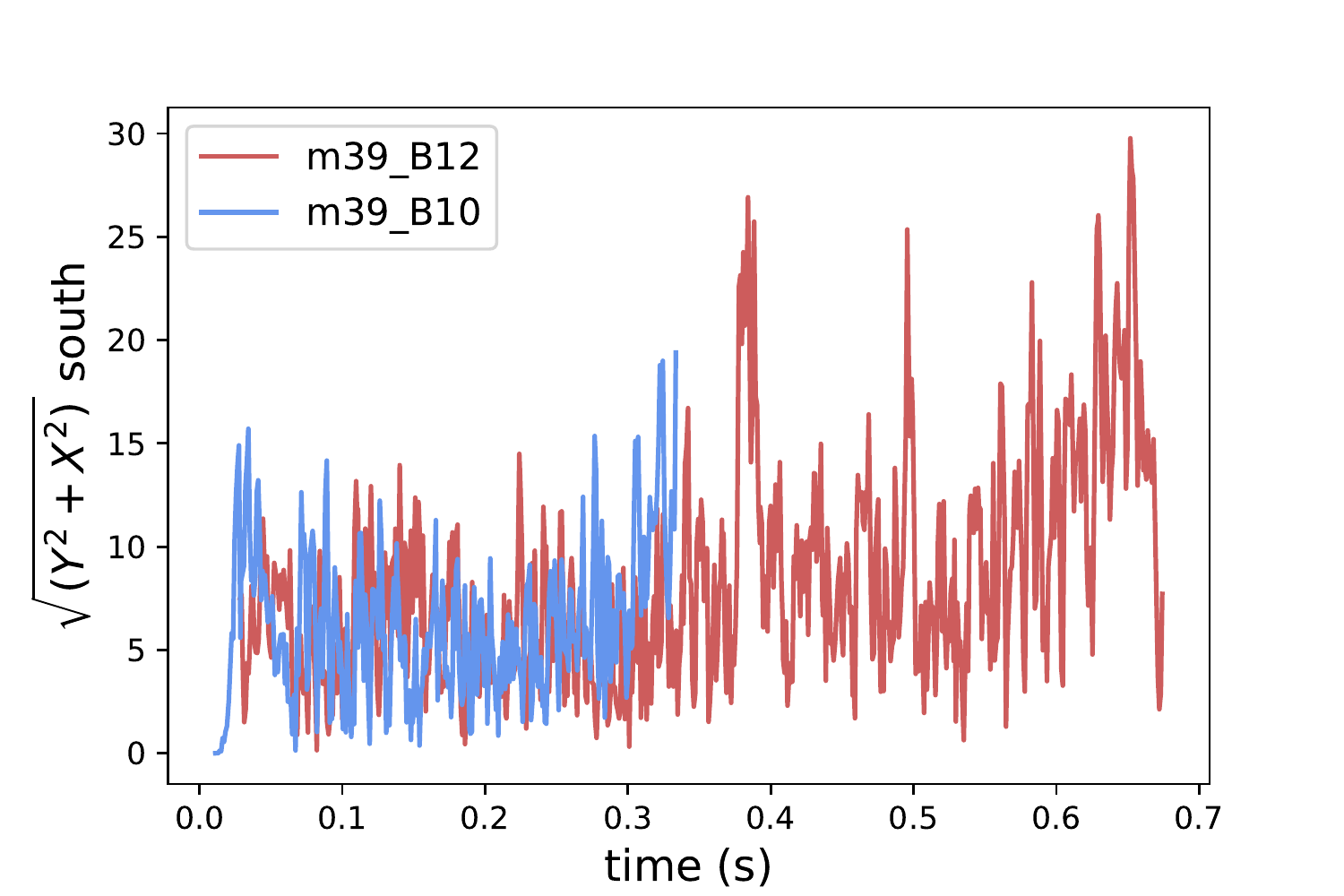}
\includegraphics[width=\columnwidth]{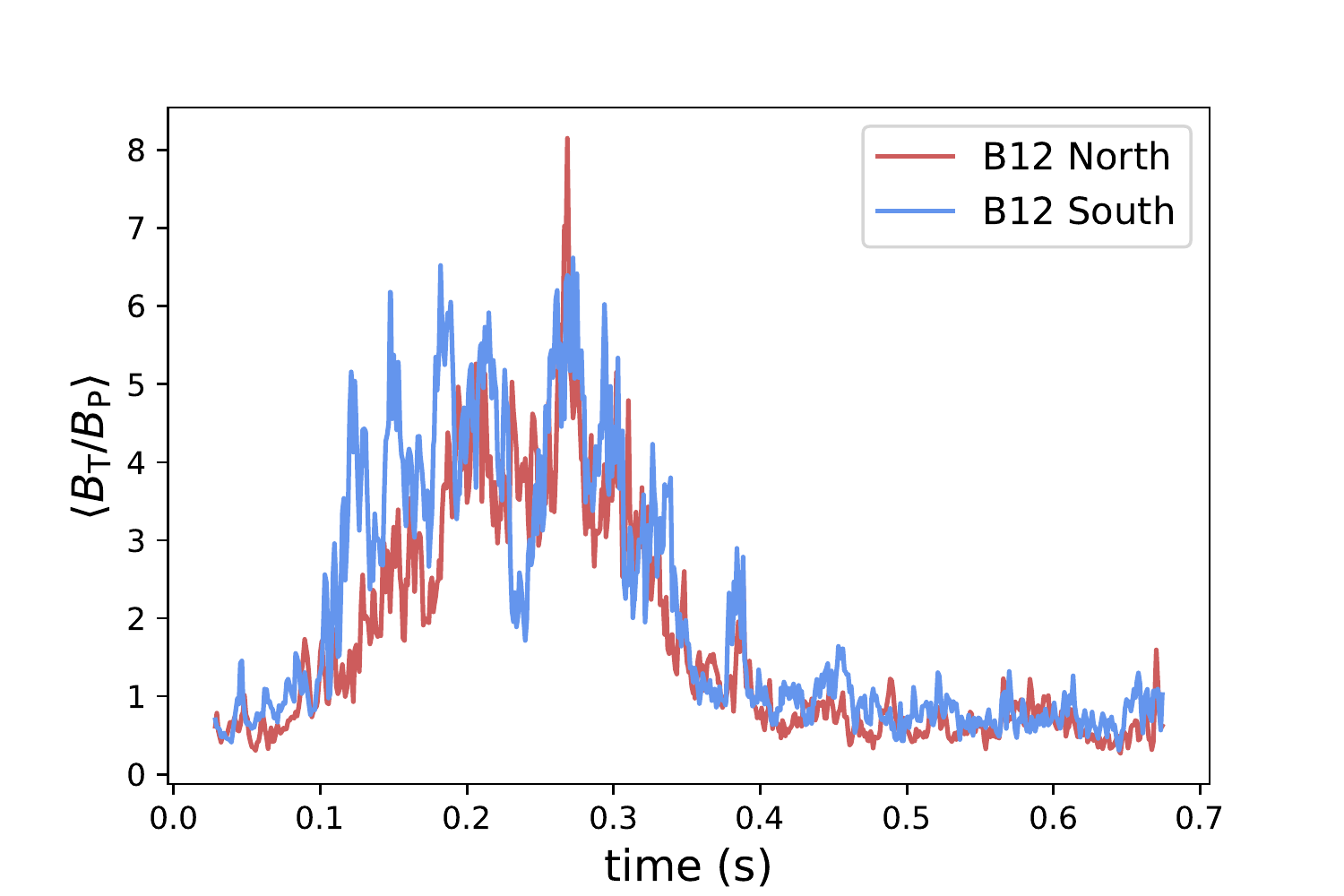}
\includegraphics[width=\columnwidth]{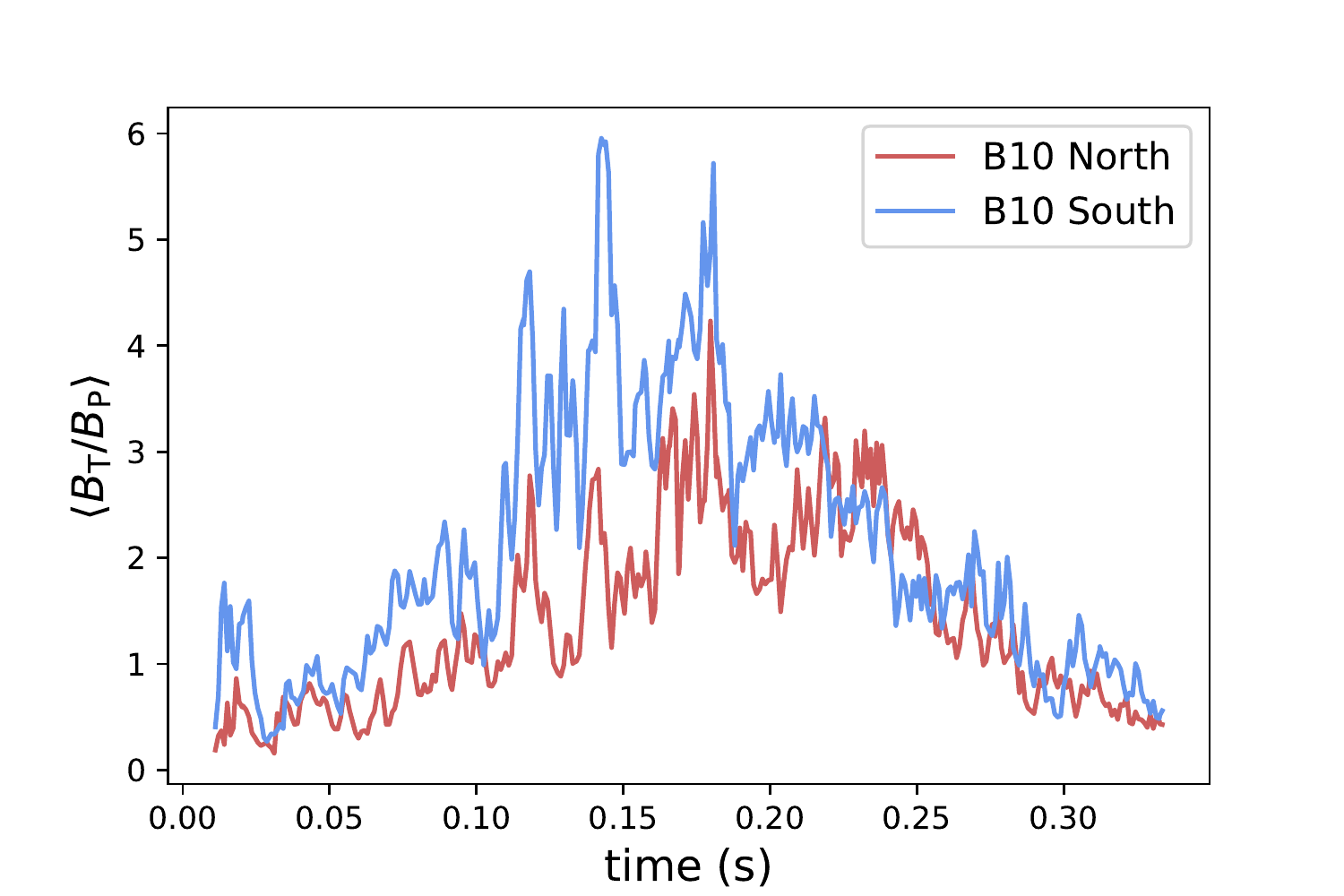}
\caption{Top row: Projected displacement
$\langle\mathbf{x}\rangle$ of the barycentre
of the magnetic field energy at $r=50\, \mathrm{km}$ in the $x$-$y$-plane for model m39\_B12 (red) and m39\_B10 (blue)
The left and right panel refer to the jets in the Northern and Southern hemisphere, respectively.
Middle row: Absolute value of the displacement from the $z$-axis, again for the jets in the Northern (left) and Southern (hemisphere). Bottom row: Toroidal-to-poloidal field ratio in the jets for model m39\_B12 (left) and
m39\_B10 (right).
}
\label{fig:jets}
\end{figure*}

The results for the jet displacement $\mathbf{x}$
and the ratio $\langle B_\mathrm{T}/B_\mathrm{P}\rangle$
are shown in Figure~\ref{fig:jets} and quantitatively confirm the presence of jet instability. For both models, the position of the barycentre rapidly varies on short time scales with frequent excursions to $5\texttt{-}10\, \mathrm{km}$  until $300\, \mathrm{s}$, corresponding to a deviation of the jet direction from the axis by $5^\circ\texttt{-}10^\circ$. Larger displacements are seen in model m39\_B12 (which has been run longer) at later times. In model  m39\_B12, the trajectories of the barycentre circle the axis, consistent with the helical deformation expected from the kink instability. The numerical values of $\langle B_\mathrm{T}/B_\mathrm{P}\rangle$ are well above unity at early times and then decrease to $\langle B_\mathrm{T}/B_\mathrm{P}\rangle\approx 1$ by $300\, \mathrm{ms}$ in m39\_B10 and by $350\, \mathrm{ms}$ in m39\_B12, which indicates that the kink instability can indeed develop on short wavelengths. It is puzzling, however, that the jet instability becomes more pronounced in model m39\_B12 at late times when $\langle B_\mathrm{T}/B_\mathrm{P}\rangle\approx 1$ than at earlier when $\langle B_\mathrm{T}/B_\mathrm{P}\rangle$ is larger. This behaviour may be related to changes in other jet parameters and the environment of the jet. It is suggestive that the phase of more vigorous jet instability coincides with the onset of the plateau in the explosion energy, but this may be a coincidence. Based on the critical wavelength 
for instability predicted by the Kruskal-Shafranov criterion, we also raise resolution effects as a potential caveat. With peak values of $\langle B_\mathrm{T}/B_\mathrm{P}\rangle \gtrsim 5$, we expect unstable modes with a wavelength of the order of the width of the jet, which amounts to just a few zones; hence the growth of the kink instability is likely somewhat affected by numerical dissipation. Moreover,
the behaviour of the kink instability may be affected by the spherical polar grid geometry. While our models confirm that spherical polar grids can qualitatively capture the kink instability at the current resolution, and are better resolved than some other models using a spherical grid geometry
\citep{2022MNRAS.512.2489O,bugli_22}, future code comparisons and resolution studies should assess the requirements for capturing the kink instability at the level required for predictive hypernova models.

Even though it may be tempting to also associate the intermittency of the jet (which is more pronounced in model m39\_B12 with slightly higher values of $\langle B_\mathrm{T}/B_\mathrm{P}\rangle$ ) to the kink instability,
we believe, however, that the intermittency may rather be regulated by the feedback of the magnetic fields on the angular momentum distribution in the PNS surface region. The intermittent weakening of the jet and an associated drop in the ratio of magnetic to thermal pressure often appears to be connected to slower and even retrograde rotation in the polar region of the PNS at the base of the jets. For the time being,
this remains a tentative hypothesis, however. Due to the uncertainty of the process behind jet intermittency it is not clear whether the higher robustness of the jet in m39\_B10 is a systematic effect or merely the result of stochastic model variations.

\begin{figure}
    \centering
    \includegraphics[width=\linewidth]{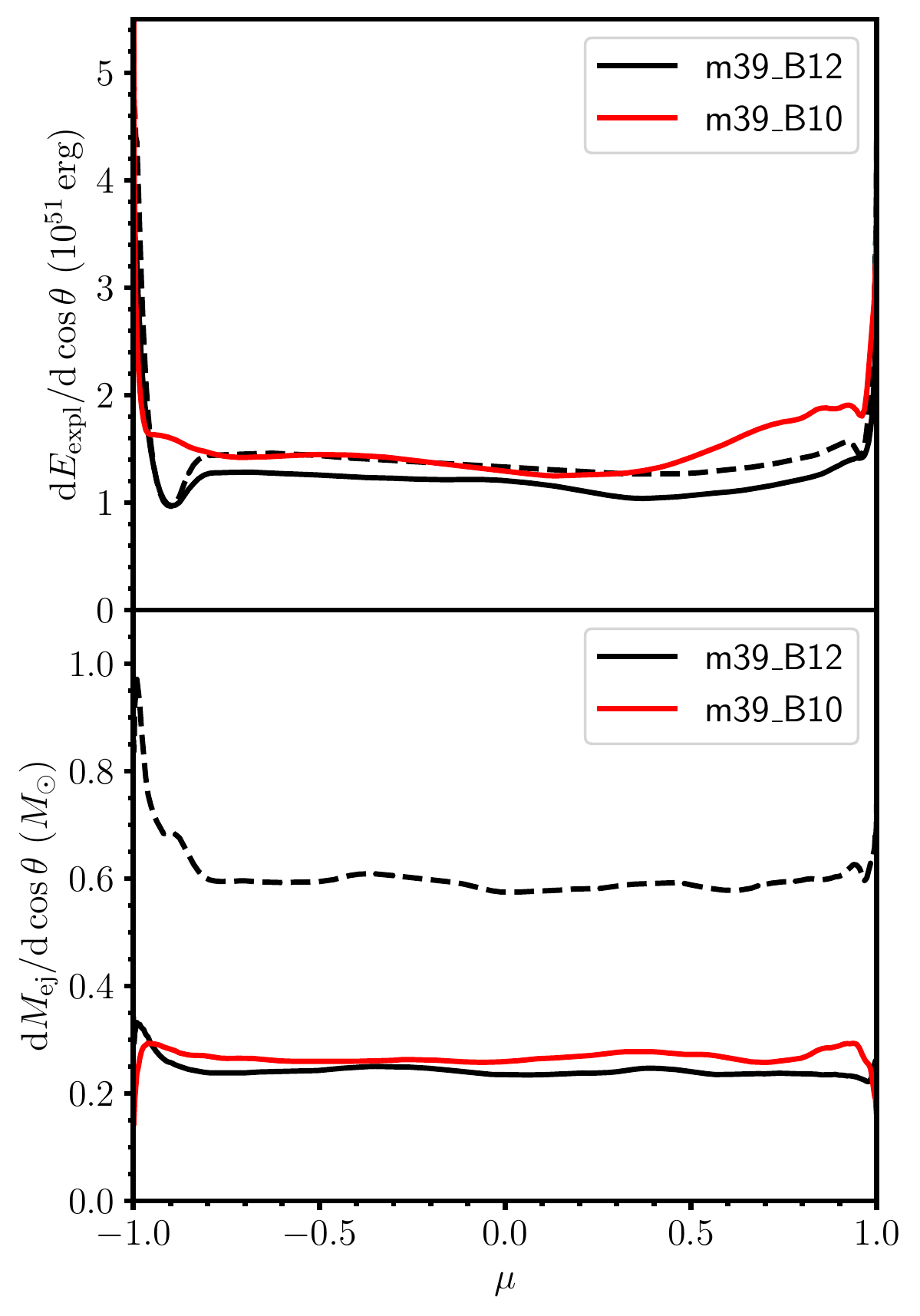}
    \caption{The differential contributions
    $\ud E_\mathrm{expl}/\ud \mu$ and $\ud M_\mathrm{ej}/\ud \mu$ to the explosion energy (top)
    and ejecta mass (bottom) for models m39\_B12 (black) and m39\_B12 (red)
    as a function of $\mu=\cos\theta$ at $330\, \mathrm{ms}$ (solid) and
    $680\, \mathrm{ms}$ (dashed, model m39\_B12 only). $\ud E_\mathrm{expl}/\ud \mu$ and $\ud M_\mathrm{ej}/\ud \mu$  illustrate the anisotropy of the explosion.
    The cores of the jets can be seen as narrow spikes in $\ud E_\mathrm{expl}/\ud \mu$ at $\mu=\pm 1$. As the southern jet entrains material, an excess in $\ud M_\mathrm{ej}/\ud \mu$ also develops in model m39\_B12 near $\mu=-1$ at late times.
    }
    \label{fig:my_label}
\end{figure}

Since the jets entrain material as they punch through the star
and do not have sharply defined boundaries, one cannot unambiguously separate the contribution of the jets to the
explosion energy and ejecta mass from the more slowly moving ejecta at lower latitudes. To show the anisotropy of the ejecta components most clearly, in Figure \ref{fig:my_label}, we show the differential explosion energy and ejecta mass $\ud E_\mathrm{expl}/\ud \mu$ and $\ud M_\mathrm{ej}/\ud \mu$ as a function of $\mu=\cos \theta$.
For a perfectly isotropic explosion, $\ud E_\mathrm{expl}/\ud \mu$ and $\ud M_\mathrm{ej}/\ud \mu$ would take on constant values. The cores of the jets can be seen as spikes near $\mu=\pm 1$. At late times, one can discern a higher ejecta mass per solid angle in a region of $\Delta\mu\approx 0.15$ or about $30^\circ$ 
around the Southern axis due to entrainment by the jet in model m39\_B12, which is consistent with the visual appearance in Figure~\ref{fig:entropy_b12}. There is also excess explosion energy and ejecta mass near the Northern axis, but the jet structures in the Northern hemisphere are so tenuous that they hardly contribute to the overall ejecta mass and explosion energy. The visible structures in the plots of $\ud E_\mathrm{expl}/\ud \mu$ and $\ud M_\mathrm{ej}/\ud \mu$ suggest that the jets carry of order 10\% of the explosion energy and ejecta in this model. For model m39\_B10, it is not yet possible to clearly identify regions around the poles with higher $\ud M_\mathrm{ej}/\ud \mu$.

%%%%%%%%%%%%%%%%%%%%%%%%%%%%%%%%%%%%%%%%%%%%%%%%%%%%%%%%%%%%
%%%%%%%%%%%%%%%%%%%%%%%%%%%%%%%%%%%%%%%%%%%%%%%%%%%%%%%%%%%%
\section{Remnant Properties}
\label{sec:pns} 

We next consider the properties of the proto-neutron star in both models. These are primarily of relevance for the behaviour of the observable transient on longer time scales, e.g., for potential energy input by millisecond-magnetar spin-down. A comparison with the Galactic neutron star population is of less interest because of the low hypernova rate in the local Universe and the lack of any observed compact remnant that could be associated with hypernova explosions.
While a hypernova origin has been considered
for Galactic supernova remnants such as
W49B \citep{lopez_13} and for the Cygnus superbubble \citep{kimura_13}, this interpretation is not uncontested, the corresponding compact remnant has either not been identified or is speculated to be a black hole in the case of W49B \citep{lopez_13}. 
However, the possibility of spin-kick alignment for jet-driven explosions is potentially of interest, since current simulations still fail to explain \citep{wongwathanarat_13,mueller_19,2020MNRAS.494.4665P,janka_22} the spin-kick alignment found by
observations \citep{johnston_05,ng_07,noutsos_12,noutsos_13}.

\begin{figure*}
\includegraphics[width=\columnwidth]{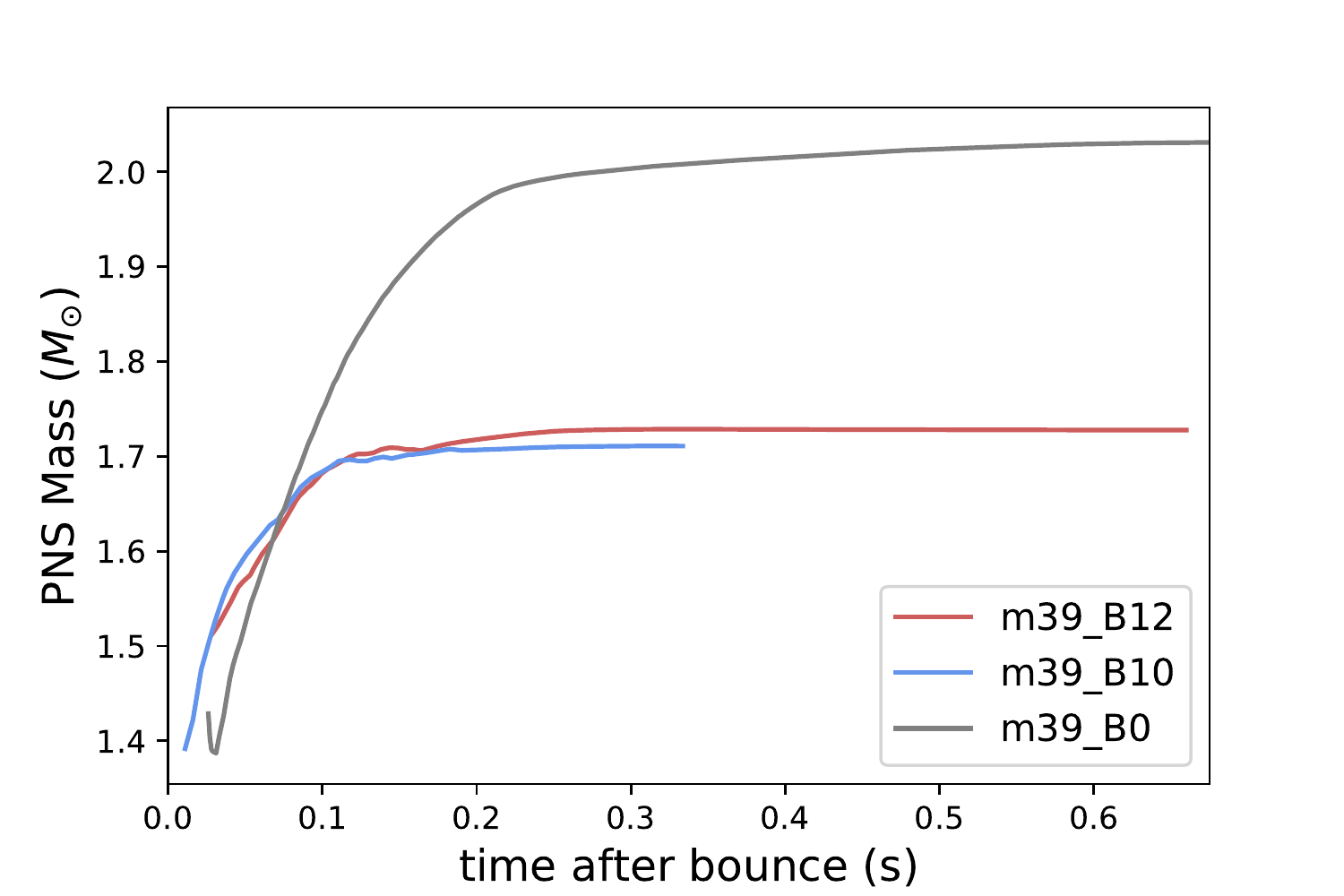}
\includegraphics[width=\columnwidth]{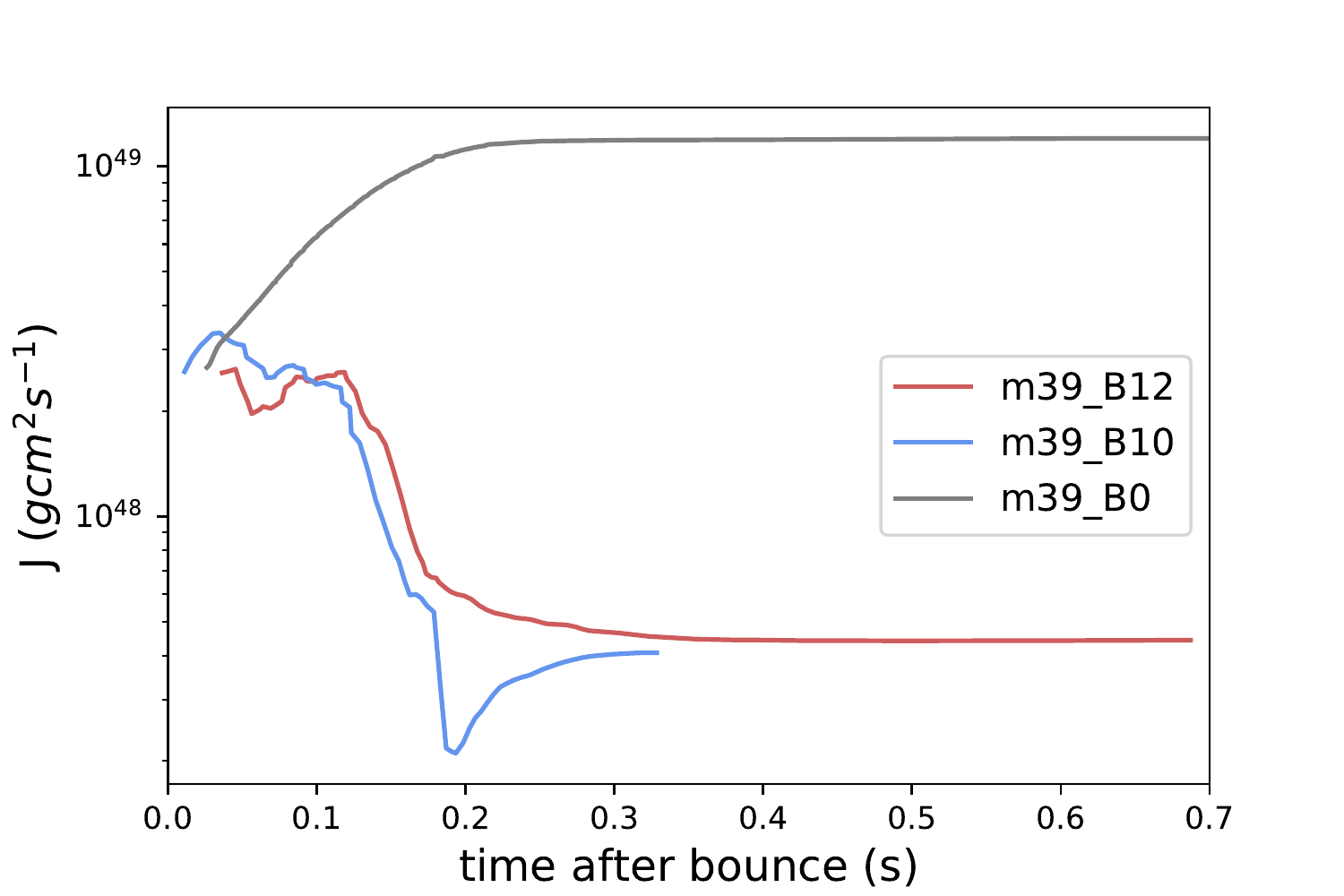}
\includegraphics[width=\columnwidth]{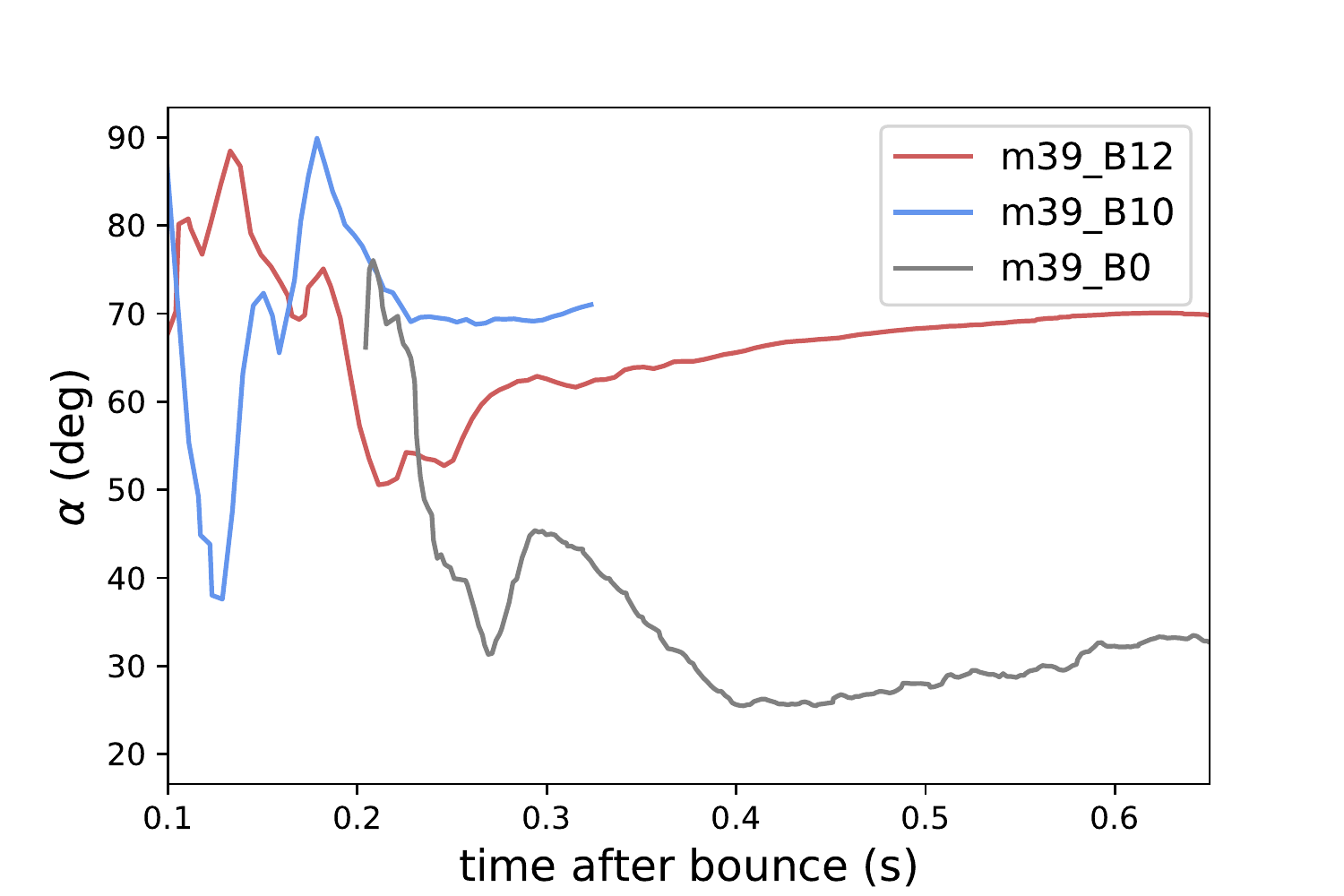}
\includegraphics[width=\columnwidth]{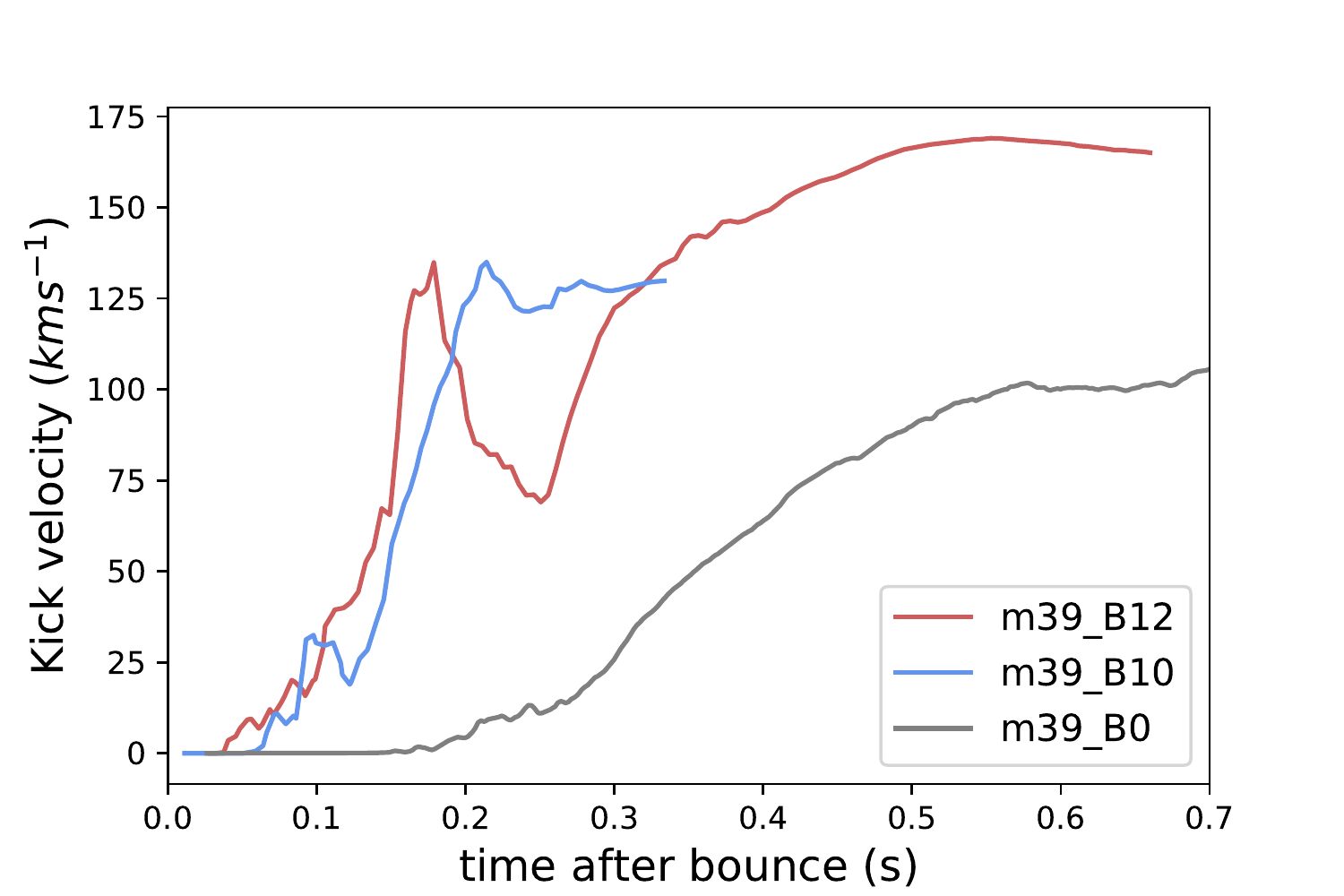}
\caption{The PNS mass (top left), angular momentum (top right), kick velocity (bottom right), and the angle $\alpha$ between the spin and kick direction (bottom left).}
\label{fig:spinkick}
\end{figure*}

In Figure~\ref{fig:spinkick}, we show the PNS mass $M$, the PNS kick velocity $\mathbf{v}_\mathrm{kick}$, its angular momentum $\mathbf{J}$,  and the angle $\alpha$ between the spin and kick direction. The (baryonic) neutron star mass is computed as the total mass above densities of $10^{11}\, \mathrm{g}\,\mathrm{cm}^{-3}$, the kick is computed from the total ejecta momentum assuming momentum conservation,
\begin{equation}
    \frac{\ud \mathrm{M} \mathbf{v}_\mathrm{kick}}{\ud t}
    =-\int\limits_\mathrm{ejecta} \rho\mathbf v\,\ud V.
\end{equation}
The PNS angular momentum is obtained by first integrating the hydrodynamic angular momentum fluxes and magnetic torques over a fiducial spherical surface of radius $50\, \mathrm{km}$
and in time, and then adding them to the initial angular momentum inside this volume at bounce,
\begin{equation}
    \frac{\ud \mathbf{J}}{\ud t}
    =\int \mathbf{r} \times \rho \mathbf{v}  \,\ud V
    -\iint(\mathbf{r}\times  \mathbf{v}) \rho v_r
    +\frac{(\mathbf{r}\times \mathbf{B}) B_r}{4\pi}\,\ud A\,\ud t.
\end{equation}
This method is used because it is preferable to estimate the angular momentum budget based on fluxes
\citep{wongwathanarat_13} instead of
relying on imperfect angular momentum conservation in grid-based hydrodynamics codes.
The final PNS is slightly more massive for model m39\_B12 than for m39\_B10 due to the earlier onset of the explosion. 
The growth of the PNS mass initially slows down earlier in the m39\_B12 model, consistent with the earlier onset of the explosion. The mass in m39\_B12 only overtakes m39\_B10 later, reflecting the lower explosion energy, which allows more residual accretion.
Correcting for the neutron star binding energy to obtain the gravitational mass $M_\mathrm{grav}$,
\begin{equation}
  M_\mathrm{grav} \approx M - 0.084 M_\odot
  \left(\frac{M_\mathrm{grav}}{M_\odot}\right)^2,
\end{equation}
puts $M_\mathrm{grav}$ to about $1.5\,M_\odot$.
Accretion onto the PNS is quenched efficiently, suggesting that the models are not likely to evolve into collapsars on short time scales despite the high mass of the progenitor. Both models end up with significantly smaller PNS masses than the m39\_B0 model, which was over $2\,M_{\odot}$ by the end of the simulation.
However, as the binding energy of the envelope
is not negligible compared to the explosion energy (with a ratio of about 1:3), substantial fallback on time scales of seconds is still possible.
For a similar ratio of explosion energy to envelope binding energy, \citet{chan_20} found
fallback of about $1\,M_\odot$ over time scales
of seconds in a 3D long-time simulation of
a supernova of a non-rotating $40\,M_\odot$ star. These results cannot be easily extrapolated to the case of a rotating progenitor, and long-time simulations are required to determine the fate of the compact remnant. The formation of a black hole with a collapsar disk of moderate mass, or at least further late-time accretion of the PNS cannot be excluded. One should also caution against extrapolating our results for a rapid magnetorotational explosion scenario to other progenitors. As mentioned in Section~\ref{sec:sim}, the
progenitor compactness is only moderately high, and rapid explosions in more compact progenitors could well lead to collapsar formation.

Note that the situation is different for model m39\_B0,
which has only reached an explosion energy
of $1.1\times 10^{51}\, \mathrm{erg}$ by the end of the simulation and whose baryonic PNS mass is already $2.04\,M_\odot$ \citep{2020MNRAS.494.4665P}. This makes black hole formation by fallback considerably more likely.

The two models only reach modest kick velocities of $165\,\mathrm{kms}^{-1}$ for model m39\_B12 and  $130\,\mathrm{kms}^{-1}$ for model m39\_B10. 
The rather low kick can be explained by the bipolar explosion geometry. It was already noted
for neutrino-driven CCSN models that bipolar explosions tend to produce lower kicks \citep{scheck_06,mueller_19} and break the trend towards  higher kicks at high explosion energies that has been observed in many neutrino-driven models \citep{mueller_19}.

\begin{figure}
    \centering
    \includegraphics[width=\linewidth]{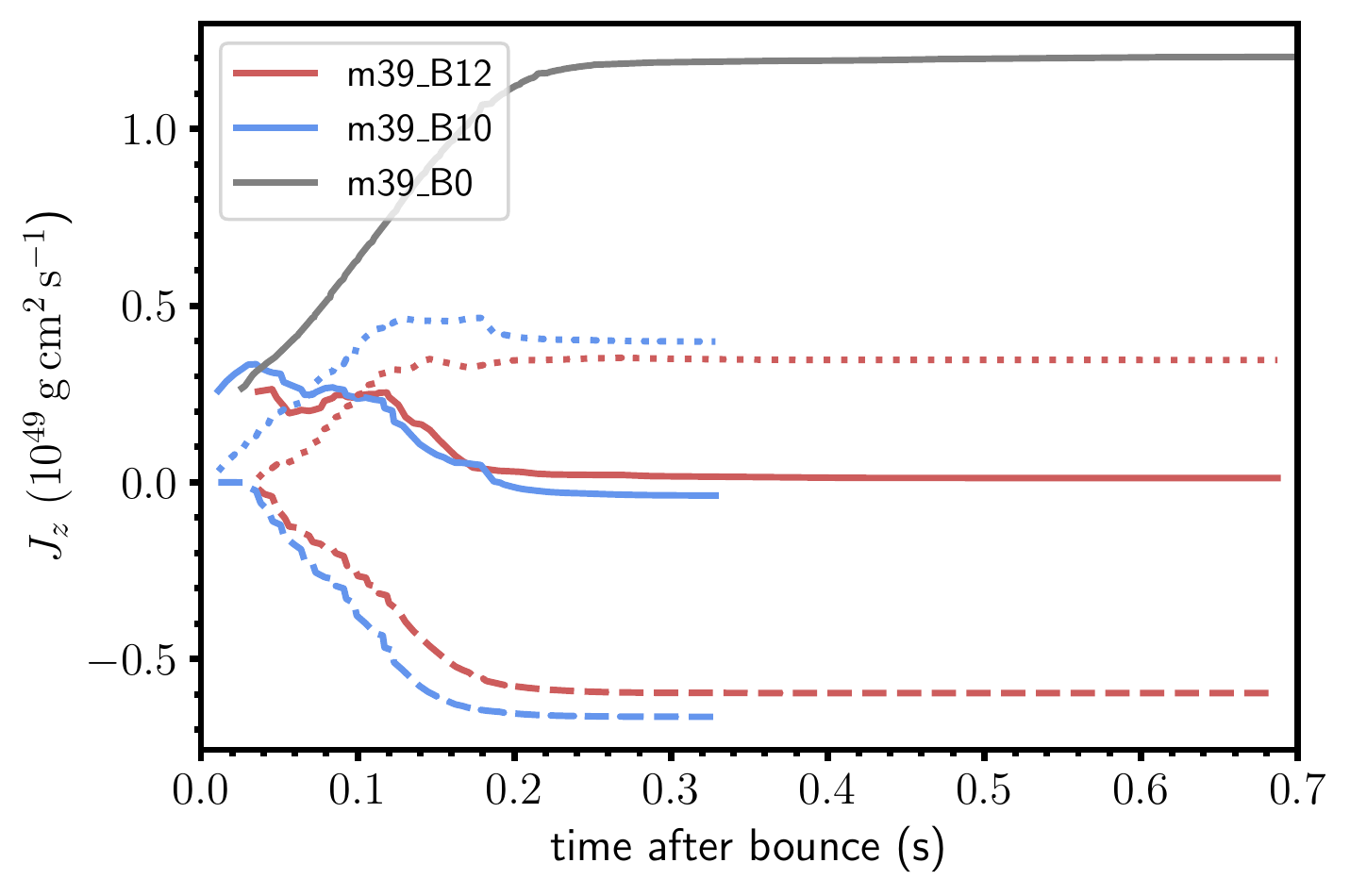}
    \caption{Time evolution of the $z$-component $J_z$ of the angular momentum for all models, together with the time integral of magnetic torques (dashed) and hydrodynamic angular momentum fluxes (dotted) for models m39\_B10 (blue) and
    m39\_B12 (red).}
    \label{fig:spin_z}
\end{figure}

Primarily due to magnetic torques, the PNS is spun down considerably in models m39\_B12 and m39\_B10. Its angular momentum is reduced by about a factor of ten to about
$4\texttt{-}5\times 10^{47}\, \mathrm{g}\,\mathrm{cm}^2 \mathrm{s}^{-1}$
at the end of the simulation in both cases.
To quantify the contribution of magnetic torques
to the spin-down, we consider the $z$-component $J_z$ of the angular momentum in Figure~\ref{fig:spin_z}, which also shows the time integral of the magnetic torques and hydrodynamic angular momentum fluxes separately. The magnetic torques initially balance and later outweigh the advective angular momentum fluxes, bringing the final PNS angular momentum down to small values.
Figure~\ref{fig:spin_z} reveals that the spin direction of the PNS is even reversed by magnetic torques in model m39\_B10; the retrograde rotation is very slow compared to the initial spin rate, however.

Assuming a final neutron star radius $R$ of $12\, \mathrm{km}$ and using the approximation of
\citet{lattimer_05} for neutron star moment of inertia $I$,
\begin{align}
    I&=0.237 M_\mathrm{grav}
    R^2 \left[1+4.2\left(\frac{M_\mathrm{grav} \, \mathrm{km}}{M_\odot R}\right)+90\left(\frac{M_\mathrm{grav} \, \mathrm{km}}{M_\odot R}\right)^4\right]
    \\
    &\approx  1.6\times 10^{45}\,\mathrm{g}\,\mathrm{cm}^2,
\end{align}
this translates into a neutron star spin period of about $20\,\mathrm{ms}$. Without substantial further accretion, about  $8\times 10^{49}\, \mathrm{erg}$ of rotational energy would thus be available for long-term powering of the explosion and the supernova light curve by magnetar spin-down. This is negligible for the overall energy budget of the explosion. More concerningly, the rapid spin-down also makes it unlikely that the classical millisecond-magnetar scenario \citep{usov_92,duncan_92} for long GRBs would work for our model. In this scenario, the rotational energy would have to be extracted and channelled into \emph{relativistic} jets with low baryon loading after the environment of the neutron star has largely been evacuated (whereas the jets emerging early on remain non-relativistic).  However, 
the rotational energy is clearly well below the  typical energy of the relativistic jet components in long GRBs $\mathord{\sim}10^{51}\, \mathrm{erg}$ \citep{woosley_06}. In a sense, the extraction of rotational energy from the PNS early on is \emph{too} efficient at early times to fit the requirements for long GRBs. 
We note that the
decrease in rotational energy of the PNS (assuming the aforementioned moment of inertia) and hence the
mechanical work exerted by the PNS on its environment during
spin-down from an initial angular momentum of about $2.5\texttt3 \times 10^{48}\,\mathrm{g}\,\mathrm{cm}^2$ 
in a short period close to $200\, \mathrm{ms}$ after bounce
roughly accounts for the rise of the explosion energy at the same time, although subsidiary energy input by neutrino heating may also contribute. By contrast, the PNS grows
to about $2\,M_\odot$ by accretion in model m39\_B0, and experiences no substantial spin-down, and hence does not release much of its potential energy into outflows.

The remaining rotational energy could, however, still have a significant impact on the transient light curve.
Judging by the magnetar-powered light curve models in Figure~31 of \citet{ertl_20},
 the rotational energy would be sufficient to generate light curves peaking at 
$\mathord{\sim} 10^{43}\, \mathrm{erg}\, \mathrm{s}^{-1}$ in bolometric luminosity
assuming strong surface fields of several $10^{14}\, \mathrm{G}$, i.e., well in the realm
of observed hypernovae \citep{woosley_99,milisavljevic_15},
but considerably below the range for superluminous supernovae.
However, it is not even clear whether the neutron star will emerge with large surface
field strengths. Estimating the ultimate surface field strengths of the neutron star is difficult based on
simulations results after a few hundred milliseconds. By the end
of the simulation, the isotropic field strength at a density of $10^{12}\, \mathrm{g}\, \mathrm{cm}^3$
is about $6.4\times 10^{14}\, \mathrm{G}$ for model m39\_B12
and $9.6\times 10^{14}\, \mathrm{G}$ for m39\_B10. The dipole field strengths at the
same radius are only $1.1\times 10^{13}\, \mathrm{G}$ and $6.6\times 10^{13}\, \mathrm{G}$, respectively, where we define isotropic field strength as 
\begin{equation}
B_\mathrm{iso} = \sqrt{ \int B^2 \, \mathrm{d}A }
\end{equation}

Remarkably, the spin and kick direction are not aligned. Because of the bipolar explosion geometry, and because ejecta outside the jets carry considerable momentum, the kick points away from the progenitor's rotation axis, whereas
the spin direction remains close to the initial rotation axis. Thus, 3D simulations of either neutrino-driven explosions or magnetorotational explosions still do not furnish a viable explanation for spin-kick alignment \citep{wongwathanarat_13,mueller_19,2020MNRAS.494.4665P,janka_22}, which may instead need to be establish by mechanisms that operate on longer time scales \citep{janka_22}.

As for the mass of the compact remnant, the possibility of fallback accretion introduces significant uncertainty on the PNS spin, kick, and surface magnetic field. Spin-up by late-time accretion can be very efficient  \citep{chan_20,stockinger_20} and could potentially trigger another activity phase
of a magnetohydrodynamic engine, possibly
with a GRB phase at late times.

\begin{figure}
\includegraphics[width=\columnwidth]{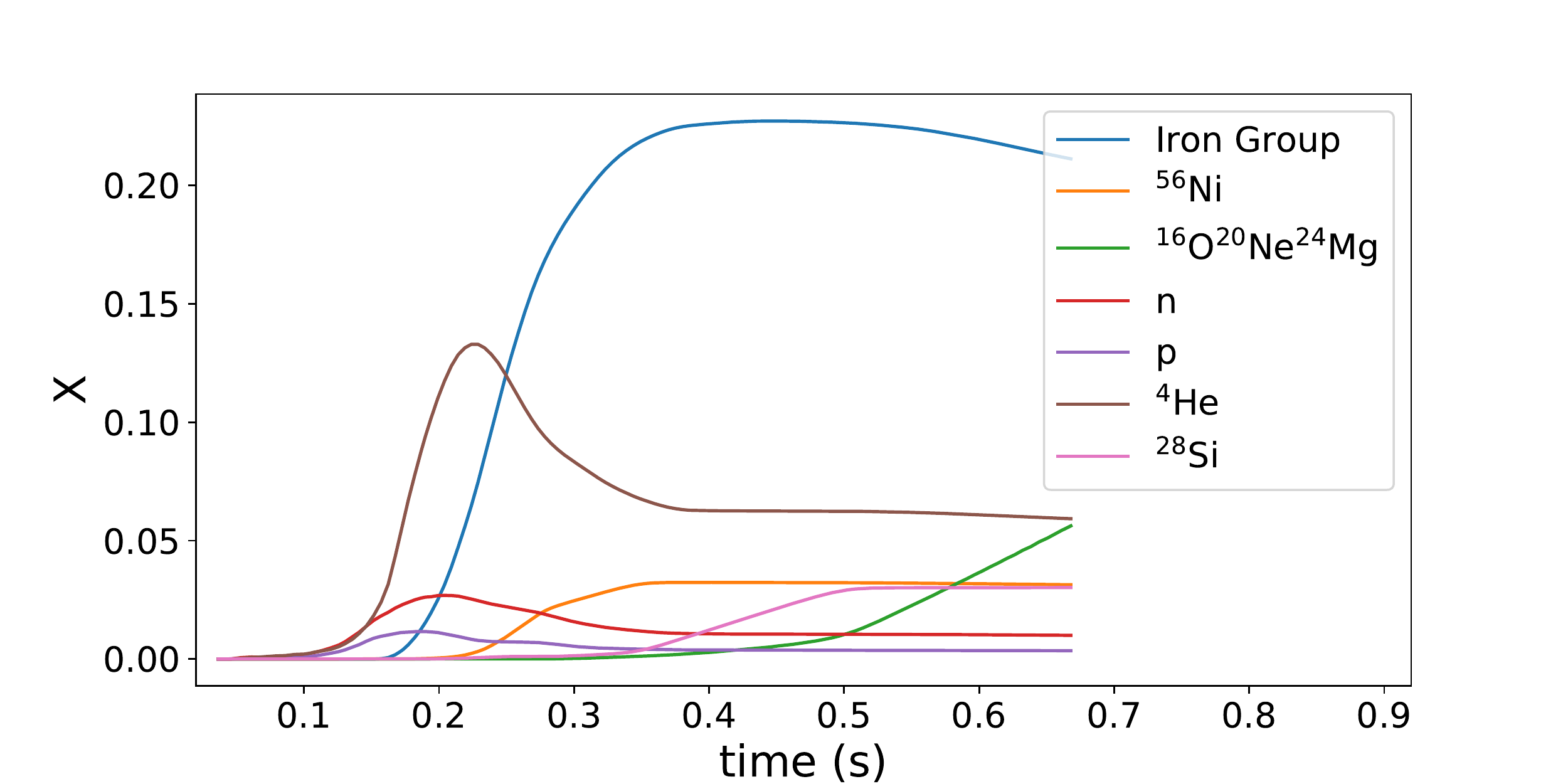}
\includegraphics[width=\columnwidth]{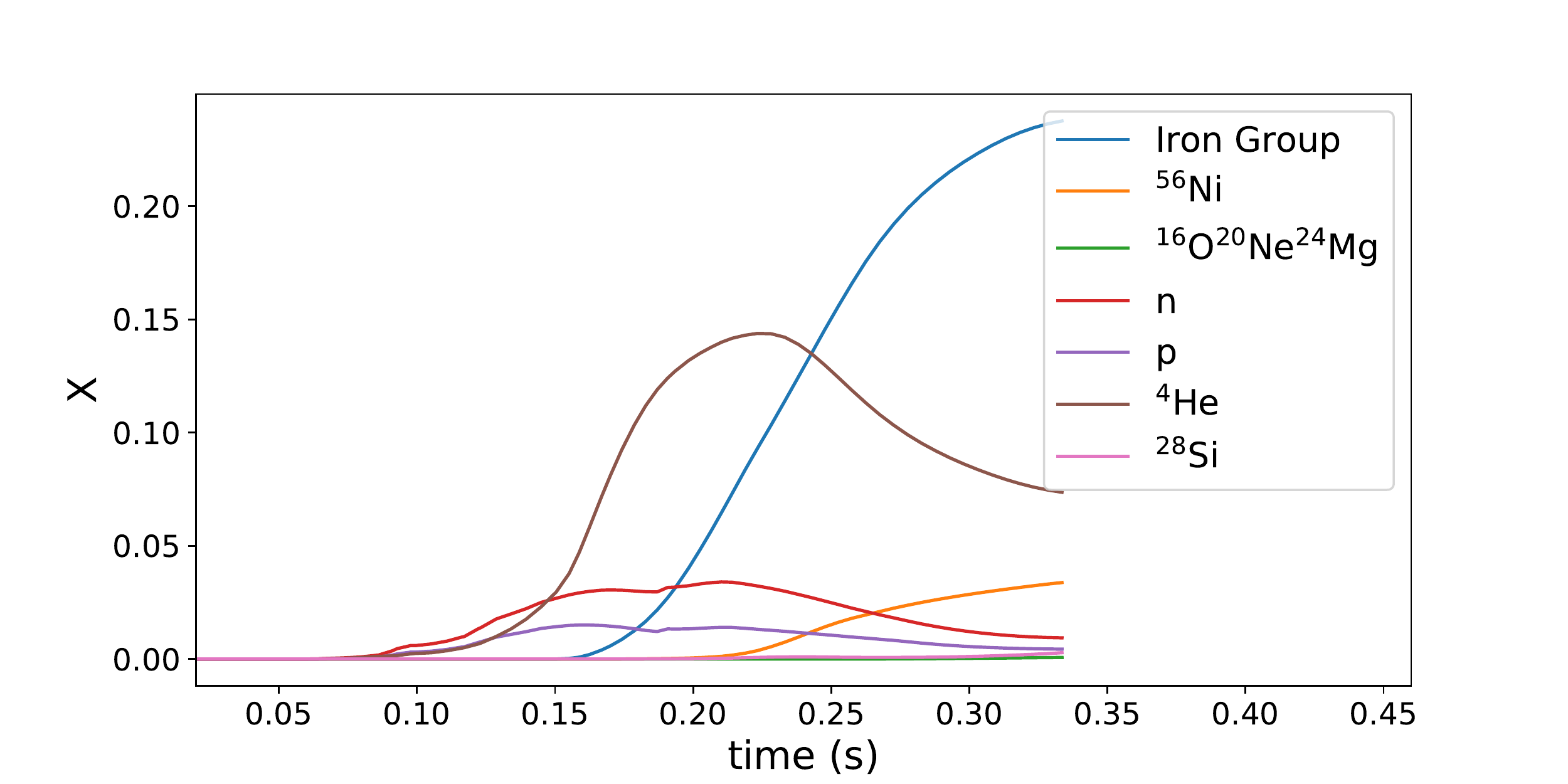}
\includegraphics[width=\columnwidth]{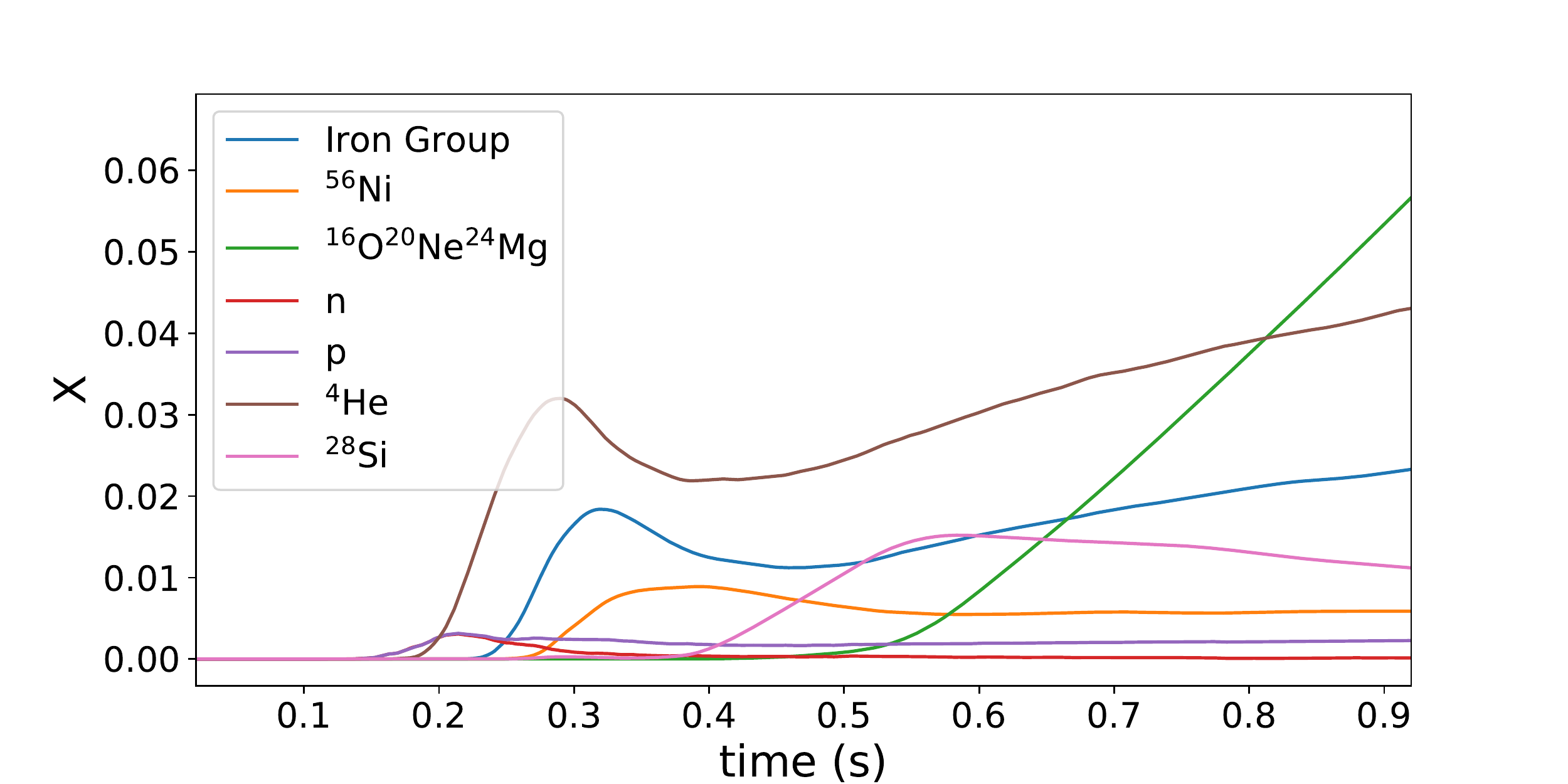}
\caption{The ejecta composition for all our models. The top panel is model m39\_B12, the middle panel is model m39\_B10, and the bottom panel is the non-magnetic model m39\_B0. }
\label{fig:ejecta}
\end{figure}

\begin{figure}
    \centering
    \includegraphics[width=\linewidth]{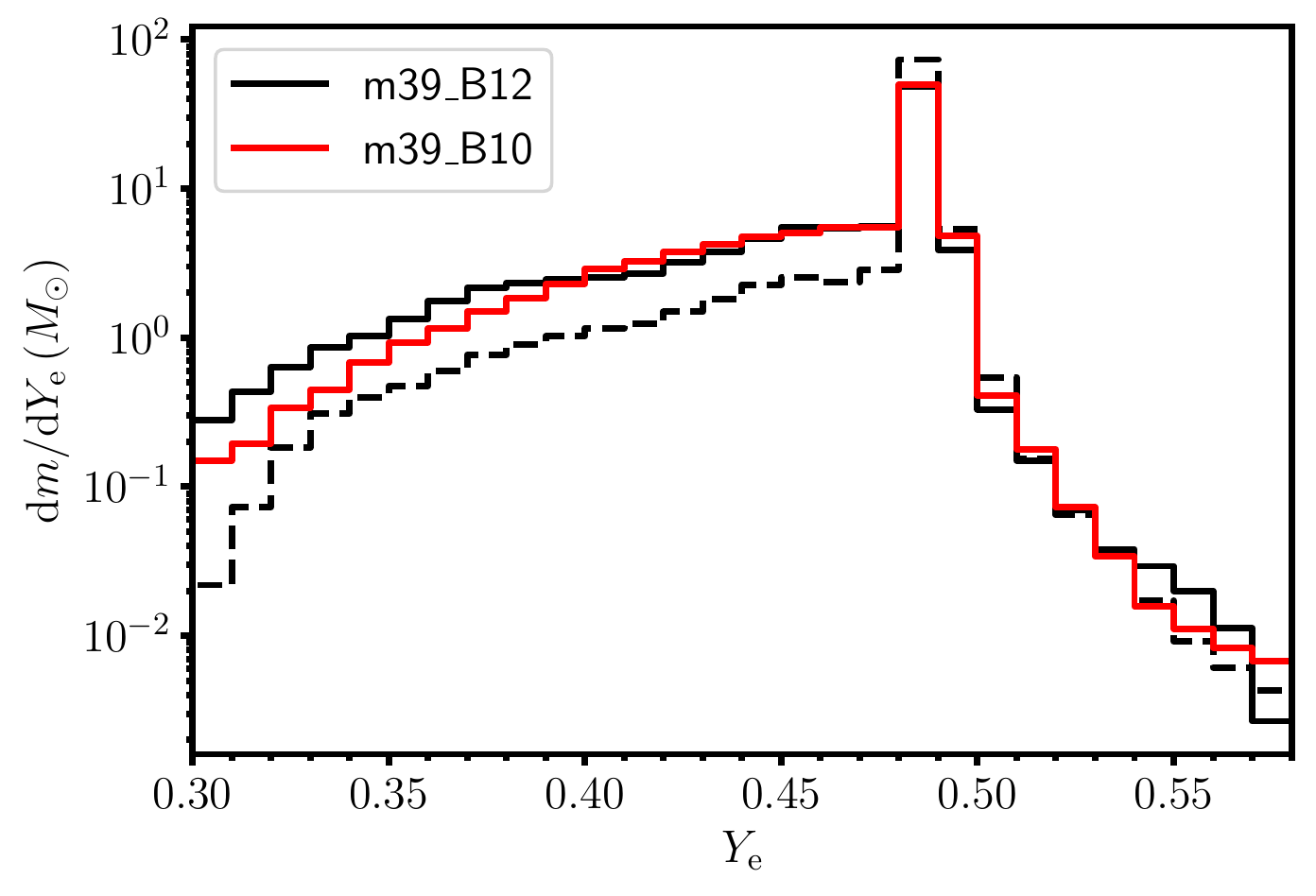}
    \caption{Distribution of the electron fraction $Y_\mathrm{e}$ in the ejecta of model m39\_B12 and m39\_B10 around
    $330\, \mathrm{ms}$ (solid) and
    $680\, \mathrm{ms}$ (dashed, model m39\_B12 only). The models are characterised by an extended neutron-rich tail down to $Y_\mathrm{e}\approx 0.3$. The peak of the distribution lies around $Y_\mathrm{e}\approx 0.49$ because the
    bulk of the material shocked at early times originates from the slightly neutron-rich silicon shell in the progenitor.}
    \label{fig:histograms}
\end{figure}

\begin{figure*}
    \centering
    \includegraphics[width=0.48\linewidth]{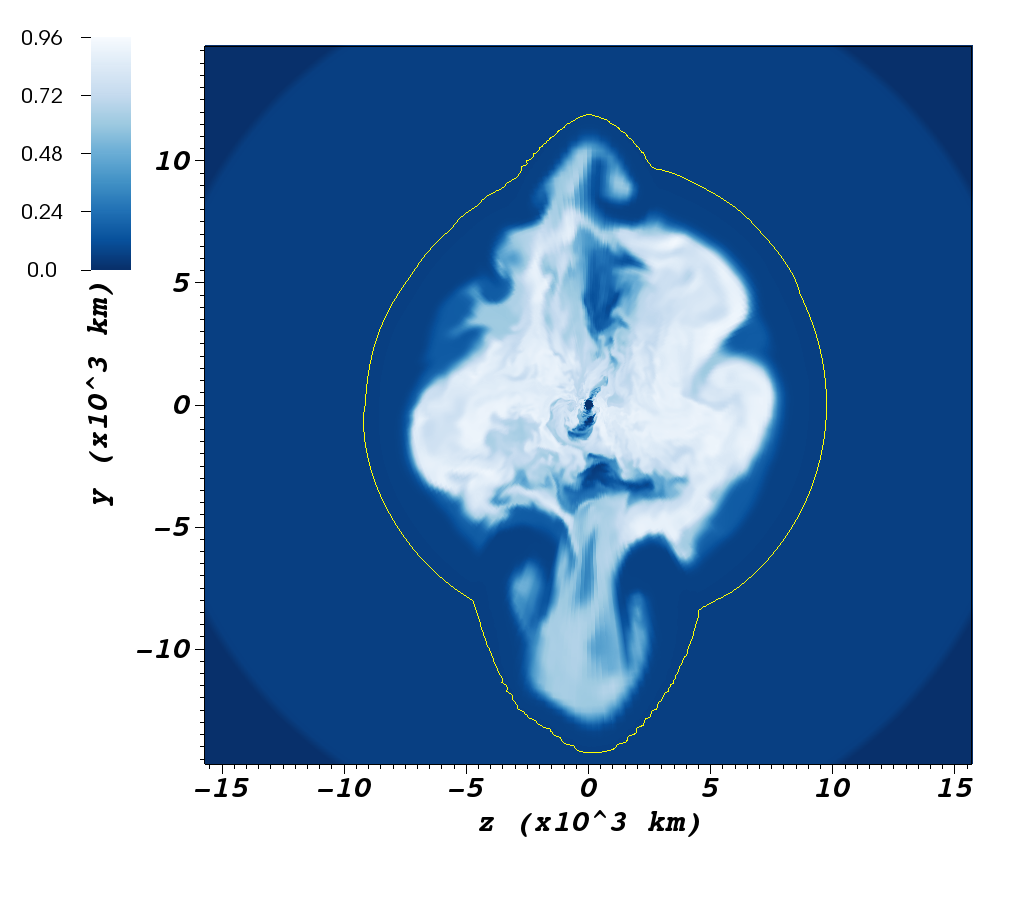}
    \hfill
    \includegraphics[width=0.48\linewidth]{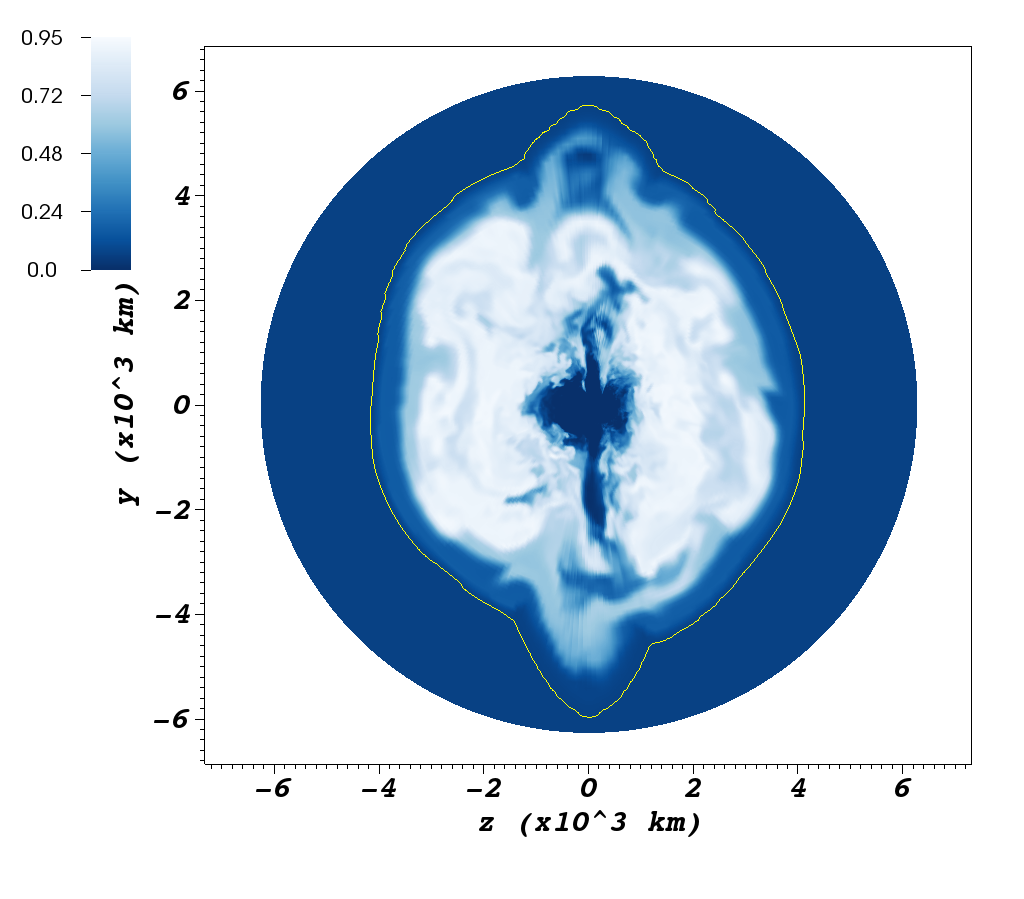}
    \caption{Iron-group mass fraction $X_\mathrm{IG}$ on meridional slices through model m39\_B12 (left) and m39\_B10 (right) at the end of the respective simulations. The shock surface is shown as a yellow line.
    Note that there is little bipolar asymmetry in the iron-group ejecta as yet. In case of model m39\_B12, a more bipolar structure is starting to merge as the jet punches through the slower ejecta.}
    \label{fig:ejecta2d}
\end{figure*}

\begin{figure}
    \centering
    \includegraphics[width=\linewidth]{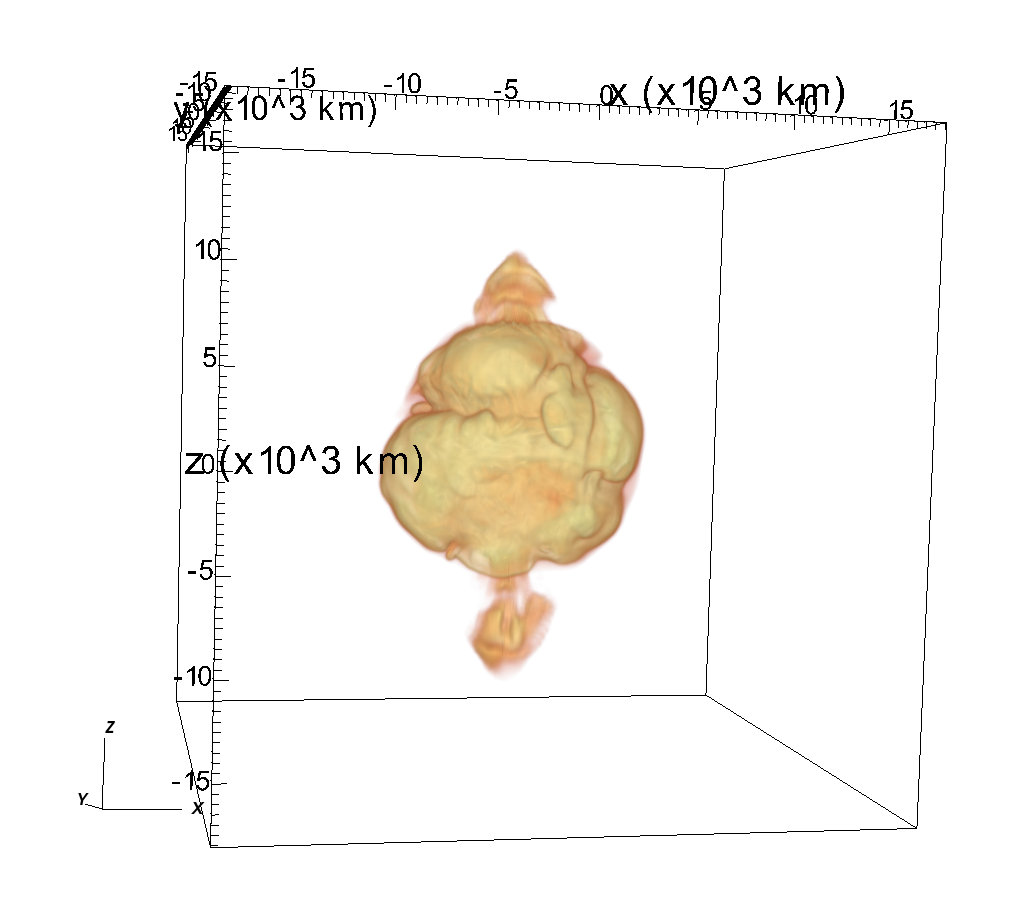}
    \caption{Volume-rendering of the distribution of iron-group ejecta at the end of the simulation for model m39\_B12. The mass concentration $\rho X_\mathrm{IG}$ of iron-group nuclei is used as opacity variable.}
    \label{fig:ejecta3d}
\end{figure}

%%%%%%%%%%%%%%%%%%%%%%%%%%%%%%%%%%%%%%%%%%%%%%%%%%%
%%%%%%%%%%%%%%%%%%%%%%%%%%%%%%%%%%%%%%%%%%%%%%%%%%%

%%%%%%%%%%%%%%%%%%%%%%%%%%%%%%%%%%%%%%%%%%%%%%%%%%%
%%%%%%%%%%%%%%%%%%%%%%%%%%%%%%%%%%%%%%%%%%%%%%%%%%%
\section{Ejecta composition}
\label{sec:ejecta}

\begin{figure}
    \centering
    \includegraphics[width=\linewidth]{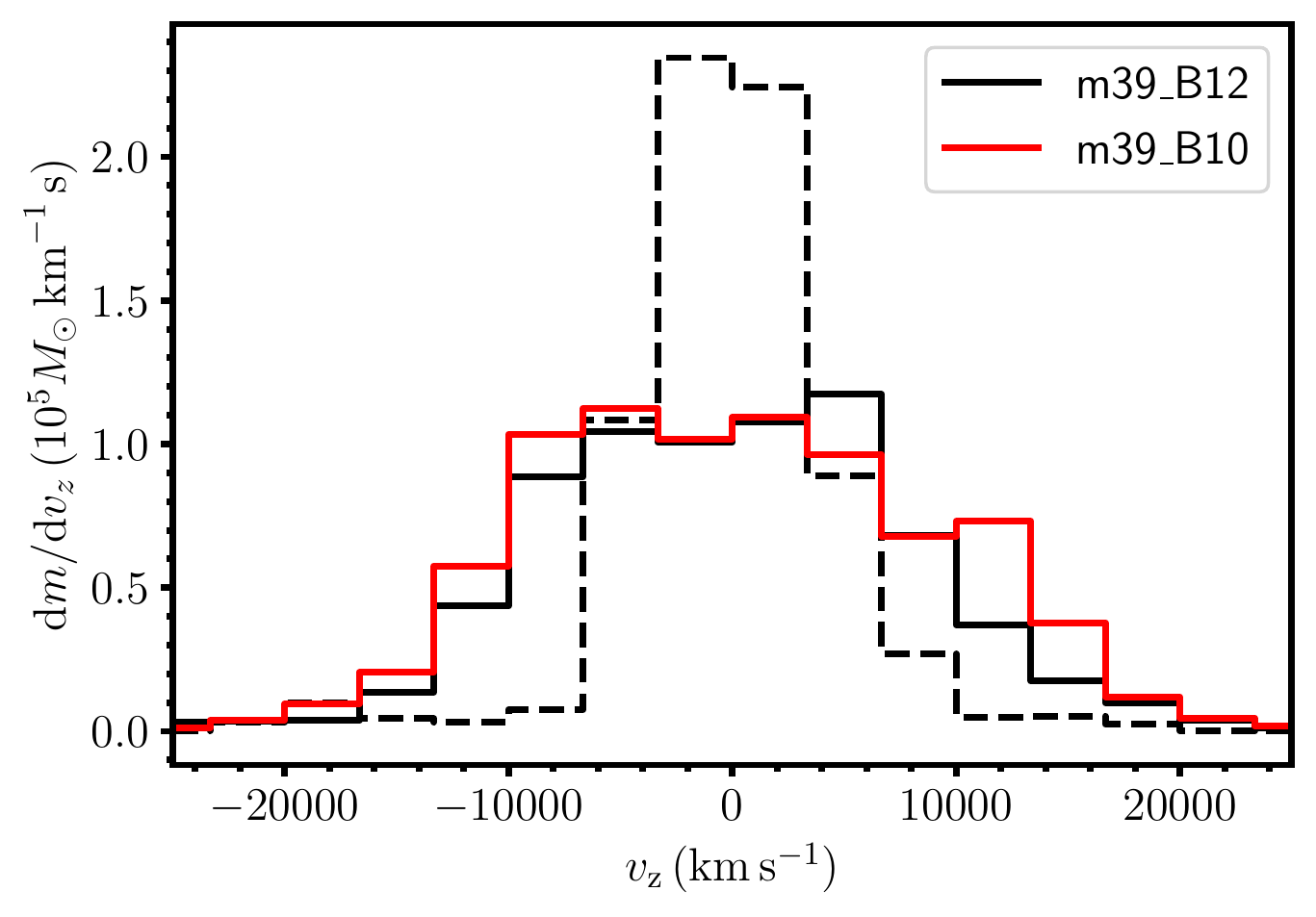}
    \caption{Distribution of the iron-group ejecta versus the velocity component $v_z$ parallel to the rotation axis in model
    m39\_B12 and m39\_B10 around
    $330\, \mathrm{ms}$ (solid) and
    $680\, \mathrm{ms}$ (dashed, model m39\_B12 only). Note that the velocity distribution is narrower in model m39\_B12 simply because the shock has expanded further and the inner ejecta have been decelerated due to the scooping up of matter by the shock.}
    \label{fig:dm_vs_dv}
\end{figure}

For a comparison of the models with the properties of observed hypernovae, the ejecta composition and geometry also needs to be considered. While detailed yields will require post-processing of the simulations using sufficiently big nuclear reaction networks, the flashing scheme in \textsc{CoCoNuT-FMT} provides a reasonable estimate for the amount of iron-group and intermediate-mass elements and their spatial distribution.

We show the time-dependent ejecta composition of our models in Figure~\ref{fig:ejecta}. The breakdown of the ejecta reflects the limitations of our on-the-fly treatment of nuclear burning. The lighter intermediate-mass elements are lumped together
(${}^{16}\mathrm{O}$,
${}^{20}\mathrm{Ne}$, and ${}^{24}\mathrm{Mg}$), and intermediate-mass elements from explosive oxygen burning are all treated as ${}^{28}\mathrm{Si}$. We provide both the total mass of iron-group nuclei and the notional mass of ${}^{56}\mathrm{Ni}$. The actual amount of ${}^{56}\mathrm{Ni}$ is more uncertain than the total amount of iron-group nuclei; production of ${}^{56}\mathrm{Ni}$ in neutrino-processed ejecta requires an electron fraction $Y_\mathrm{e}\gtrsim 0.49$ \citep{hartmann_85,wanajo_18} and is therefore very sensitive to the differences in the electron neutrino and antineutrino emission and therefore considerably affected even by minor uncertainties in the neutrino transport. Further uncertainties come from the initial conditions in the stellar model, as the steller models were calculated using a 21 isotope nuclear network. 

The magnetic models m39\_B12 and m39\_B10 eject
considerable amounts of iron-group material of
more than $0.2\,M_\odot$. The ejecta mass of iron-group
material already asymptotes by about $\sim 300$\,ms after bounce. In model m39\_B12, about
$0.03\,M_\odot$ of ${}^{28}\mathrm{Si}$ are added subsequently by explosive oxygen burning, and after $0.5\,\mathrm{s}$, the shock scoops up oxygen, neon and magnesium. 

The amount of ejected iron-group material is considerably larger than in the non-magnetic model model m39\_B0. However, only 
$0.03\,M_\odot$ are ejected as ${}^{56}\mathrm{Ni}$, which reflects that most of the iron-group ejecta are neutron-rich. Compared to model m39\_B0, the ratio of the ${}^{56}\mathrm{Ni}$ yield to the total iron-group yield is smaller in models m39\_B12 and m39\_B10 (although the total mass of ${}^{56}\mathrm{Ni}$  is larger). This is a result of more neutron-rich conditions in the magnetohydrodynamics models overall .
In Figure~\ref{fig:histograms}, we show histograms of the $Y_\mathrm{e}$-distribution in the ejecta for the two magnetohydrodynamic simulations. Like in other magnetorotational explosion models \citep{2007ApJ...664..416B,2012ApJ...750L..22W,moesta_14b,moesta_18,reichert_22,Varma2021}, we find considerable amount of rather neutron-rich material down to $Y_\mathrm{e}\approx 0.3$, which avoids strong protonisation by neutrino processes due to fast ejection. The bulk of the material (including most of the iron-group ejecta) has an electron fraction close to or slightly below $Y_\mathrm{e}\approx 0.49$. 
Different from the typical situation in neutrino-driven models \citep{wanajo_18,harris_17,sieverding_20},
the distribution of $Y_\mathrm{e}$ peaks below 
$Y_\mathrm{e}=0.5$ because the explosion develops before the infall of the Si/O shell interface. Early on the shock therefore unbinds materials from the silicon shell that has already undergone deleptonisation during hydrostatic burning before core collapse.
Figures~\ref{fig:ejecta} and
\ref{fig:histograms} show that the 
ejecta composition and nucleosynthesis conditions in the two magnetohydrodynamic models are rather similar. Differences in the ``final'' yields and $Y_\mathrm{e}$-distribution
are mostly due to the different run time of the simulations and less so to dissimilar explosion dynamics.

This actually makes it rather unlikely that a major fraction of the iron-group ejecta could end up as $^{56}\mathrm{Ni}$ even if we account for uncertainties in the electron fraction of the neutrino-processed ejecta. Due to the sensitivity of the NSE composition to the difference in electron neutrino and antineutrino  luminosities and mean energies. Given these uncertainties, more material with
$Y_\mathrm{e}>0.49$ might be ejected, but
such uncertainties would not affect the 
considerable amount of iron-group material 
produced by explosive burning of silicon shell material. It is therefore doubtful whether the
two magnetorotational explosion models are not compatible with the nickel masses
of several $0.1\,M_\odot$ inferred from hypernova light curves\footnote{Note that the inference of nickel masses from observed hypernovae light curves rests on the assumption that the light curves are powered by radioactive decay and not by another energy source (e.g., magnetar spin-down, \citealp{woosley_2010}). } and nebular spectroscopy \citep{iwamoto_98,mazzali_01,nakamura_01,maeda_02,woosley_06,mazzali_07}.

Moreover, the ejecta geometry in the magnetorotational explosion models is somewhat at odds with the observational constraints on hypernovae from
spectroscopy and spectropolarimetry, which indicate
a bipolar distribution of iron-group material 
and
a toroidal geometry of the ejected oxygen \citep{mazzali_01,maeda_2008,tanaka_2017}.
In Figures~\ref{fig:ejecta2d} and
\ref{fig:ejecta3d}, we show the iron-group
mass fraction on meridional slices through models m39\_B12 and m39\_B10 at the end of each simulation. In model m39\_B10, there is hardly any indication of a bipolar distribution of iron-group material, which may, however, be due to the fact that the jet has hardly caught up with the expanding shock. In model m39\_B12, we find some indication of emerging bipolar structures. This can be seen more clearly in a volume-rendering of the density of iron-group material in Figure~\ref{fig:ejecta3d}, which shows jet-driven ``arrowheads'' of iron-group material. Even in model m39\_B12, the jet heads only contain a small part of the iron-group ejecta, and
most of the iron-group material is still organised in a fairly spherical structure. 
The iron-group elements in the ``arrowheads'' do not originate from the thin, highly-collimated high entropy structures or from explosive burning in the bow-shock. They primarily originates from entrainment of material surrounding the jet. When the jet starts to bulge out the shock, post-shock temperatures are already too low to add more iron-group material directly by explosive burning.

The lack of a clear bipolar structure can also be seen by considering the distribution of the iron-group material in velocity space.
Figure~\ref{fig:dm_vs_dv} shows the distribution of the velocity component $v_z$ parallel to the rotation axis in the iron-group ejecta (which would be the line-of-sight velocity for an observer in the polar direction).
In model m39\_B12, the bulk of the iron-group material has velocities $v_z$ well below $10,000\,\mathrm{km}\,\mathrm{s}^{-1}$
by the end of the simulation. Only minute amounts of iron-group material reach 
the large velocities well in excess of $10,000\, \mathrm{km}\, \mathrm{s}^{-1}$  observed in hypernovae \citep{modjaz_16}.
Large ejecta velocities in m39\_B10 are
merely due to the shorter simulation time.

If these models were to be compatible with
observed hypernovae, a more bipolar distribution of iron-group material would have to emerge later. It cannot be excluded that late-time mixing instabilities could redistribute the iron-group material appreciably, but at present it is not certainly not obvious that the magnetorotational explosions match the characteristics of observed hypernovae.

%%%%%%%%%%%%%%%%%%%%%%%%%%%%%%%%%%%%%%%%%%%%%%%%%%%
%%%%%%%%%%%%%%%%%%%%%%%%%%%%%%%%%%%%%%%%%%%%%%%%%%%
\section{Gravitational Waves}
\label{sec:grav_waves} 
\subsection{Amplitudes and Spectrograms}
\begin{figure*}
\includegraphics[width=\textwidth]{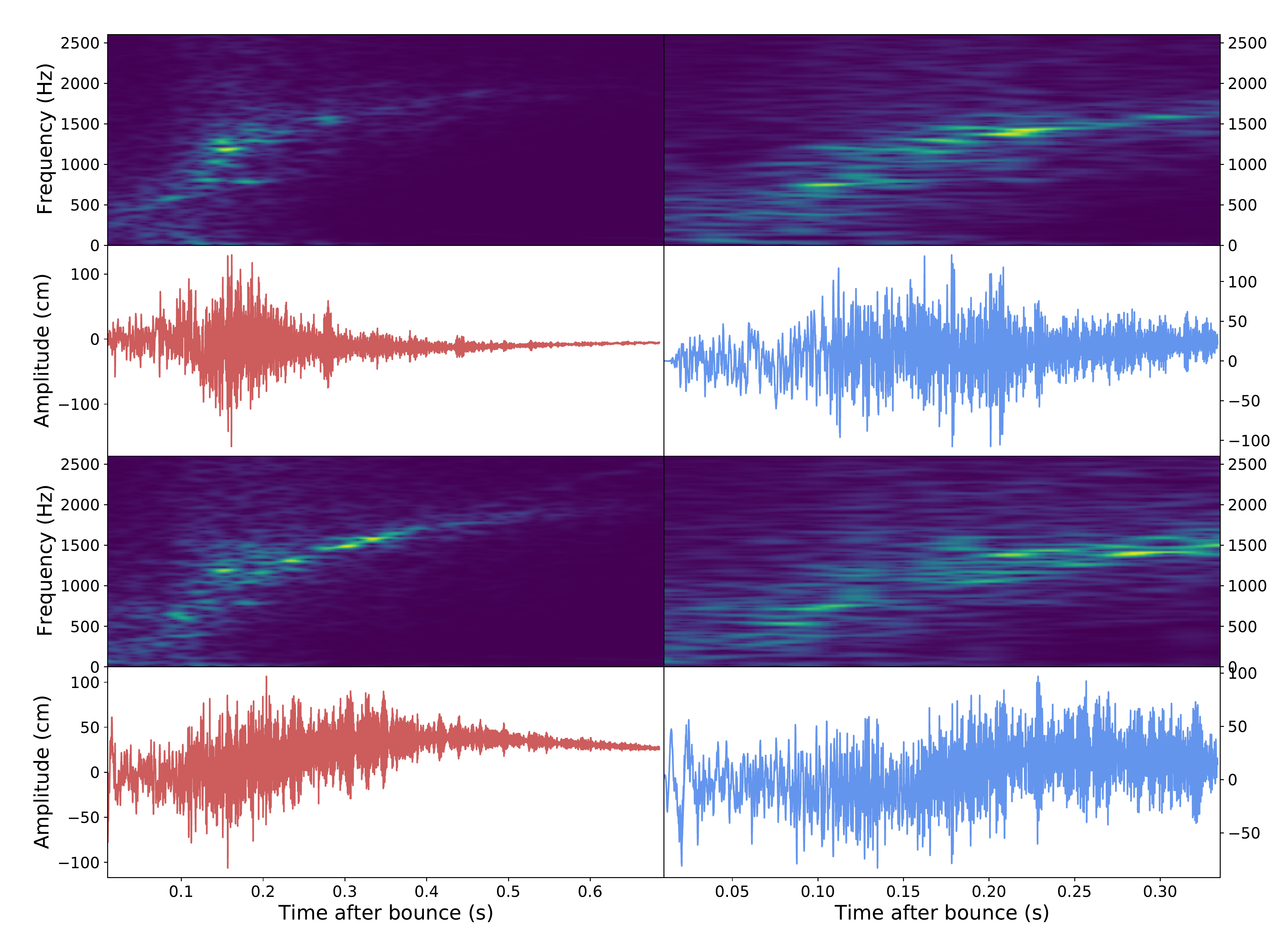}
\caption{Amplitudes $h D$ and spectrograms of the plus polarisation mode of our two models as measured at the pole and the equator. Top left: Model m39\_B12 at the pole. Bottom left: Model m39\_B12 at the equator. Top right: Model m39\_B10 at the pole. Bottom right: Model m39\_B10 at the equator. The GW amplitudes are high for less than half a second and reach frequencies of $\sim 2,000$\,Hz. The GW signals are similar in the cross polarisation. 
}
\label{fig:gravwaves}
\end{figure*}

For each of our models, the time series and spectrograms 
(evaluted using \textsc{Matplotlib}'s \texttt{specgram} function) of the GW emission are shown in Figure~\ref{fig:gravwaves}. We show distance-independent amplitudes for the plus polarisation mode, i.e., the product of strain $h D$ of strain $h$ and observer distance $D$. 
The cross polarisation amplitudes and spectrograms are very similar.
At the pole, model m39\_B12  reaches a maximum amplitude of 165\,cm in the plus polarisation mode, and an amplitude of 152\,cm in the cross polarisation. At the equator, the amplitudes are slightly smaller, namely 106\,cm in the plus polarisation and 121\,cm in the cross polarisation. For model m39\_B10 model, the maximum amplitude at the pole  is 133\,cm in the plus polarisation and 152\,cm  in the cross polarisation. At the equator, this model also has smaller amplitudes, with 96\,cm in the plus polarisation and 109\,cm in the cross polarisation. Most of the detectable GW amplitudes occur within the first 400\,ms and decrease significantly around the same time that the explosion energy starts to reach its final value. 

These amplitudes are significantly \emph{higher} than in the 2D simulations of these models in \citet{Jardine_2021}, which only reached amplitudes of 76\,cm for m39\_B10, and 62\,cm for m39\_B12 (disregarding the late-time tail signal). They found that model m39\_B12 has a smaller GW amplitude than m39\_B10 (which exploded late in 2D and hence had a longer phase of accretion-phase GW emission), which is the opposite of what we see in our 3D models.  On the other hand, GW emission in  m39\_B12 tapers off more clearly at late times in our 3D simulation compared to the 2D model of \citet{Jardine_2021}. The GW amplitudes are also significantly larger than 
the maximum amplitudes of about $40 \, \mathrm{cm}$ in
the non-magnetised model m39\_B0, which is reflective of the more powerful explosions in the MHD models. 
Model m39\_B12  develops a small tail in the plus polarisation only at the equator. Model m39\_B10 also develops a very small tail, but it is visible at all observer directions. The tail also differs from the 2D case, as \citet{Jardine_2021} show a significantly larger tail for model m39\_B12 in 2D. The difference to the 2D case can be explained by the more modest bipolar deformation of the shock.

A dominant high-frequency emission band is clearly visible in the spectrograms of both models. It is evident somewhat more clearly than in the 2D counterpart of  m39\_B12 in \citet{Jardine_2021} that the structure of the oscillation modes underlying GW emission is not \emph{qualitatively} changed
from the familiar picture in neutrino-driven CCSNe,
where GW emission is predominantly due to the $f$-mode or the $^{2}g_2$-mode \citep{2018ApJ...861...10M, torres-forne_universal_2019, 2021PhRvD.104l3009S}. 
The high-frequency band reaches up $\sim2,000$\,Hz by the end of the simulation in model m39\_B12 and about $\sim$1,500\,Hz in model m39\_B10. The frequencies are higher than in our recent neutrino-driven simulations, which is due to the different treatment of gravity used in our MHD simulations; as noted by \citet{mueller_13} GW frequencies are systematically overestimated by $\mathord{\sim}20\%$ in pseudo-Newtonian simulations due to the omission of metric terms in the equations of hydrodynamics.

\begin{figure}
\includegraphics[width=\columnwidth]{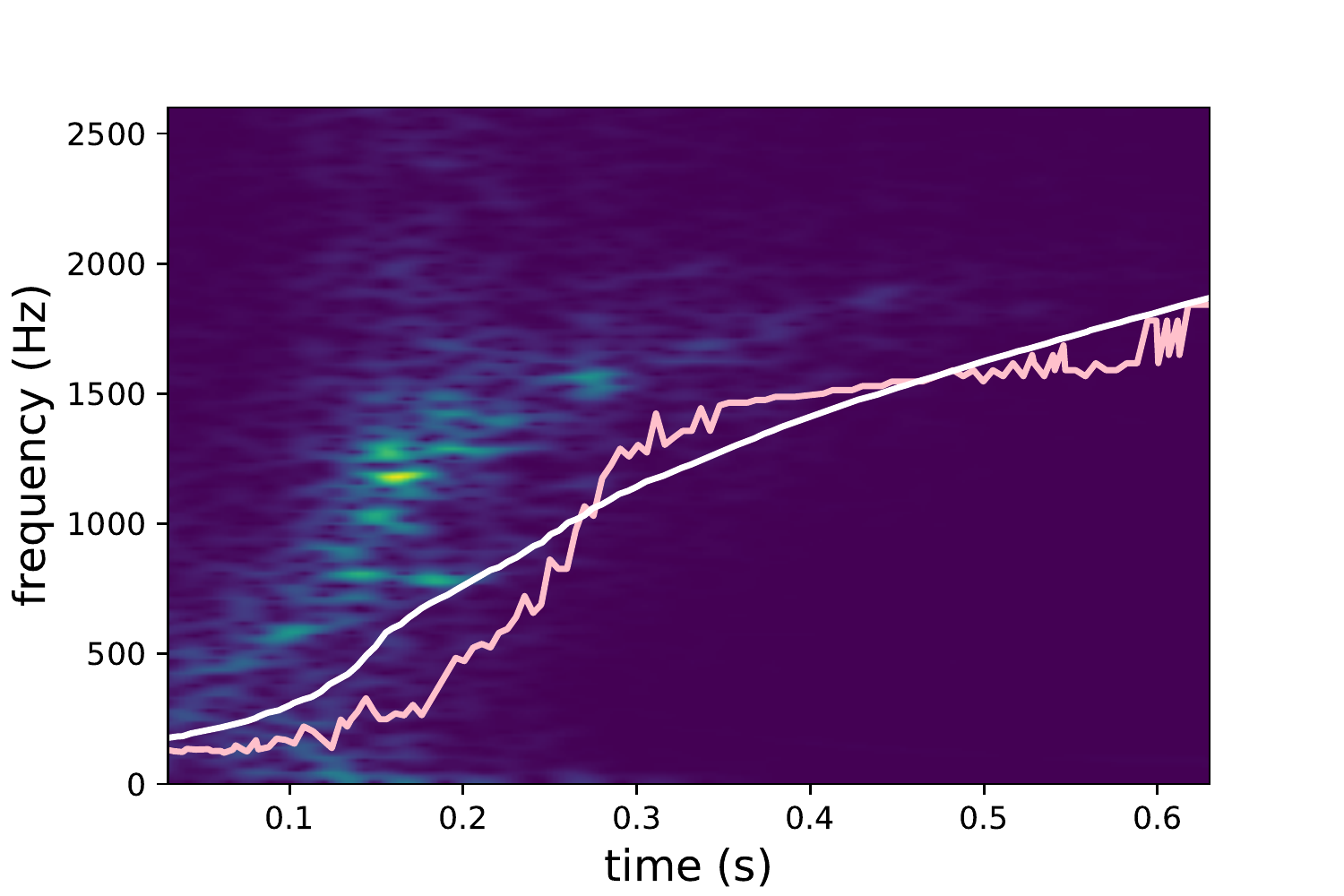}
\includegraphics[width=\columnwidth]{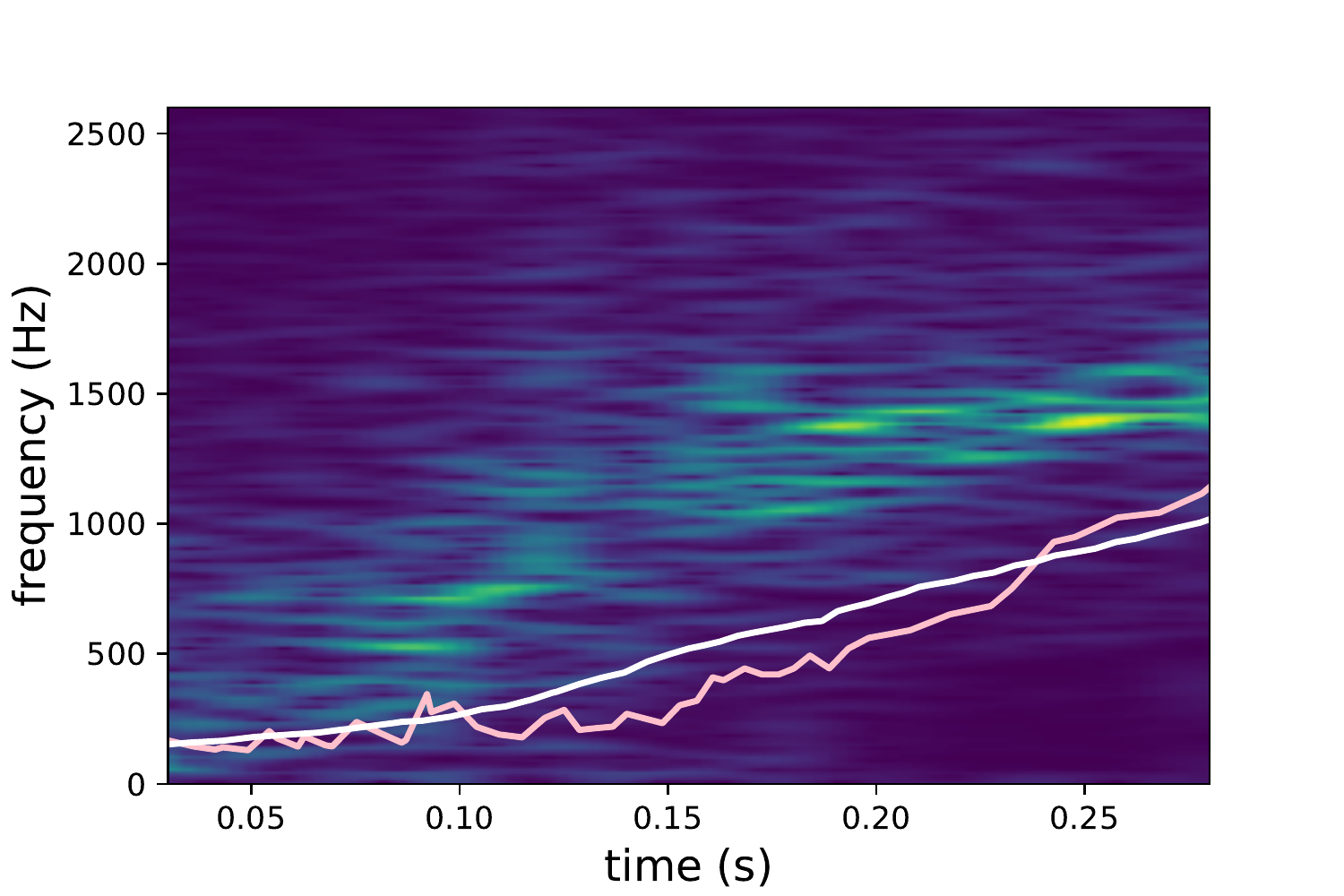}
\caption{Comparison of the GW spectrograms of the plus polarisation of model m39\_B12 (top) and  m39\_B10 (bottom) 
to the analytic relation for the f/g-mode frequency given by
Equation~(\ref{eq:fpeaks}) (white lines) and the $^{2}g_2$ universal relation from \citet{torres-forne_universal_2019} (pink lines). The analytic relation does not provide a good fit when magnetic fields and rapid progenitor rotation are included. }
\label{fig:gw_freqs}
\end{figure}

Similar to \citet{2020MNRAS.494.4665P}, we do, however, find that the high-frequency emission band deviates quantitatively from the relations established for neutrino-driven supernovae of non-rotating progenitors.
In the 2D case, \citet{Jardine_2021} still found that the time dependence of the high-frequency f/g-mode emission band was well matched by the formula of \citet{mueller_13},
\begin{align}\label{eq:fpeaks}
 f_{\text {peak }}  \approx \frac{1}{2 \pi} \frac{G M_{\text {by}}}{R_{\text {PNS}}^{2}} \sqrt{1.1 \frac{m_\mathrm{n}}{\left\langle E_{\bar{v}_{e}}\right\rangle}}, 
 \end{align}
 where $R_{\text{PNS}}$ is the radius of the PNS, $E_{\bar{v}_{e}}$ is the electron antineutrino mean energy, $m_\mathrm{n}$ is the neutron mass and $M_{\text {by}}$ is the baryonic mass of the PNS. As shown Figure~\ref{fig:gw_freqs}, we find this relation does not match the high-frequency emission in 3D models. We also show the $^{2}g_2$ mode universal relation from \citet{torres-forne_universal_2019}, which also does not match. The predicted frequencies are lower than what is observed in our models by $\sim 500$\,Hz.
The general shape of the time-dependence is somewhat similar, but there is clear quantitative mismatch. For m39\_B12, the actual frequency trajectory has something more of an S-shape that bends up noticeably just before $0.2\, \mathrm{s}$ after bounce. As discussed in \citet{2020MNRAS.494.4665P,Jardine_2021}, the impact of rotation and magnetic fields on PNS oscillation eigenfunctions and eigenfrequencies is not trivial and can just be stated as an empirical finding at this point. In future, linear eigenmode analysis \citep{torres-forne_universal_2019, 2021PhRvD.104l3009S} will need to be extended to include rotation and magnetic fields to understand this effect. For now it remains important to consider that empirical scaling laws for the frequency of the dominant f/g-mode become uncertain in the presence of rapid rotation and magnetic fields and should be applied with care to prospective CCSN GW signals if rapid core rotation is determined from the bounce signal \citep{2014PhRvD..90d4001A} or the observed explosion is very energetic and possibly magnetorotational in origin.

The emission frequency at the time of strongest emission is of critical importance for the prospects of detecting a CCSN in GWs. In Table~\ref{tab:detection}, we therefore show the frequency at which the GW amplitude of the signal peaks. This peak frequency is between 1100\,Hz and 1400\,Hz for both models in different observer directions. We note that the peak frequencies, the overall width of the spectrum, and the  time-frequency signal morphologies are very similar to the 2D case \citep{Jardine_2021}. Importantly, this  confirms that there is no significant amount of extra power above
$2000\,\mathrm{Hz}$, which might accidentally be cut out by current choices of cut-off frequencies in event analysis \citep{2020PhRvD.101h4002A, 2021PhRvD.104l2004A}.

%%%%%%%%%%%%%%%%%%%%%%%%%%%%%%%%%%%%%%%%%%%%%%%%%%
%%%%%%%%%%%%%%%%%%%%%%%%%%%%%%%%%%%%%%%%%%%%%%%%%%%
\subsection{Detection Prospects}
\label{sec:detect}

\begin{figure}
\includegraphics[width=\columnwidth]{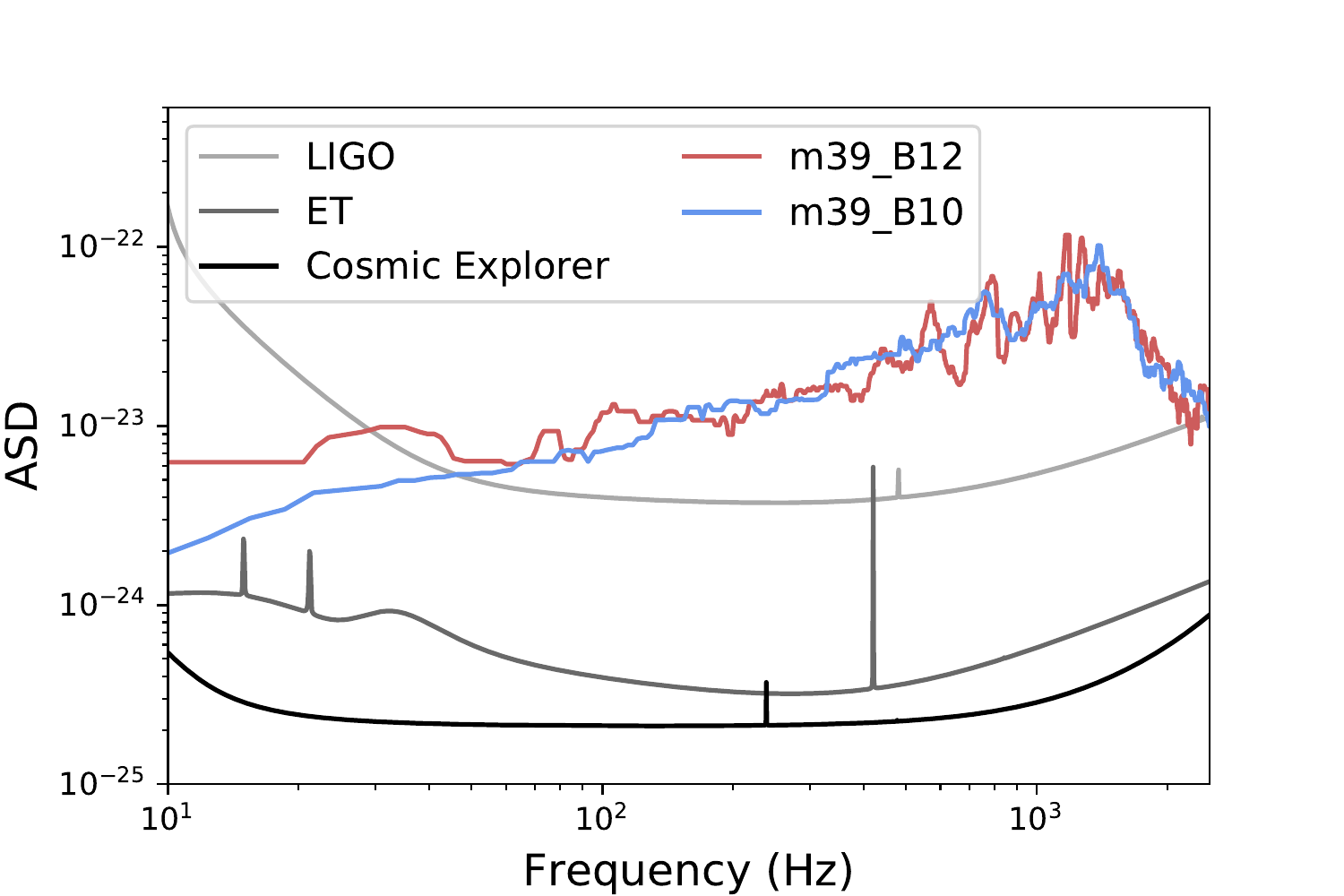}
\caption{The amplitude spectral density of our models as seen from the pole for a distance of 100\,kpc. The GW signals peak frequency is reasonably high at over 1000\,Hz, but still within the sensitive frequency band of current GW detectors and the future detectors Einstein Telescope and Cosmic Explorer. }
\label{fig:asd}
\end{figure}

In this section, we estimate the distances out to which our models could be detected by current and future GW detectors. In Table~\ref{tab:detection}, we show the maximum detectable distances for each of our models for two different observer directions.  

The maximum detectable distance is given by the distance needed for an optimal signal-to-noise ratio (SNR) of 8. 
The optimal SNR is defined as,
\begin{equation}
\rho^{2} = \int^{\infty}_{0} 4 \frac{|\tilde{h}|^{2}}{S(f)} df 
\end{equation}
where $\tilde{h}$ is the frequency-domain GW signal, and $S(f)$ is the GW detectors noise power spectral density (PSD). We give values for a single Advanced LIGO detector \citep{2015CQGra..32g4001L}, Einstein Telescope \citep{2011CQGra..28i4013H} and Cosmic Explorer \citep{2021arXiv210909882E} detectors. The PSD curves we use for the different detectors are shown in Figure \ref{fig:asd}. 
We also determine the maximum detection distances for two different networks of detectors. The first is a network of LIGO, Virgo \citep{2015CQGra..32b4001A} and KAGRA \citep{2021PTEP.2021eA102A} detectors, and the second is a network of Cosmic Explorer detectors 
where the network SNR is defined as,
 \begin{equation}
 \rho_{\mathrm{net}}^{2} =  \rho_{1}^{2}  +  \rho_{2}^{2} 
 \end{equation}
 where $\rho_{1}$ and $\rho_{2}$ are the individual detector SNRs. 

This method does not account for the short-duration transient noise artefacts found in real GW detector noise. We also do not account for the limitations of specific search algorithms. For example, \citet{Szczepanczyk_2021}, show that the current main morphology-independent search for GW bursts can detect CCSNe down to SNRs of $10-20$. 
The core bounce and late time GW emission are not included in our GW models, which will also make the detection distances smaller than if we had full waveforms. 

The detection distances for all our models are shown in Table~\ref{tab:detection}. Model m39\_B10 has the largest detection distances with up 260\,kpc in a single LIGO detector, and up to 2.63\,Mpc in a single Einstein Telescope detector. Note that model m39\_B10 only covers half the simulation time of model m39\_B12, so it is likely that the difference in detection distances would have grown if the models were continued for longer. 

The detection distances are twice as large as those predicted for the same models in 2D in \citet{Jardine_2021}. This is partially due to the larger GW amplitudes in the 3D models and partially due to having both GW polarisation modes in 3D. As expected, the models have larger maximum detection distances than previous 3D simulations of non-rotating neutrino-driven explosions \citep{2019MNRAS.487.1178P, 2020MNRAS.494.4665P}. There are no GW detection studies of long-duration 3D MHD models that we can compare to our results, but it should be noted that based on the post-bounce emission alone our models give comparable detection distance
to MHD simulations of the core-bounce signal only \citep{2020PhRvD.101h4002A, Szczepanczyk_2021}. Different from the bounce signal, the detectability of the post-bounce signal does not depend significantly on the observer direction for our models and is therefore, in a sense, more robust.
The pseudo-Newtonian gravity used in our simulations has likely increased the GW frequency by about 20\%. This will have resulted in a decrease in the predicted detection distances, as the GW detectors have a better sensitivity at the lower frequencies.

\begin{table*}
\centering
 \begin{tabular}{||c c c c c c c c c c c c c||} 
 \hline
  Model & Observer direction & LIGO &  & ET &  & CE &  & LVK &   & 2xCE  &   & Frequency at peak \\ 
   name & (degrees of latitude) & distance &  & distance &  & distance &  & distance &  & distance & & emission  \\
 \hline\hline
  m39\_B12 & 90 & 211\,kpc &  & 2.15\,Mpc &  & 3.66\,Mpc &  & 331.6\,kpc &  & 5.12\,Mpc &  &  1395\,Hz \\
 \hline
  m39\_B12 & 0  & 243\,kpc &  & 2.40\,Mpc &  & 4.11\,Mpc &  & 375\,kpc &  & 5.81\,Mpc &  & 1159\,Hz \\ 
 \hline
  m39\_B10 & 90 & 226\,kpc &  & 2.34\,Mpc &  & 3.96\,Mpc &  & 358\,kpc &  & 5.60\,Mpc &  & 1288\,Hz \\
 \hline
  m39\_B10 & 0  & 260\,kpc &  & 2.63\,Mpc &  & 4.47\,Mpc &  & 405\,kpc &  & 6.32\,Mpc &  & 1337\,Hz \\
 \hline
\end{tabular}
\caption{The maximum detection distances for individual LIGO, Einstein Telescope and Cosmic Explorer detectors. The LVK and 2xCE columns show the distance needed for a network signal to noise ratio of 8 in a network of 2 LIGO detectors, Virgo, and Kagra and a network of two Cosmic Explorer detectors, respectively. }
\label{tab:detection}
\end{table*}

%%%%%%%%%%%%%%%%%%%%%%%%%%%%%%%%%%%%%%%%%%%%%%%%%%%%
%%%%%%%%%%%%%%%%%%%%%%%%%%%%%%%%%%%%%%%%%%%%%%%%%%%%
\section{Conclusions}
\label{sec:conclusion}

In this paper, we perform 3D MHD CCSNe simulations of two $39\,M_{\odot}$ progenitor models with rapid rotation and two different initial magnetic field strengths of $10^{10}$\,G and $10^{12}$\,G. We also compare our results to our previous simulation of this progenitor model without magnetic fields. 

Inclusion of strong magnetic fields resulted in shock revival by the magnetorotational mechanism, which occurred faster than our previous neutrino-driven explosion of this progenitor model with no magnetic fields. Both of the models show strong jets, which are, however, later affected by the kink instability. Model m39\_B12  develops a jet $\sim 150$\,ms after core bounce, but the jet in the North direction quickly disappears and only a jet in the South direction remains. The jets in model m39\_B10 start a little later, but continues to grow larger in both directions up to the end of the simulation. Interestingly, the explosion is not primarily driven by the jets; the shock already expands more or less isotropically to about $1000\, \mathrm{km}$ before fast jets start to emerge. Both models reach high explosion energies that level off only a few hundred milliseconds after the core-bounce. The explosion energy of about $3\times 10^{51}\, \mathrm{erg}$ lies at the low end of the hypernova distribution.

Both of our new models have a significantly smaller final PNS mass than the m39\_B0 model, with a final mass cut within the silicon shell of the progenitor. We do not observe any spin kick alignment in these models despite the alignment of the jets with the rotation axis.
Importantly, the proto-neutron star is spun down on short time scales, so that its final rotation period is expected to be about $20\, \mathrm{ms}$. This would not leave enough rotational energy to launch a relativistic gamma-ray burst jet after the environment of the neutron star has been evacuated, but might still be sufficient to lead to a brightening of the light curve by energy input from pulsar spin-down.

The magnetorotational explosion models produce more than $0.2\,M_\odot$ of iron-group material, which is superficially compatible with values inferred for observed hypernovae based on their light curves and spectroscopy. However, much of the iron-group material is not in the form of $^{56}\mathrm{Ni}$ in our simulations since it is made by explosive burning of slightly neutron-rich material from the silicon shell. If one accepts the explosion dynamics of the models, the resulting nucleosynthesis should be relatively robust as it is not subject to uncertainties in the neutrino irradiation, unlike ejecta from close to the surface of the proto-neutron star. The iron-group material is also distributed quite isotropically by the end of the simulations. The bipolar distribution observed in hypernovae \citep{mazzali_01,maeda_2008,tanaka_2017} would have to emerge later by mixing instabilities.

We calculate the GW emission for our models. Both models have high amplitudes that are significantly larger than the non-magnetic model m39\_B0. The GW signals reach high frequencies of over 1000\,Hz. We show that our models do not fit previous relations for the GW frequency due to the magnetic fields and rapid rotation. The models could be detected up to a maximum distance of 260\,kpc in Advanced LIGO and 4.47\,Mpc in Cosmic Explorer. 

While our models confirm that 3D magnetorotational explosion models are capable of reaching hypernova energies \citep{obergaulinger_21,bugli_21}, they clearly do not yet fit important characteristics of observed broad-lined Type Ic supernovae or magnetar-powered superluminous supernovae, and the newly born neutron star also rotates too slowly to enable a long-gamma ray burst in the millisecond magnetar scenario. They rather suggest that a scenario with a later onset of the explosion (due to weaker initial field or slow build-up of strong, large-scale magnetic fields after collapse) may be preferable over a rapid magnetorotational explosion scenario. Allowing longer accretion before a magnetorotational engine starts to operate, either within the millisecond magnetar or the collapsar scenario, would considerably increase the energy reservoir available for the explosion and might obviate the problem of inefficient nickel production.
Clearly, two models for a single progenitor star are not sufficient to rule out the millisecond-magnetar scenario for hypernovae and long gamma-ray bursts. Nor can it be
excluded that some rapid magnetorotational explosions of moderately high explosion energy
are simply observed in a different guise and not as broad-lined Ic supernovae; e.g., 
it may worth scrutinising explosions with a high Ni/Fe ratio in the nebular phase
more closely, as these could be explained by ejection of silicon-shell material
\citep{jerkstrand_15a,jerkstrand_15b}. 
At the very least our results suggest that requirements for plausible hypernova explosion models need to be investigated in greater detail.
A wider exploration of stellar progenitor models
\citep{obergaulinger_22} and initial field configurations \citep{bugli_21} may reveal a viable sweet spot for
genuine hypernova explosions.
Moreover, our results suggest that the evolution of magnetorotational explosions needs to be studied over longer time scales of seconds. The energetics of our models suggest the potential for fallback at later stages, which may still lead to the formation of a black hole and collapsar disk, or again spin up the PNS to trigger another ``engine'' phase that might further increase the explosion energy and launch a GRB jet. At any rate, fitting the various constraints on optical light curves and spectroscopy,
gamma-ray and X-ray signatures, and nucleosynthesis constraints remains a formidable challenge that also
requires a better connection between the short- and long-term evolution of magnetorotational explosions and
the observational data.

%%%%%%%%%%%%%%%%%%%%%%%%%%%%%%%%%%%%%%%%%%%%%%%%%%%%%%
%%%%%%%%%%%%%%%%%%%%%%%%%%%%%%%%%%%%%%%%%%%%%%%%%%%%%%
\section*{Acknowledgements}

Authors JP and BM are supported by the Australian Research Council (ARC) Centre of Excellence (CoE) for Gravitational Wave Discovery (OzGrav) project number CE170100004.  JP is supported by the ARC Discovery Early Career Researcher Award (DECRA) project number DE210101050.  BM is supported by ARC Future Fellowship FT160100035.  We acknowledge computer time allocations from Astronomy Australia Limited's ASTAC scheme, the National Computational Merit Allocation Scheme (NCMAS), and from an Australasian Leadership Computing Grant. Some of this work was performed on the Gadi supercomputer with the assistance of resources and services from the National Computational Infrastructure (NCI), which is supported by the Australian Government, and through support by an Australasian Leadership Computing Grant.  Some of this work was performed on the OzSTAR national facility at Swinburne University of Technology.  OzSTAR is funded by Swinburne University of Technology and the National Collaborative Research Infrastructure Strategy (NCRIS). D.R.A-D. acknowledges support by the Stavros Niarchos Foundation (SNF) and the Hellenic Foundation for Research and Innovation (H.F.R.I.) under the 2nd Call of “Science and Society” Action Always strive for excellence - ``Theodoros Papazoglou'' (Project Number: 01431).

%%%%%%%%%%%%%%%%%%%%%%%%%%%%%%%%%%%%%%%%%%%%%%%%%%
\section*{Data Availability}

The data from our simulations will be made available upon reasonable requests made to the authors.



\bibliographystyle{mnras}
\bibliography{example} % if your bibtex file is called example.bib

\begin{thebibliography}{}
\makeatletter
\relax
\def\mn@urlcharsother{\let\do\@makeother \do\$\do\&\do\#\do\^\do\_\do\%\do\~}
\def\mn@doi{\begingroup\mn@urlcharsother \@ifnextchar [ {\mn@doi@}
  {\mn@doi@[]}}
\def\mn@doi@[#1]#2{\def\@tempa{#1}\ifx\@tempa\@empty \href
  {http://dx.doi.org/#2} {doi:#2}\else \href {http://dx.doi.org/#2} {#1}\fi
  \endgroup}
\def\mn@eprint#1#2{\mn@eprint@#1:#2::\@nil}
\def\mn@eprint@arXiv#1{\href {http://arxiv.org/abs/#1} {{\tt arXiv:#1}}}
\def\mn@eprint@dblp#1{\href {http://dblp.uni-trier.de/rec/bibtex/#1.xml}
  {dblp:#1}}
\def\mn@eprint@#1:#2:#3:#4\@nil{\def\@tempa {#1}\def\@tempb {#2}\def\@tempc
  {#3}\ifx \@tempc \@empty \let \@tempc \@tempb \let \@tempb \@tempa \fi \ifx
  \@tempb \@empty \def\@tempb {arXiv}\fi \@ifundefined
  {mn@eprint@\@tempb}{\@tempb:\@tempc}{\expandafter \expandafter \csname
  mn@eprint@\@tempb\endcsname \expandafter{\@tempc}}}

\bibitem[\protect\citeauthoryear{{Abbott} et~al.,}{{Abbott}
  et~al.}{2019}]{2019PhRvX...9c1040A}
{Abbott} B.~P.,  et~al., 2019, \mn@doi [Physical Review X]
  {10.1103/PhysRevX.9.031040}, \href
  {https://ui.adsabs.harvard.edu/abs/2019PhRvX...9c1040A} {9, 031040}

\bibitem[\protect\citeauthoryear{{Abbott} et~al.,}{{Abbott}
  et~al.}{2020}]{2020PhRvD.101h4002A}
{Abbott} B.~P.,  et~al., 2020, \mn@doi [\prd] {10.1103/PhysRevD.101.084002},
  \href {https://ui.adsabs.harvard.edu/abs/2020PhRvD.101h4002A} {101, 084002}

\bibitem[\protect\citeauthoryear{{Abbott} et~al.,}{{Abbott}
  et~al.}{2021a}]{2020arXiv201014527A}
{Abbott} R.,  et~al., 2021a, \mn@doi [Physical Review X]
  {10.1103/PhysRevX.11.021053}, \href
  {https://ui.adsabs.harvard.edu/abs/2021PhRvX..11b1053A} {11, 021053}

\bibitem[\protect\citeauthoryear{{Abbott} et~al.,}{{Abbott}
  et~al.}{2021b}]{2021PhRvD.104l2004A}
{Abbott} R.,  et~al., 2021b, \mn@doi [\prd] {10.1103/PhysRevD.104.122004},
  \href {https://ui.adsabs.harvard.edu/abs/2021PhRvD.104l2004A} {104, 122004}

\bibitem[\protect\citeauthoryear{{Abdikamalov}, {Gossan}, {DeMaio}  \&
  {Ott}}{{Abdikamalov} et~al.}{2014}]{2014PhRvD..90d4001A}
{Abdikamalov} E.,  {Gossan} S.,  {DeMaio} A.~M.,   {Ott} C.~D.,  2014, \mn@doi
  [\prd] {10.1103/PhysRevD.90.044001}, \href
  {https://ui.adsabs.harvard.edu/abs/2014PhRvD..90d4001A} {90, 044001}

\bibitem[\protect\citeauthoryear{{Abdikamalov}, {Pagliaroli}  \&
  {Radice}}{{Abdikamalov} et~al.}{2020}]{2020arXiv201004356A}
{Abdikamalov} E.,  {Pagliaroli} G.,   {Radice} D.,  2020, arXiv e-prints, \href
  {https://ui.adsabs.harvard.edu/abs/2020arXiv201004356A} {p. arXiv:2010.04356}

\bibitem[\protect\citeauthoryear{{Acernese} et~al.,}{{Acernese}
  et~al.}{2015}]{2015CQGra..32b4001A}
{Acernese} F.,  et~al., 2015, \mn@doi [Classical and Quantum Gravity]
  {10.1088/0264-9381/32/2/024001}, \href
  {https://ui.adsabs.harvard.edu/abs/2015CQGra..32b4001A} {32, 024001}

\bibitem[\protect\citeauthoryear{{Aguilera-Dena}, {Langer}, {Moriya}  \&
  {Schootemeijer}}{{Aguilera-Dena} et~al.}{2018}]{2018ApJ...858..115A}
{Aguilera-Dena} D.~R.,  {Langer} N.,  {Moriya} T.~J.,   {Schootemeijer} A.,
  2018, \mn@doi [\apj] {10.3847/1538-4357/aabfc1}, \href
  {https://ui.adsabs.harvard.edu/abs/2018ApJ...858..115A} {858, 115}

\bibitem[\protect\citeauthoryear{{Aguilera-Dena}, {Langer}, {Antoniadis}  \&
  {M{\"u}ller}}{{Aguilera-Dena} et~al.}{2020}]{aguilera_20}
{Aguilera-Dena} D.~R.,  {Langer} N.,  {Antoniadis} J.,   {M{\"u}ller} B.,
  2020, \mn@doi [\apj] {10.3847/1538-4357/abb138}, \href
  {https://ui.adsabs.harvard.edu/abs/2020ApJ...901..114A} {901, 114}

\bibitem[\protect\citeauthoryear{{Akiyama}, {Wheeler}, {Meier}  \&
  {Lichtenstadt}}{{Akiyama} et~al.}{2003}]{akiyama_03}
{Akiyama} S.,  {Wheeler} J.~C.,  {Meier} D.~L.,   {Lichtenstadt} I.,  2003,
  \mn@doi [\apj] {10.1086/344135}, \href
  {https://ui.adsabs.harvard.edu/abs/2003ApJ...584..954A} {584, 954}

\bibitem[\protect\citeauthoryear{{Akutsu} et~al.,}{{Akutsu}
  et~al.}{2021}]{2021PTEP.2021eA102A}
{Akutsu} T.,  et~al., 2021, \mn@doi [Progress of Theoretical and Experimental
  Physics] {10.1093/ptep/ptab018}, \href
  {https://ui.adsabs.harvard.edu/abs/2021PTEP.2021eA102A} {2021, 05A102}

\bibitem[\protect\citeauthoryear{Aloy \& Obergaulinger}{Aloy \&
  Obergaulinger}{2020}]{10.1093/mnras/staa3273}
Aloy M.~{\'A}.,  Obergaulinger M.,  2020, Monthly Notices of the Royal
  Astronomical Society, 500, 4365

\bibitem[\protect\citeauthoryear{{Andresen}, {M{\"u}ller}, {Janka}, {Summa},
  {Gill}  \& {Zanolin}}{{Andresen} et~al.}{2019}]{andresen_19}
{Andresen} H.,  {M{\"u}ller} E.,  {Janka} H.~T.,  {Summa} A.,  {Gill} K.,
  {Zanolin} M.,  2019, \mn@doi [\mnras] {10.1093/mnras/stz990}, \href
  {https://ui.adsabs.harvard.edu/abs/2019MNRAS.486.2238A} {486, 2238}

\bibitem[\protect\citeauthoryear{{Begelman}}{{Begelman}}{1998}]{1998ApJ...493..291B}
{Begelman} M.~C.,  1998, \mn@doi [\apj] {10.1086/305119}, \href
  {https://ui.adsabs.harvard.edu/abs/1998ApJ...493..291B} {493, 291}

\bibitem[\protect\citeauthoryear{{Bisnovatyi-Kogan}, {Popov}  \&
  {Samokhin}}{{Bisnovatyi-Kogan} et~al.}{1976}]{bisnovatyi_76}
{Bisnovatyi-Kogan} G.~S.,  {Popov} I.~P.,   {Samokhin} A.~A.,  1976, \mn@doi
  [\apss] {10.1007/BF00646184}, \href
  {https://ui.adsabs.harvard.edu/abs/1976Ap&SS..41..287B} {41, 287}

\bibitem[\protect\citeauthoryear{{Blanchet}, {Damour}  \&
  {Schaefer}}{{Blanchet} et~al.}{1990}]{blanchet_90}
{Blanchet} L.,  {Damour} T.,   {Schaefer} G.,  1990, \mnras, \href
  {http://adsabs.harvard.edu/abs/1990MNRAS.242..289B} {242, 289}

\bibitem[\protect\citeauthoryear{{Blondin} \& {Mezzacappa}}{{Blondin} \&
  {Mezzacappa}}{2006}]{2006ApJ...642..401B}
{Blondin} J.~M.,  {Mezzacappa} A.,  2006, \mn@doi [\apj] {10.1086/500817},
  \href {http://adsabs.harvard.edu/abs/2006ApJ...642..401B} {642, 401}

\bibitem[\protect\citeauthoryear{{Blondin}, {Mezzacappa}  \&
  {DeMarino}}{{Blondin} et~al.}{2003}]{0004-637X-584-2-971}
{Blondin} J.~M.,  {Mezzacappa} A.,   {DeMarino} C.,  2003, \mn@doi [\apj]
  {10.1086/345812}, \href
  {https://ui.adsabs.harvard.edu/abs/2003ApJ...584..971B} {584, 971}

\bibitem[\protect\citeauthoryear{{Bugli}, {Guilet}, {Obergaulinger},
  {Cerd{\'a}-Dur{\'a}n}  \& {Aloy}}{{Bugli} et~al.}{2020}]{2020MNRAS.492...58B}
{Bugli} M.,  {Guilet} J.,  {Obergaulinger} M.,  {Cerd{\'a}-Dur{\'a}n} P.,
  {Aloy} M.~A.,  2020, \mn@doi [\mnras] {10.1093/mnras/stz3483}, \href
  {https://ui.adsabs.harvard.edu/abs/2020MNRAS.492...58B} {492, 58}

\bibitem[\protect\citeauthoryear{{Bugli}, {Guilet}  \& {Obergaulinger}}{{Bugli}
  et~al.}{2021}]{bugli_21}
{Bugli} M.,  {Guilet} J.,   {Obergaulinger} M.,  2021, \mn@doi [\mnras]
  {10.1093/mnras/stab2161}, \href
  {https://ui.adsabs.harvard.edu/abs/2021MNRAS.507..443B} {507, 443}

\bibitem[\protect\citeauthoryear{{Bugli}, {Guilet}, {Foglizzo}  \&
  {Obergaulinger}}{{Bugli} et~al.}{2022}]{bugli_22}
{Bugli} M.,  {Guilet} J.,  {Foglizzo} T.,   {Obergaulinger} M.,  2022, arXiv
  e-prints, \href {https://ui.adsabs.harvard.edu/abs/2022arXiv221005012B} {p.
  arXiv:2210.05012}

\bibitem[\protect\citeauthoryear{{Buras}, {Rampp}, {Janka}  \&
  {Kifonidis}}{{Buras} et~al.}{2006a}]{Buras2006b}
{Buras} R.,  {Rampp} M.,  {Janka} H.~T.,   {Kifonidis} K.,  2006a, \mn@doi
  [\aap] {10.1051/0004-6361:20053783}, \href
  {https://ui.adsabs.harvard.edu/abs/2006A&A...447.1049B} {447, 1049}

\bibitem[\protect\citeauthoryear{{Buras}, {Janka}, {Rampp}  \&
  {Kifonidis}}{{Buras} et~al.}{2006b}]{Buras2006}
{Buras} R.,  {Janka} H.~T.,  {Rampp} M.,   {Kifonidis} K.,  2006b, \mn@doi
  [\aap] {10.1051/0004-6361:20054654}, \href
  {https://ui.adsabs.harvard.edu/abs/2006A&A...457..281B} {457, 281}

\bibitem[\protect\citeauthoryear{{Burrows} \& {Vartanyan}}{{Burrows} \&
  {Vartanyan}}{2021}]{burrows_21}
{Burrows} A.,  {Vartanyan} D.,  2021, \mn@doi [\nat]
  {10.1038/s41586-020-03059-w}, \href
  {https://ui.adsabs.harvard.edu/abs/2021Natur.589...29B} {589, 29}

\bibitem[\protect\citeauthoryear{{Burrows}, {Dessart}, {Livne}, {Ott}  \&
  {Murphy}}{{Burrows} et~al.}{2007}]{2007ApJ...664..416B}
{Burrows} A.,  {Dessart} L.,  {Livne} E.,  {Ott} C.~D.,   {Murphy} J.,  2007,
  \mn@doi [\apj] {10.1086/519161}, \href
  {https://ui.adsabs.harvard.edu/abs/2007ApJ...664..416B} {664, 416}

\bibitem[\protect\citeauthoryear{{Chan}, {M{\"u}ller}  \& {Heger}}{{Chan}
  et~al.}{2020}]{chan_20}
{Chan} C.,  {M{\"u}ller} B.,   {Heger} A.,  2020, \mn@doi [\mnras]
  {10.1093/mnras/staa1431}, \href
  {https://ui.adsabs.harvard.edu/abs/2020MNRAS.495.3751C} {495, 3751}

\bibitem[\protect\citeauthoryear{{Dedner}, {Kemm}, {Kr{\"o}ner}, {Munz},
  {Schnitzer}  \& {Wesenberg}}{{Dedner} et~al.}{2002}]{2002JCoPh.175..645D}
{Dedner} A.,  {Kemm} F.,  {Kr{\"o}ner} D.,  {Munz} C.~D.,  {Schnitzer} T.,
  {Wesenberg} M.,  2002, \mn@doi [Journal of Computational Physics]
  {10.1006/jcph.2001.6961}, \href
  {https://ui.adsabs.harvard.edu/abs/2002JCoPh.175..645D} {175, 645}

\bibitem[\protect\citeauthoryear{{Dimmelmeier}, {Ott}, {Marek}  \&
  {Janka}}{{Dimmelmeier} et~al.}{2008}]{2008PhRvD..78f4056D}
{Dimmelmeier} H.,  {Ott} C.~D.,  {Marek} A.,   {Janka} H.~T.,  2008, \mn@doi
  [\prd] {10.1103/PhysRevD.78.064056}, \href
  {https://ui.adsabs.harvard.edu/abs/2008PhRvD..78f4056D} {78, 064056}

\bibitem[\protect\citeauthoryear{{Duncan} \& {Thompson}}{{Duncan} \&
  {Thompson}}{1992}]{duncan_92}
{Duncan} R.~C.,  {Thompson} C.,  1992, \mn@doi [\apjl] {10.1086/186413}, \href
  {https://ui.adsabs.harvard.edu/abs/1992ApJ...392L...9D} {392, L9}

\bibitem[\protect\citeauthoryear{{Edwards}}{{Edwards}}{2021}]{2021PhRvD.103b4025E}
{Edwards} M.~C.,  2021, \mn@doi [\prd] {10.1103/PhysRevD.103.024025}, \href
  {https://ui.adsabs.harvard.edu/abs/2021PhRvD.103b4025E} {103, 024025}

\bibitem[\protect\citeauthoryear{{Eichler}}{{Eichler}}{1993}]{1993ApJ...419..111E}
{Eichler} D.,  1993, \mn@doi [\apj] {10.1086/173464}, \href
  {https://ui.adsabs.harvard.edu/abs/1993ApJ...419..111E} {419, 111}

\bibitem[\protect\citeauthoryear{{Ertl}, {Woosley}, {Sukhbold}  \&
  {Janka}}{{Ertl} et~al.}{2020}]{ertl_20}
{Ertl} T.,  {Woosley} S.~E.,  {Sukhbold} T.,   {Janka} H.~T.,  2020, \mn@doi
  [\apj] {10.3847/1538-4357/ab6458}, \href
  {https://ui.adsabs.harvard.edu/abs/2020ApJ...890...51E} {890, 51}

\bibitem[\protect\citeauthoryear{{Evans} et~al.,}{{Evans}
  et~al.}{2021}]{2021arXiv210909882E}
{Evans} M.,  et~al., 2021, arXiv e-prints, \href
  {https://ui.adsabs.harvard.edu/abs/2021arXiv210909882E} {p. arXiv:2109.09882}

\bibitem[\protect\citeauthoryear{{Finn}}{{Finn}}{1989}]{finn_89}
{Finn} L.~S.,  1989, in {Evans} C.~R.,  {Finn} L.~S.,   {Hobill} D.~W.,  eds,
  Frontiers in Numerical Relativity. Cambridge University Press, Cambridge
  (UK), pp 126--145

\bibitem[\protect\citeauthoryear{{Finn} \& {Evans}}{{Finn} \&
  {Evans}}{1990}]{finn_90}
{Finn} L.~S.,  {Evans} C.~R.,  1990, \mn@doi [\apj] {10.1086/168497}, \href
  {http://adsabs.harvard.edu/abs/1990ApJ...351..588F} {351, 588}

\bibitem[\protect\citeauthoryear{{Foglizzo}, {Galletti}, {Scheck}  \&
  {Janka}}{{Foglizzo} et~al.}{2007}]{2007ApJ...654.1006F}
{Foglizzo} T.,  {Galletti} P.,  {Scheck} L.,   {Janka} H.-T.,  2007, \mn@doi
  [\apj] {10.1086/509612}, \href
  {http://adsabs.harvard.edu/abs/2007ApJ...654.1006F} {654, 1006}

\bibitem[\protect\citeauthoryear{{Fuller}, {Klion}, {Abdikamalov}  \&
  {Ott}}{{Fuller} et~al.}{2015}]{2015MNRAS.450..414F}
{Fuller} J.,  {Klion} H.,  {Abdikamalov} E.,   {Ott} C.~D.,  2015, \mn@doi
  [\mnras] {10.1093/mnras/stv698}, \href
  {https://ui.adsabs.harvard.edu/abs/2015MNRAS.450..414F} {450, 414}

\bibitem[\protect\citeauthoryear{{Gal-Yam}}{{Gal-Yam}}{2017}]{gal-yam_17}
{Gal-Yam} A.,  2017, in {Alsabti} A.~W.,  {Murdin} P.,  eds, , Handbook of
  Supernovae.
p.~195, \mn@doi{10.1007/978-3-319-21846-5_35}

\bibitem[\protect\citeauthoryear{{Gossan}, {Sutton}, {Stuver}, {Zanolin},
  {Gill}  \& {Ott}}{{Gossan} et~al.}{2016}]{2016PhRvD..93d2002G}
{Gossan} S.~E.,  {Sutton} P.,  {Stuver} A.,  {Zanolin} M.,  {Gill} K.,   {Ott}
  C.~D.,  2016, \mn@doi [\prd] {10.1103/PhysRevD.93.042002}, \href
  {https://ui.adsabs.harvard.edu/abs/2016PhRvD..93d2002G} {93, 042002}

\bibitem[\protect\citeauthoryear{{Grimmett}, {M{\"u}ller}, {Heger}, {Banerjee}
  \& {Obergaulinger}}{{Grimmett} et~al.}{2021a}]{grimmett_21}
{Grimmett} J.~J.,  {M{\"u}ller} B.,  {Heger} A.,  {Banerjee} P.,
  {Obergaulinger} M.,  2021a, \mn@doi [\mnras] {10.1093/mnras/staa3819}, \href
  {https://ui.adsabs.harvard.edu/abs/2021MNRAS.501.2764G} {501, 2764}

\bibitem[\protect\citeauthoryear{{Grimmett}, {M{\"u}ller}, {Heger}, {Banerjee}
  \& {Obergaulinger}}{{Grimmett} et~al.}{2021b}]{2021MNRAS.501.2764G}
{Grimmett} J.~J.,  {M{\"u}ller} B.,  {Heger} A.,  {Banerjee} P.,
  {Obergaulinger} M.,  2021b, \mn@doi [\mnras] {10.1093/mnras/staa3819}, \href
  {https://ui.adsabs.harvard.edu/abs/2021MNRAS.501.2764G} {501, 2764}

\bibitem[\protect\citeauthoryear{{Gurski}}{{Gurski}}{2004}]{Gurski_2004}
{Gurski} K.~F.,  2004, \mn@doi [SIAM Journal on Scientific Computing]
  {10.1137/S1064827502407962}

\bibitem[\protect\citeauthoryear{{Harris}, {Hix}, {Chertkow}, {Lee}, {Lentz}
  \& {Messer}}{{Harris} et~al.}{2017}]{harris_17}
{Harris} J.~A.,  {Hix} W.~R.,  {Chertkow} M.~A.,  {Lee} C.~T.,  {Lentz} E.~J.,
   {Messer} O.~E.~B.,  2017, \mn@doi [\apj] {10.3847/1538-4357/aa76de}, \href
  {https://ui.adsabs.harvard.edu/abs/2017ApJ...843....2H} {843, 2}

\bibitem[\protect\citeauthoryear{{Hartmann}, {Woosley}  \& {El Eid}}{{Hartmann}
  et~al.}{1985}]{hartmann_85}
{Hartmann} D.,  {Woosley} S.~E.,   {El Eid} M.~F.,  1985, \mn@doi [\apj]
  {10.1086/163580}, \href
  {https://ui.adsabs.harvard.edu/abs/1985ApJ...297..837H} {297, 837}

\bibitem[\protect\citeauthoryear{{Hild} et~al.,}{{Hild}
  et~al.}{2011}]{2011CQGra..28i4013H}
{Hild} S.,  et~al., 2011, \mn@doi [Classical and Quantum Gravity]
  {10.1088/0264-9381/28/9/094013}, \href
  {https://ui.adsabs.harvard.edu/abs/2011CQGra..28i4013H} {28, 094013}

\bibitem[\protect\citeauthoryear{{Iwamoto} et~al.,}{{Iwamoto}
  et~al.}{1998}]{iwamoto_98}
{Iwamoto} K.,  et~al., 1998, \mn@doi [\nat] {10.1038/27155}, \href
  {https://ui.adsabs.harvard.edu/abs/1998Natur.395..672I} {395, 672}

\bibitem[\protect\citeauthoryear{{Janka}, {Wongwathanarat}  \&
  {Kramer}}{{Janka} et~al.}{2022}]{janka_22}
{Janka} H.-T.,  {Wongwathanarat} A.,   {Kramer} M.,  2022, \mn@doi [\apj]
  {10.3847/1538-4357/ac403c}, \href
  {https://ui.adsabs.harvard.edu/abs/2022ApJ...926....9J} {926, 9}

\bibitem[\protect\citeauthoryear{{Jardine}, {Powell}  \&
  {M{\"u}ller}}{{Jardine} et~al.}{2022}]{Jardine_2021}
{Jardine} R.,  {Powell} J.,   {M{\"u}ller} B.,  2022, \mn@doi [\mnras]
  {10.1093/mnras/stab3763}, \href
  {https://ui.adsabs.harvard.edu/abs/2022MNRAS.510.5535J} {510, 5535}

\bibitem[\protect\citeauthoryear{{Jerkstrand} et~al.,}{{Jerkstrand}
  et~al.}{2015a}]{jerkstrand_15a}
{Jerkstrand} A.,  et~al., 2015a, \mn@doi [\mnras] {10.1093/mnras/stv087}, \href
  {https://ui.adsabs.harvard.edu/abs/2015MNRAS.448.2482J} {448, 2482}

\bibitem[\protect\citeauthoryear{{Jerkstrand} et~al.,}{{Jerkstrand}
  et~al.}{2015b}]{jerkstrand_15b}
{Jerkstrand} A.,  et~al., 2015b, \mn@doi [\apj] {10.1088/0004-637X/807/1/110},
  \href {https://ui.adsabs.harvard.edu/abs/2015ApJ...807..110J} {807, 110}

\bibitem[\protect\citeauthoryear{{Johnston}, {Hobbs}, {Vigeland}, {Kramer},
  {Weisberg}  \& {Lyne}}{{Johnston} et~al.}{2005}]{johnston_05}
{Johnston} S.,  {Hobbs} G.,  {Vigeland} S.,  {Kramer} M.,  {Weisberg} J.~M.,
  {Lyne} A.~G.,  2005, \mn@doi [\mnras] {10.1111/j.1365-2966.2005.09669.x},
  \href {http://adsabs.harvard.edu/abs/2005MNRAS.364.1397J} {364, 1397}

\bibitem[\protect\citeauthoryear{{Kalogera} et~al.,}{{Kalogera}
  et~al.}{2019}]{kalogera_19}
{Kalogera} V.,  et~al., 2019, \baas, \href
  {https://ui.adsabs.harvard.edu/abs/2019BAAS...51c.239K} {51, 239}

\bibitem[\protect\citeauthoryear{{Kimura}, {Tsunemi}, {Tomida}, {Sugizaki},
  {Ueno}, {Hanayama}, {Yoshidome}  \& {Sasaki}}{{Kimura}
  et~al.}{2013}]{kimura_13}
{Kimura} M.,  {Tsunemi} H.,  {Tomida} H.,  {Sugizaki} M.,  {Ueno} S.,
  {Hanayama} T.,  {Yoshidome} K.,   {Sasaki} M.,  2013, \mn@doi [\pasj]
  {10.1093/pasj/65.1.14}, \href
  {https://ui.adsabs.harvard.edu/abs/2013PASJ...65...14K} {65, 14}

\bibitem[\protect\citeauthoryear{{Kobayashi}, {Umeda}, {Nomoto}, {Tominaga}  \&
  {Ohkubo}}{{Kobayashi} et~al.}{2006}]{kobayashi_06}
{Kobayashi} C.,  {Umeda} H.,  {Nomoto} K.,  {Tominaga} N.,   {Ohkubo} T.,
  2006, \mn@doi [\apj] {10.1086/508914}, \href
  {https://ui.adsabs.harvard.edu/abs/2006ApJ...653.1145K} {653, 1145}

\bibitem[\protect\citeauthoryear{{Kruskal} \& {Tuck}}{{Kruskal} \&
  {Tuck}}{1958}]{kruskal_58}
{Kruskal} M.,  {Tuck} J.~L.,  1958, \mn@doi [Proceedings of the Royal Society
  of London Series A] {10.1098/rspa.1958.0079}, \href
  {https://ui.adsabs.harvard.edu/abs/1958RSPSA.245..222K} {245, 222}

\bibitem[\protect\citeauthoryear{{Kuroda}, {Takiwaki}  \& {Kotake}}{{Kuroda}
  et~al.}{2014}]{2014PhRvD..89d4011K}
{Kuroda} T.,  {Takiwaki} T.,   {Kotake} K.,  2014, \mn@doi [\prd]
  {10.1103/PhysRevD.89.044011}, \href
  {https://ui.adsabs.harvard.edu/abs/2014PhRvD..89d4011K} {89, 044011}

\bibitem[\protect\citeauthoryear{{Kuroda}, {Arcones}, {Takiwaki}  \&
  {Kotake}}{{Kuroda} et~al.}{2020}]{2020ApJ...896..102K}
{Kuroda} T.,  {Arcones} A.,  {Takiwaki} T.,   {Kotake} K.,  2020, \mn@doi
  [\apj] {10.3847/1538-4357/ab9308}, \href
  {https://ui.adsabs.harvard.edu/abs/2020ApJ...896..102K} {896, 102}

\bibitem[\protect\citeauthoryear{{LIGO Scientific Collaboration} et~al.,}{{LIGO
  Scientific Collaboration} et~al.}{2015}]{2015CQGra..32g4001L}
{LIGO Scientific Collaboration} et~al., 2015, \mn@doi [Classical and Quantum
  Gravity] {10.1088/0264-9381/32/7/074001}, \href
  {https://ui.adsabs.harvard.edu/abs/2015CQGra..32g4001L} {32, 074001}

\bibitem[\protect\citeauthoryear{{Lattimer} \& {Schutz}}{{Lattimer} \&
  {Schutz}}{2005}]{lattimer_05}
{Lattimer} J.~M.,  {Schutz} B.~F.,  2005, \mn@doi [\apj] {10.1086/431543},
  \href {http://adsabs.harvard.edu/abs/2005ApJ...629..979L} {629, 979}

\bibitem[\protect\citeauthoryear{{Lattimer} \& {Swesty}}{{Lattimer} \&
  {Swesty}}{1991}]{1991NuPhA.535..331L}
{Lattimer} J.~M.,  {Swesty} D.~F.,  1991, \mn@doi [\nphysa]
  {10.1016/0375-9474(91)90452-C}, \href
  {https://ui.adsabs.harvard.edu/abs/1991NuPhA.535..331L} {535, 331}

\bibitem[\protect\citeauthoryear{{Lopez}, {Ramirez-Ruiz}, {Castro}  \&
  {Pearson}}{{Lopez} et~al.}{2013}]{lopez_13}
{Lopez} L.~A.,  {Ramirez-Ruiz} E.,  {Castro} D.,   {Pearson} S.,  2013, \mn@doi
  [\apj] {10.1088/0004-637X/764/1/50}, \href
  {https://ui.adsabs.harvard.edu/abs/2013ApJ...764...50L} {764, 50}

\bibitem[\protect\citeauthoryear{{MacFadyen} \& {Woosley}}{{MacFadyen} \&
  {Woosley}}{1999}]{macfadyen_99}
{MacFadyen} A.~I.,  {Woosley} S.~E.,  1999, \mn@doi [\apj] {10.1086/307790},
  \href {https://ui.adsabs.harvard.edu/abs/1999ApJ...524..262M} {524, 262}

\bibitem[\protect\citeauthoryear{{Maeda}, {Nakamura}, {Nomoto}, {Mazzali},
  {Patat}  \& {Hachisu}}{{Maeda} et~al.}{2002}]{maeda_02}
{Maeda} K.,  {Nakamura} T.,  {Nomoto} K.,  {Mazzali} P.~A.,  {Patat} F.,
  {Hachisu} I.,  2002, \mn@doi [\apj] {10.1086/324487}, \href
  {https://ui.adsabs.harvard.edu/abs/2002ApJ...565..405M} {565, 405}

\bibitem[\protect\citeauthoryear{Maeda et~al.,}{Maeda
  et~al.}{2008}]{maeda_2008}
Maeda K.,  et~al., 2008, \mn@doi [Science] {10.1126/science.1149437}, 319, 1220

\bibitem[\protect\citeauthoryear{{Mazzali}, {Nomoto}, {Patat}  \&
  {Maeda}}{{Mazzali} et~al.}{2001}]{mazzali_01}
{Mazzali} P.~A.,  {Nomoto} K.,  {Patat} F.,   {Maeda} K.,  2001, \mn@doi [\apj]
  {10.1086/322420}, \href
  {https://ui.adsabs.harvard.edu/abs/2001ApJ...559.1047M} {559, 1047}

\bibitem[\protect\citeauthoryear{{Mazzali} et~al.,}{{Mazzali}
  et~al.}{2007}]{mazzali_07}
{Mazzali} P.~A.,  et~al., 2007, \mn@doi [\apj] {10.1086/521873}, \href
  {https://ui.adsabs.harvard.edu/abs/2007ApJ...670..592M} {670, 592}

\bibitem[\protect\citeauthoryear{{Milisavljevic} et~al.,}{{Milisavljevic}
  et~al.}{2015}]{milisavljevic_15}
{Milisavljevic} D.,  et~al., 2015, \mn@doi [\apj] {10.1088/0004-637X/799/1/51},
  \href {https://ui.adsabs.harvard.edu/abs/2015ApJ...799...51M} {799, 51}

\bibitem[\protect\citeauthoryear{{Miyoshi} \& {Kusano}}{{Miyoshi} \&
  {Kusano}}{2005}]{2005JCoPh.208..315M}
{Miyoshi} T.,  {Kusano} K.,  2005, \mn@doi [Journal of Computational Physics]
  {10.1016/j.jcp.2005.02.017}, \href
  {https://ui.adsabs.harvard.edu/abs/2005JCoPh.208..315M} {208, 315}

\bibitem[\protect\citeauthoryear{{Modjaz}, {Liu}, {Bianco}  \&
  {Graur}}{{Modjaz} et~al.}{2016}]{modjaz_16}
{Modjaz} M.,  {Liu} Y.~Q.,  {Bianco} F.~B.,   {Graur} O.,  2016, \mn@doi [\apj]
  {10.3847/0004-637X/832/2/108}, \href
  {https://ui.adsabs.harvard.edu/abs/2016ApJ...832..108M} {832, 108}

\bibitem[\protect\citeauthoryear{{Morozova}, {Radice}, {Burrows}  \&
  {Vartanyan}}{{Morozova} et~al.}{2018}]{2018ApJ...861...10M}
{Morozova} V.,  {Radice} D.,  {Burrows} A.,   {Vartanyan} D.,  2018, \mn@doi
  [\apj] {10.3847/1538-4357/aac5f1}, \href
  {https://ui.adsabs.harvard.edu/abs/2018ApJ...861...10M} {861, 10}

\bibitem[\protect\citeauthoryear{{M{\"o}sta} et~al.,}{{M{\"o}sta}
  et~al.}{2014}]{moesta_14b}
{M{\"o}sta} P.,  et~al., 2014, \mn@doi [\apjl] {10.1088/2041-8205/785/2/L29},
  \href {http://adsabs.harvard.edu/abs/2014ApJ...785L..29M} {785, L29}

\bibitem[\protect\citeauthoryear{{M{\"o}sta}, {Ott}, {Radice}, {Roberts},
  {Schnetter}  \& {Haas}}{{M{\"o}sta} et~al.}{2015}]{moesta_15}
{M{\"o}sta} P.,  {Ott} C.~D.,  {Radice} D.,  {Roberts} L.~F.,  {Schnetter} E.,
   {Haas} R.,  2015, \mn@doi [\nat] {10.1038/nature15755}, \href
  {https://ui.adsabs.harvard.edu/abs/2015Natur.528..376M} {528, 376}

\bibitem[\protect\citeauthoryear{{M{\"o}sta}, {Roberts}, {Halevi}, {Ott},
  {Lippuner}, {Haas}  \& {Schnetter}}{{M{\"o}sta} et~al.}{2018}]{moesta_18}
{M{\"o}sta} P.,  {Roberts} L.~F.,  {Halevi} G.,  {Ott} C.~D.,  {Lippuner} J.,
  {Haas} R.,   {Schnetter} E.,  2018, \mn@doi [\apj]
  {10.3847/1538-4357/aad6ec}, \href
  {https://ui.adsabs.harvard.edu/abs/2018ApJ...864..171M} {864, 171}

\bibitem[\protect\citeauthoryear{{M{\"u}ller}}{{M{\"u}ller}}{2020}]{mueller_20}
{M{\"u}ller} B.,  2020, \mn@doi [Living Reviews in Computational Astrophysics]
  {10.1007/s41115-020-0008-5}, \href
  {https://ui.adsabs.harvard.edu/abs/2020LRCA....6....3M} {6, 3}

\bibitem[\protect\citeauthoryear{{M{\"u}ller} \& {Janka}}{{M{\"u}ller} \&
  {Janka}}{2014}]{2014ApJ...788...82M}
{M{\"u}ller} B.,  {Janka} H.-T.,  2014, \mn@doi [\apj]
  {10.1088/0004-637X/788/1/82}, \href
  {https://ui.adsabs.harvard.edu/abs/2014ApJ...788...82M} {788, 82}

\bibitem[\protect\citeauthoryear{{M{\"u}ller} \& {Janka}}{{M{\"u}ller} \&
  {Janka}}{2015}]{2015MNRAS.448.2141M}
{M{\"u}ller} B.,  {Janka} H.~T.,  2015, \mn@doi [\mnras]
  {10.1093/mnras/stv101}, \href
  {https://ui.adsabs.harvard.edu/abs/2015MNRAS.448.2141M} {448, 2141}

\bibitem[\protect\citeauthoryear{{M{\"u}ller} \& {Varma}}{{M{\"u}ller} \&
  {Varma}}{2020}]{2020MNRAS.498L.109M}
{M{\"u}ller} B.,  {Varma} V.,  2020, \mn@doi [\mnras] {10.1093/mnrasl/slaa137},
  \href {https://ui.adsabs.harvard.edu/abs/2020MNRAS.498L.109M} {498, L109}

\bibitem[\protect\citeauthoryear{{M{\"u}ller}, {Dimmelmeier}  \&
  {M{\"u}ller}}{{M{\"u}ller} et~al.}{2008}]{mueller_08}
{M{\"u}ller} B.,  {Dimmelmeier} H.,   {M{\"u}ller} E.,  2008, \mn@doi [\aap]
  {10.1051/0004-6361:200809609}, \href
  {https://ui.adsabs.harvard.edu/abs/2008A&A...489..301M} {489, 301}

\bibitem[\protect\citeauthoryear{{M{\"u}ller}, {Janka}  \&
  {Marek}}{{M{\"u}ller} et~al.}{2013}]{mueller_13}
{M{\"u}ller} B.,  {Janka} H.-T.,   {Marek} A.,  2013, \mn@doi [\apj]
  {10.1088/0004-637X/766/1/43}, \href
  {https://ui.adsabs.harvard.edu/abs/2013ApJ...766...43M} {766, 43}

\bibitem[\protect\citeauthoryear{{M{\"u}ller} et~al.,}{{M{\"u}ller}
  et~al.}{2019}]{mueller_19}
{M{\"u}ller} B.,  et~al., 2019, \mn@doi [\mnras] {10.1093/mnras/stz216}, \href
  {https://ui.adsabs.harvard.edu/abs/2019MNRAS.484.3307M} {484, 3307}

\bibitem[\protect\citeauthoryear{Müller, Janka  \& Dimmelmeier}{Müller
  et~al.}{2010}]{M_ller_2010}
Müller B.,  Janka H.-T.,   Dimmelmeier H.,  2010, \mn@doi [The Astrophysical
  Journal Supplement Series] {10.1088/0067-0049/189/1/104}, 189, 104

\bibitem[\protect\citeauthoryear{{Nakamura}, {Mazzali}, {Nomoto}  \&
  {Iwamoto}}{{Nakamura} et~al.}{2001}]{nakamura_01}
{Nakamura} T.,  {Mazzali} P.~A.,  {Nomoto} K.,   {Iwamoto} K.,  2001, \mn@doi
  [\apj] {10.1086/319784}, \href
  {https://ui.adsabs.harvard.edu/abs/2001ApJ...550..991N} {550, 991}

\bibitem[\protect\citeauthoryear{{Ng} \& {Romani}}{{Ng} \&
  {Romani}}{2007}]{ng_07}
{Ng} C.-Y.,  {Romani} R.~W.,  2007, \mn@doi [\apj] {10.1086/513597}, \href
  {http://adsabs.harvard.edu/abs/2007ApJ...660.1357N} {660, 1357}

\bibitem[\protect\citeauthoryear{{Nomoto}, {Tominaga}, {Umeda}, {Kobayashi}  \&
  {Maeda}}{{Nomoto} et~al.}{2006}]{nomoto_06}
{Nomoto} K.,  {Tominaga} N.,  {Umeda} H.,  {Kobayashi} C.,   {Maeda} K.,  2006,
  \mn@doi [\nphysa] {10.1016/j.nuclphysa.2006.05.008}, \href
  {https://ui.adsabs.harvard.edu/abs/2006NuPhA.777..424N} {777, 424}

\bibitem[\protect\citeauthoryear{{Noutsos}, {Kramer}, {Carr}  \&
  {Johnston}}{{Noutsos} et~al.}{2012}]{noutsos_12}
{Noutsos} A.,  {Kramer} M.,  {Carr} P.,   {Johnston} S.,  2012, \mn@doi
  [\mnras] {10.1111/j.1365-2966.2012.21083.x}, \href
  {http://adsabs.harvard.edu/abs/2012MNRAS.423.2736N} {423, 2736}

\bibitem[\protect\citeauthoryear{{Noutsos}, {Schnitzeler}, {Keane}, {Kramer}
  \& {Johnston}}{{Noutsos} et~al.}{2013}]{noutsos_13}
{Noutsos} A.,  {Schnitzeler} D.~H.~F.~M.,  {Keane} E.~F.,  {Kramer} M.,
  {Johnston} S.,  2013, \mn@doi [\mnras] {10.1093/mnras/stt047}, \href
  {http://adsabs.harvard.edu/abs/2013MNRAS.430.2281N} {430, 2281}

\bibitem[\protect\citeauthoryear{{O'Connor} \& {Ott}}{{O'Connor} \&
  {Ott}}{2011}]{oconnor_11}
{O'Connor} E.,  {Ott} C.~D.,  2011, \mn@doi [\apj]
  {10.1088/0004-637X/730/2/70}, \href
  {https://ui.adsabs.harvard.edu/abs/2011ApJ...730...70O} {730, 70}

\bibitem[\protect\citeauthoryear{{Obergaulinger} \& {Aloy}}{{Obergaulinger} \&
  {Aloy}}{2017}]{obergaulinger_17}
{Obergaulinger} M.,  {Aloy} M.~{\'A}.,  2017, \mn@doi [\mnras]
  {10.1093/mnrasl/slx046}, \href
  {https://ui.adsabs.harvard.edu/abs/2017MNRAS.469L..43O} {469, L43}

\bibitem[\protect\citeauthoryear{{Obergaulinger} \& {Aloy}}{{Obergaulinger} \&
  {Aloy}}{2020}]{2020MNRAS.492.4613O}
{Obergaulinger} M.,  {Aloy} M.~{\'A}.,  2020, \mn@doi [\mnras]
  {10.1093/mnras/staa096}, \href
  {https://ui.adsabs.harvard.edu/abs/2020MNRAS.492.4613O} {492, 4613}

\bibitem[\protect\citeauthoryear{{Obergaulinger} \& {Aloy}}{{Obergaulinger} \&
  {Aloy}}{2021a}]{2020arXiv200807205O}
{Obergaulinger} M.,  {Aloy} M.~{\'A}.,  2021a, \mn@doi [\mnras]
  {10.1093/mnras/stab295}, \href
  {https://ui.adsabs.harvard.edu/abs/2021MNRAS.503.4942O} {503, 4942}

\bibitem[\protect\citeauthoryear{{Obergaulinger} \& {Aloy}}{{Obergaulinger} \&
  {Aloy}}{2021b}]{obergaulinger_21}
{Obergaulinger} M.,  {Aloy} M.~{\'A}.,  2021b, \mn@doi [\mnras]
  {10.1093/mnras/stab295}, \href
  {https://ui.adsabs.harvard.edu/abs/2021MNRAS.503.4942O} {503, 4942}

\bibitem[\protect\citeauthoryear{{Obergaulinger} \& {Aloy}}{{Obergaulinger} \&
  {Aloy}}{2022a}]{2022MNRAS.512.2489O}
{Obergaulinger} M.,  {Aloy} M.~{\'A}.,  2022a, \mn@doi [\mnras]
  {10.1093/mnras/stac613}, \href
  {https://ui.adsabs.harvard.edu/abs/2022MNRAS.512.2489O} {512, 2489}

\bibitem[\protect\citeauthoryear{{Obergaulinger} \& {Aloy}}{{Obergaulinger} \&
  {Aloy}}{2022b}]{obergaulinger_22}
{Obergaulinger} M.,  {Aloy} M.~{\'A}.,  2022b, \mn@doi [\mnras]
  {10.1093/mnras/stac613}, \href
  {https://ui.adsabs.harvard.edu/abs/2022MNRAS.512.2489O} {512, 2489}

\bibitem[\protect\citeauthoryear{{Obergaulinger}, {Aloy}  \&
  {M{\"u}ller}}{{Obergaulinger} et~al.}{2006}]{obergaulinger_06}
{Obergaulinger} M.,  {Aloy} M.~A.,   {M{\"u}ller} E.,  2006, \mn@doi [\aap]
  {10.1051/0004-6361:20054306}, \href
  {https://ui.adsabs.harvard.edu/abs/2006A&A...450.1107O} {450, 1107}

\bibitem[\protect\citeauthoryear{{Obergaulinger}, {Janka}  \&
  {Aloy}}{{Obergaulinger} et~al.}{2014}]{obergaulinger_14}
{Obergaulinger} M.,  {Janka} H.-T.,   {Aloy} M.~A.,  2014, \mn@doi [\mnras]
  {10.1093/mnras/stu1969}, \href
  {http://adsabs.harvard.edu/abs/2014MNRAS.445.3169O} {445, 3169}

\bibitem[\protect\citeauthoryear{{Pajkos}, {Couch}, {Pan}  \&
  {O'Connor}}{{Pajkos} et~al.}{2019}]{pajkos_19}
{Pajkos} M.~A.,  {Couch} S.~M.,  {Pan} K.-C.,   {O'Connor} E.~P.,  2019,
  \mn@doi [\apj] {10.3847/1538-4357/ab1de2}, \href
  {https://ui.adsabs.harvard.edu/abs/2019ApJ...878...13P} {878, 13}

\bibitem[\protect\citeauthoryear{{Paxton}, {Bildsten}, {Dotter}, {Herwig},
  {Lesaffre}  \& {Timmes}}{{Paxton} et~al.}{2011}]{2011ApJS..192....3P}
{Paxton} B.,  {Bildsten} L.,  {Dotter} A.,  {Herwig} F.,  {Lesaffre} P.,
  {Timmes} F.,  2011, \mn@doi [\apjs] {10.1088/0067-0049/192/1/3}, \href
  {https://ui.adsabs.harvard.edu/abs/2011ApJS..192....3P} {192, 3}

\bibitem[\protect\citeauthoryear{{Paxton} et~al.,}{{Paxton}
  et~al.}{2013}]{paxton_13}
{Paxton} B.,  et~al., 2013, \mn@doi [\apjs] {10.1088/0067-0049/208/1/4}, \href
  {https://ui.adsabs.harvard.edu/abs/2013ApJS..208....4P} {208, 4}

\bibitem[\protect\citeauthoryear{{Paxton} et~al.,}{{Paxton}
  et~al.}{2015}]{paxton_15}
{Paxton} B.,  et~al., 2015, \mn@doi [\apjs] {10.1088/0067-0049/220/1/15}, \href
  {https://ui.adsabs.harvard.edu/abs/2015ApJS..220...15P} {220, 15}

\bibitem[\protect\citeauthoryear{{Paxton} et~al.,}{{Paxton}
  et~al.}{2018}]{paxton_18}
{Paxton} B.,  et~al., 2018, \mn@doi [\apjs] {10.3847/1538-4365/aaa5a8}, \href
  {https://ui.adsabs.harvard.edu/abs/2018ApJS..234...34P} {234, 34}

\bibitem[\protect\citeauthoryear{{Powell} \& {M{\"u}ller}}{{Powell} \&
  {M{\"u}ller}}{2019}]{2019MNRAS.487.1178P}
{Powell} J.,  {M{\"u}ller} B.,  2019, \mn@doi [\mnras] {10.1093/mnras/stz1304},
  \href {https://ui.adsabs.harvard.edu/abs/2019MNRAS.487.1178P} {487, 1178}

\bibitem[\protect\citeauthoryear{{Powell} \& {M{\"u}ller}}{{Powell} \&
  {M{\"u}ller}}{2020}]{2020MNRAS.494.4665P}
{Powell} J.,  {M{\"u}ller} B.,  2020, \mn@doi [\mnras]
  {10.1093/mnras/staa1048}, \href
  {https://ui.adsabs.harvard.edu/abs/2020MNRAS.494.4665P} {494, 4665}

\bibitem[\protect\citeauthoryear{{Rampp} \& {Janka}}{{Rampp} \&
  {Janka}}{2002}]{2002A&A...396..361R}
{Rampp} M.,  {Janka} H.~T.,  2002, \mn@doi [\aap] {10.1051/0004-6361:20021398},
  \href {https://ui.adsabs.harvard.edu/abs/2002A&A...396..361R} {396, 361}

\bibitem[\protect\citeauthoryear{{Raynaud}, {Cerd{\'a}-Dur{\'a}n}  \&
  {Guilet}}{{Raynaud} et~al.}{2022}]{2021arXiv210312445R}
{Raynaud} R.,  {Cerd{\'a}-Dur{\'a}n} P.,   {Guilet} J.,  2022, \mn@doi [\mnras]
  {10.1093/mnras/stab3109}, \href
  {https://ui.adsabs.harvard.edu/abs/2022MNRAS.509.3410R} {509, 3410}

\bibitem[\protect\citeauthoryear{{Reichert}, {Obergaulinger}, {Eichler}, {Aloy}
   \& {Arcones}}{{Reichert} et~al.}{2021}]{2021MNRAS.tmp...72R}
{Reichert} M.,  {Obergaulinger} M.,  {Eichler} M.,  {Aloy} M.~{\'A}.,
  {Arcones} A.,  2021, \mn@doi [\mnras] {10.1093/mnras/stab029}, \href
  {https://ui.adsabs.harvard.edu/abs/2021MNRAS.tmp...72R} {}

\bibitem[\protect\citeauthoryear{{Reichert}, {Obergaulinger}, {Aloy}, {Gabler},
  {Arcones}  \& {Thielemann}}{{Reichert} et~al.}{2022}]{reichert_22}
{Reichert} M.,  {Obergaulinger} M.,  {Aloy} M.~{\'A}.,  {Gabler} M.,  {Arcones}
  A.,   {Thielemann} F.~K.,  2022, \mn@doi [\mnras] {10.1093/mnras/stac3185},
  \href {https://ui.adsabs.harvard.edu/abs/2022MNRAS.tmp.2984R} {}

\bibitem[\protect\citeauthoryear{{Richers}, {Ott}, {Abdikamalov}, {O'Connor}
  \& {Sullivan}}{{Richers} et~al.}{2017}]{2017PhRvD..95f3019R}
{Richers} S.,  {Ott} C.~D.,  {Abdikamalov} E.,  {O'Connor} E.,   {Sullivan} C.,
   2017, \mn@doi [\prd] {10.1103/PhysRevD.95.063019}, \href
  {https://ui.adsabs.harvard.edu/abs/2017PhRvD..95f3019R} {95, 063019}

\bibitem[\protect\citeauthoryear{{Scheck}, {Kifonidis}, {Janka}  \&
  {M{\"u}ller}}{{Scheck} et~al.}{2006}]{scheck_06}
{Scheck} L.,  {Kifonidis} K.,  {Janka} H.~T.,   {M{\"u}ller} E.,  2006, \mn@doi
  [\aap] {10.1051/0004-6361:20064855}, \href
  {https://ui.adsabs.harvard.edu/abs/2006A&A...457..963S} {457, 963}

\bibitem[\protect\citeauthoryear{{Scheidegger}, {Fischer}, {Whitehouse}  \&
  {Liebend{\"o}rfer}}{{Scheidegger} et~al.}{2008}]{Scheidegger2008}
{Scheidegger} S.,  {Fischer} T.,  {Whitehouse} S.~C.,   {Liebend{\"o}rfer} M.,
  2008, \mn@doi [\aap] {10.1051/0004-6361:20078577}, \href
  {https://ui.adsabs.harvard.edu/abs/2008A&A...490..231S} {490, 231}

\bibitem[\protect\citeauthoryear{{Shibagaki}, {Kuroda}, {Kotake}  \&
  {Takiwaki}}{{Shibagaki} et~al.}{2020}]{shibagaki_20}
{Shibagaki} S.,  {Kuroda} T.,  {Kotake} K.,   {Takiwaki} T.,  2020, \mn@doi
  [\mnras] {10.1093/mnrasl/slaa021}, \href
  {https://ui.adsabs.harvard.edu/abs/2020MNRAS.493L.138S} {493, L138}

\bibitem[\protect\citeauthoryear{{Sieverding}, {M{\"u}ller}  \&
  {Qian}}{{Sieverding} et~al.}{2020}]{sieverding_20}
{Sieverding} A.,  {M{\"u}ller} B.,   {Qian} Y.~Z.,  2020, \mn@doi [\apj]
  {10.3847/1538-4357/abc61b}, \href
  {https://ui.adsabs.harvard.edu/abs/2020ApJ...904..163S} {904, 163}

\bibitem[\protect\citeauthoryear{{Sotani} \& {Takiwaki}}{{Sotani} \&
  {Takiwaki}}{2016}]{sotani_16}
{Sotani} H.,  {Takiwaki} T.,  2016, \mn@doi [\prd]
  {10.1103/PhysRevD.94.044043}, \href
  {https://ui.adsabs.harvard.edu/abs/2016PhRvD..94d4043S} {94, 044043}

\bibitem[\protect\citeauthoryear{{Sotani}, {Takiwaki}  \& {Togashi}}{{Sotani}
  et~al.}{2021}]{2021PhRvD.104l3009S}
{Sotani} H.,  {Takiwaki} T.,   {Togashi} H.,  2021, \mn@doi [\prd]
  {10.1103/PhysRevD.104.123009}, \href
  {https://ui.adsabs.harvard.edu/abs/2021PhRvD.104l3009S} {104, 123009}

\bibitem[\protect\citeauthoryear{{Stockinger} et~al.,}{{Stockinger}
  et~al.}{2020}]{stockinger_20}
{Stockinger} G.,  et~al., 2020, \mn@doi [\mnras] {10.1093/mnras/staa1691},
  \href {https://ui.adsabs.harvard.edu/abs/2020MNRAS.496.2039S} {496, 2039}

\bibitem[\protect\citeauthoryear{{Suwa}, {Takiwaki}, {Kotake}  \&
  {Sato}}{{Suwa} et~al.}{2007}]{Suwa2007}
{Suwa} Y.,  {Takiwaki} T.,  {Kotake} K.,   {Sato} K.,  2007, \mn@doi [\pasj]
  {10.1093/pasj/59.4.771}, \href
  {https://ui.adsabs.harvard.edu/abs/2007PASJ...59..771S} {59, 771}

\bibitem[\protect\citeauthoryear{{Szczepa{\'n}czyk} et~al.,}{{Szczepa{\'n}czyk}
  et~al.}{2021}]{Szczepanczyk_2021}
{Szczepa{\'n}czyk} M.~J.,  et~al., 2021, \mn@doi [\prd]
  {10.1103/PhysRevD.104.102002}, \href
  {https://ui.adsabs.harvard.edu/abs/2021PhRvD.104j2002S} {104, 102002}

\bibitem[\protect\citeauthoryear{{Takahashi} \& {Langer}}{{Takahashi} \&
  {Langer}}{2021}]{takahashi_21}
{Takahashi} K.,  {Langer} N.,  2021, \mn@doi [\aap]
  {10.1051/0004-6361/202039253}, \href
  {https://ui.adsabs.harvard.edu/abs/2021A&A...646A..19T} {646, A19}

\bibitem[\protect\citeauthoryear{{Takiwaki} \& {Kotake}}{{Takiwaki} \&
  {Kotake}}{2011}]{takiwaki_11}
{Takiwaki} T.,  {Kotake} K.,  2011, \mn@doi [\apj]
  {10.1088/0004-637X/743/1/30}, \href
  {https://ui.adsabs.harvard.edu/abs/2011ApJ...743...30T} {743, 30}

\bibitem[\protect\citeauthoryear{{Tanaka}, {Maeda}, {Mazzali}, {Kawabata}  \&
  {Nomoto}}{{Tanaka} et~al.}{2017}]{tanaka_2017}
{Tanaka} M.,  {Maeda} K.,  {Mazzali} P.~A.,  {Kawabata} K.~S.,   {Nomoto} K.,
  2017, \mn@doi [\apj] {10.3847/1538-4357/aa6035}, \href
  {https://ui.adsabs.harvard.edu/abs/2017ApJ...837..105T} {837, 105}

\bibitem[\protect\citeauthoryear{{The LIGO Scientific Collaboration}
  et~al.,}{{The LIGO Scientific Collaboration}
  et~al.}{2021}]{2021arXiv211103606T}
{The LIGO Scientific Collaboration} et~al., 2021, arXiv e-prints, \href
  {https://ui.adsabs.harvard.edu/abs/2021arXiv211103606T} {p. arXiv:2111.03606}

\bibitem[\protect\citeauthoryear{{Torres-Forn{\'e}}, {Cerd{\'a}-Dur{\'a}n},
  {Passamonti}  \& {Font}}{{Torres-Forn{\'e}} et~al.}{2018}]{torres_forne_18}
{Torres-Forn{\'e}} A.,  {Cerd{\'a}-Dur{\'a}n} P.,  {Passamonti} A.,   {Font}
  J.~A.,  2018, \mn@doi [\mnras] {10.1093/mnras/stx3067}, \href
  {https://ui.adsabs.harvard.edu/abs/2018MNRAS.474.5272T} {474, 5272}

\bibitem[\protect\citeauthoryear{{Torres-Forn{\'e}}, {Cerd{\'a}-Dur{\'a}n},
  {Obergaulinger}, {M{\"u}ller}  \& {Font}}{{Torres-Forn{\'e}}
  et~al.}{2019}]{torres-forne_universal_2019}
{Torres-Forn{\'e}} A.,  {Cerd{\'a}-Dur{\'a}n} P.,  {Obergaulinger} M.,
  {M{\"u}ller} B.,   {Font} J.~A.,  2019, \mn@doi [\prl]
  {10.1103/PhysRevLett.123.051102}, \href
  {https://ui.adsabs.harvard.edu/abs/2019PhRvL.123e1102T} {123, 051102}

\bibitem[\protect\citeauthoryear{{Usov}}{{Usov}}{1992}]{usov_92}
{Usov} V.~V.,  1992, \mn@doi [\nat] {10.1038/357472a0}, \href
  {https://ui.adsabs.harvard.edu/abs/1992Natur.357..472U} {357, 472}

\bibitem[\protect\citeauthoryear{{Varma} \& {M{\"u}ller}}{{Varma} \&
  {M{\"u}ller}}{2021}]{2021arXiv210100213V}
{Varma} V.,  {M{\"u}ller} B.,  2021, \mn@doi [\mnras] {10.1093/mnras/stab883},
  \href {https://ui.adsabs.harvard.edu/abs/2021MNRAS.504..636V} {504, 636}

\bibitem[\protect\citeauthoryear{{Varma}, {M{\"u}ller}  \&
  {Obergaulinger}}{{Varma} et~al.}{2021}]{Varma2021}
{Varma} V.,  {M{\"u}ller} B.,   {Obergaulinger} M.,  2021, \mn@doi [\mnras]
  {10.1093/mnras/stab2983}, \href
  {https://ui.adsabs.harvard.edu/abs/2021MNRAS.508.6033V} {508, 6033}

\bibitem[\protect\citeauthoryear{{Varma}, {Mueller}  \& {Schneider}}{{Varma}
  et~al.}{2022}]{varma_22}
{Varma} V.,  {Mueller} B.,   {Schneider} F. R.~N.,  2022, arXiv e-prints, \href
  {https://ui.adsabs.harvard.edu/abs/2022arXiv220411009V} {p. arXiv:2204.11009}

\bibitem[\protect\citeauthoryear{{Wanajo}, {M{\"u}ller}, {Janka}  \&
  {Heger}}{{Wanajo} et~al.}{2018}]{wanajo_18}
{Wanajo} S.,  {M{\"u}ller} B.,  {Janka} H.-T.,   {Heger} A.,  2018, \mn@doi
  [\apj] {10.3847/1538-4357/aa9d97}, \href
  {https://ui.adsabs.harvard.edu/abs/2018ApJ...852...40W} {852, 40}

\bibitem[\protect\citeauthoryear{{Winteler}, {K{\"a}ppeli}, {Perego},
  {Arcones}, {Vasset}, {Nishimura}, {Liebend{\"o}rfer}  \&
  {Thielemann}}{{Winteler} et~al.}{2012}]{2012ApJ...750L..22W}
{Winteler} C.,  {K{\"a}ppeli} R.,  {Perego} A.,  {Arcones} A.,  {Vasset} N.,
  {Nishimura} N.,  {Liebend{\"o}rfer} M.,   {Thielemann} F.~K.,  2012, \mn@doi
  [\apjl] {10.1088/2041-8205/750/1/L22}, \href
  {https://ui.adsabs.harvard.edu/abs/2012ApJ...750L..22W} {750, L22}

\bibitem[\protect\citeauthoryear{{Wongwathanarat}, {Janka}  \&
  {M{\"u}ller}}{{Wongwathanarat} et~al.}{2013}]{wongwathanarat_13}
{Wongwathanarat} A.,  {Janka} H.~T.,   {M{\"u}ller} E.,  2013, \mn@doi [\aap]
  {10.1051/0004-6361/201220636}, \href
  {https://ui.adsabs.harvard.edu/abs/2013A&A...552A.126W} {552, A126}

\bibitem[\protect\citeauthoryear{{Woosley}}{{Woosley}}{2010}]{woosley_2010}
{Woosley} S.~E.,  2010, \mn@doi [\apjl] {10.1088/2041-8205/719/2/L204}, \href
  {https://ui.adsabs.harvard.edu/abs/2010ApJ...719L.204W} {719, L204}

\bibitem[\protect\citeauthoryear{{Woosley} \& {Bloom}}{{Woosley} \&
  {Bloom}}{2006}]{woosley_06}
{Woosley} S.~E.,  {Bloom} J.~S.,  2006, \mn@doi [\araa]
  {10.1146/annurev.astro.43.072103.150558}, \href
  {https://ui.adsabs.harvard.edu/abs/2006ARA&A..44..507W} {44, 507}

\bibitem[\protect\citeauthoryear{{Woosley}, {Eastman}  \& {Schmidt}}{{Woosley}
  et~al.}{1999}]{woosley_99}
{Woosley} S.~E.,  {Eastman} R.~G.,   {Schmidt} B.~P.,  1999, \mn@doi [\apj]
  {10.1086/307131}, \href
  {https://ui.adsabs.harvard.edu/abs/1999ApJ...516..788W} {516, 788}

\makeatother
\end{thebibliography}

%\begin{thebibliography}{99}
%\end{thebibliography}

%%%%%%%%%%%%%%%%%%%%%%%%%%%%%%%%%%%%%%%%%%%%%%%%%%

%%%%%%%%%%%%%%%%% APPENDICES %%%%%%%%%%%%%%%%%%%%%
%
%\appendix

%%%%%%%%%%%%%%%%%%%%%%%%%%%%%%%%%%%%%%%%%%%%%%%%%%

% Don't change these lines
\bsp	% typesetting comment
\label{lastpage}
\end{document}